\documentclass[twocolumn]{aastex701}

\usepackage{bmpsize}
\usepackage{comment}
\usepackage{adjustbox}
\usepackage{amsmath}
\usepackage{booktabs}
\usepackage{appendix}
\usepackage{soul}

\newcommand{\nsne}{68}

\newcommand{\hotpants}{{\tt Hotpants}}

\bibpunct[, ]{(}{)}{;}{a}{,}{,}

\begin{document}

\title{Expanding the High-$z$~Supernova Frontier:\\ ``Wide-Area'' JWST Discoveries from the First Two Years of COSMOS-Web\\}

\author[0000-0003-2238-1572]{Ori D.Fox}
\affiliation{Space Telescope Science Institute, 3700 San Martin Drive, Baltimore, MD 21218, USA}
\email{ofox@stsci.edu}

\author[0000-0002-4410-5387]{Armin Rest}
\affiliation{Space Telescope Science Institute, 3700 San Martin Drive, Baltimore, MD 21218, USA}
\affiliation{William H. Miller III Department of Physics \& Astronomy, Johns Hopkins University, 3701 San Martin Dr, Baltimore, MD 21218, USA}
\email{arest@stsci.edu}

\author[0000-0002-2361-7201]{Justin~D.~R.~Pierel}
\affiliation{Space Telescope Science Institute, 3700 San Martin Drive, Baltimore, MD 21218, USA}
\altaffiliation{NASA Einstein Fellow}
\email{jpierel@stsci.edu}

\author[0000-0003-4263-2228]{David~A.~Coulter}
\affiliation{Space Telescope Science Institute, 3700 San Martin Drive, Baltimore, MD 21218, USA}
\email{dcoulter@stsci.edu}

\author[0000-0002-0930-6466]{Caitlin M. Casey}
\email{cmcasey@ucsb.edu}
\affiliation{Department of Physics, University of California, Santa Barbara, Santa Barbara, CA 93106, USA}
\affiliation{Cosmic Dawn Center (DAWN), Jagtvej 128, DK2200 Copenhagen N, Denmark}

\author[0000-0001-9187-3605]{Jeyhan S. Kartaltepe}
\email{jeyhan@astro.rit.edu}
\affiliation{Laboratory for Multiwavelength Astrophysics, School of Physics and Astronomy, Rochester Institute of Technology, 84 Lomb Memorial Drive, Rochester, NY 14623, USA}
\email{jeyhan@astro.rit.edu}

\author[0000-0003-3596-8794]{Hollis B. Akins}
\email{hollis.akins@gmail.com}
\altaffiliation{NSF Graduate Research Fellow}
\affiliation{Department of Astronomy, The University of Texas at Austin, 2515 Speedway Blvd Stop C1400, Austin, TX 78712, USA}

\author[0000-0002-3560-8599]{Maximilien Franco}
\email{maximilien.franco@cea.fr}
\affiliation{Université Paris-Saclay, Université Paris Cité, CEA, CNRS, AIM, 91191 Gif-sur-Yvette, France}
\email{francomaximilien@gmail.com}

\author[0000-0003-0209-674X]{Mike~Engesser}
\affiliation{Space Telescope Science Institute, 3700 San Martin Drive, Baltimore, MD 21218, USA}
\email{mengesser@stsci.edu}

\author[0000-0003-2037-4619]{Conor~Larison}
\affiliation{Space Telescope Science Institute, 3700 San Martin Drive, Baltimore, MD 21218, USA}
\email{clarison@stsci.edu}

\author[0000-0003-1169-1954]{Takashi~J.~Moriya}
\affiliation{National Astronomical Observatory of Japan, National Institutes of Natural Sciences, 2-21-1 Osawa, Mitaka, Tokyo 181-8588, Japan} 
\affiliation{Graduate Institute for Advanced Studies, SOKENDAI, 2-21-1 Osawa, Mitaka, Tokyo 181-8588, Japan}
\affiliation{School of Physics and Astronomy, Monash University, Clayton, VIC 3800, Australia}
\email{takashi.moriya@nao.ac.jp}

\author[0000-0001-9171-5236]{Robert M. Quimby} 
\affiliation{Department of Astronomy and Mount Laguna Observatory, \\ San Diego State University, San Diego, CA 92182, USA} 
\affiliation{Kavli Institute for the Physics and Mathematics of the Universe (WPI), The University of Tokyo Institutes for Advanced Study, The University of Tokyo, Kashiwa, Chiba 277-8583, Japan}
\email{rquimby@sdsu.edu}

\author[0000-0002-7087-0701]{Marko Shuntov}
\email{marko.shuntov@nbi.ku.dk}
\affiliation{Cosmic Dawn Center (DAWN), Jagtvej 128, DK2200 Copenhagen N, Denmark}
\affiliation{Niels Bohr Institute, University of Copenhagen, Jagtvej 128, DK-2200, Copenhagen, Denmark}
\affiliation{University of Geneva, 24 rue du Général-Dufour, 1211 Genève 4, Switzerland}
\email{marko.shuntov@nbi.ku.dk}

\author[0000-0003-2445-3891]{Matthew~R.~Siebert}
\affiliation{Space Telescope Science Institute, 3700 San Martin Drive, Baltimore, MD 21218, USA}
\email{msiebert@stsci.edu}

\author[0000-0002-4781-9078]{Christa DeCoursey}
\affiliation{Steward Observatory, University of Arizona, 933 N. Cherry Ave, Tucson, AZ 85721, USA}
\email{cndecoursey@arizona.edu}

\author[0000-0001-5281-731X]{Rodrigo Angulo}
\affiliation{William H. Miller III Department of Physics \& Astronomy, Johns Hopkins University, 3701 San Martin Dr, Baltimore, MD 21218, USA}
\email{rangulo1@jhu.edu}

\author[0000-0002-7566-6080]{James~M.~DerKacy}
\affiliation{Space Telescope Science Institute, 3700 San Martin Drive, Baltimore, MD 21218, USA}
\email{jderkacy@stsci.edu}

\author[0000-0003-4761-2197]{Nicole E. Drakos}
\email{ndrakos@hawaii.edu}
\affiliation{Department of Physics and Astronomy, University of Hawaii, Hilo, 200 W Kawili St, Hilo, HI 96720, USA}
\email{ndrakos@hawaii.edu}

\author[0000-0003-1344-9475]{Eiichi Egami}
\affiliation{Steward Observatory, University of Arizona, 933 N. Cherry Ave, Tucson, AZ 85721, USA}
\email{egami@arizona.edu}

 \author[0000-0001-8519-1130]{Steven L. Finkelstein}
 \affiliation{Department of Astronomy, The University of Texas at Austin, 2515 Speedway Blvd Stop C1400, Austin, TX 78712, USA} 
\affiliation{Cosmic Frontier Center, The University of Texas at Austin, Austin, TX, USA} 
\email{stevenf@astro.as.utexas.edu}

\author[0009-0000-0272-5468]{Carter Flayhart}
\email{cbf2388@rit.edu}
\affiliation{Laboratory for Multiwavelength Astrophysics, School of Physics and Astronomy, Rochester Institute of Technology, 84 Lomb Memorial Drive, Rochester, NY 14623, USA}

\author[0000-0001-7201-5066]{Seiji Fujimoto}
\affiliation{David A. Dunlap Department of Astronomy and Astrophysics, University of Toronto, 50 St. George Street, Toronto, Ontario, M5S 3H4, Canada} \affiliation{Dunlap Institute for Astronomy and Astrophysics, 50 St. George Street, Toronto, Ontario, M5S 3H4, Canada}
\email{seiji.fujimoto@utoronto.ca}

\author[0000-0003-0209-9246]{Estefania~Padilla~Gonzalez}
\affiliation{Space Telescope Science Institute, 3700 San Martin Drive, Baltimore, MD 21218, USA}
\email{epadill7@jh.edu}

\author[0000-0002-5060-1379]{Massimo~Griggio}
\affiliation{Space Telescope Science Institute, 3700 San Martin Drive, Baltimore, MD 21218, USA}
\email{mgriggio@stsci.edu}

\author[0000-0003-0129-2079]{Santosh Harish}
\email{harish.santosh@gmail.com}
\affiliation{Space Telescope Science Institute, 3700 San Martin Drive, Baltimore, MD 21218, USA}
\affiliation{Laboratory for Multiwavelength Astrophysics, School of Physics and Astronomy, Rochester Institute of Technology, 84 Lomb Memorial Drive, Rochester, NY 14623, USA}

\author[0000-0002-7303-4397]{Olivier Ilbert}
\email{olivier.ilbert@lam.fr}
\affiliation{Aix Marseille Univ, CNRS, CNES, LAM, Marseille, France}
\email{olivier.ilbert@lam.fr}

\author[orcid=0000-0000-0000-0001,sname='Inayoshi']{Kohei Inayoshi}
\altaffiliation{}
\affiliation{Kavli Institute for Astronomy and Astrophysics, Peking University, Beijing 100871, China}
\email{inayoshi@pku.edu.cn}  

\author[0000-0002-6610-2048]{Anton~M.~Koekemoer}
\affiliation{Space Telescope Science Institute, 3700 San Martin Drive, Baltimore, MD 21218, USA}
\email{koekemoer@stsci.edu}

\author[0000-0002-5588-9156]{Vasily Kokorev}
\affiliation{Department of Astronomy, The University of Texas at Austin, 2515 Speedway Blvd Stop C1400, Austin, TX 78712, USA}
\email{vasily.kokorev.astro@gmail.com}

\author[0009-0008-5926-818X]{Clotilde Laigle}\affiliation{Institut d’Astrophysique de Paris, UMR 7095, CNRS, and Sorbonne Université, 98bis boulevard Arago, 75014 Paris, France}
\email{clotilde.laigle@physics.ox.ac.uk}

\author[0000-0003-3216-7190]{Erini~Lambrides}
\affiliation{NASA-Goddard Space Flight Center, Greenbelt MD, 20771, USA}
\email{erini.lambrides@nasa.gov}

\author[0000-0003-2366-8858]{Rebecca L. Larson}
\affiliation{Space Telescope Science Institute, 3700 San Martin Drive, Baltimore, MD 21218, USA}
\email{rlarson@stsci.edu}

\author[0000-0002-0514-5650]{Xiaolong Li}
\affiliation{William H. Miller III Department of Physics \& Astronomy, Johns Hopkins University, 3701 San Martin Dr, Baltimore, MD 21218, USA}
\email{xli360@jhu.edu}

\author[0000-0001-9773-7479]{Daizhong Liu}
\email{dzliu@pmo.ac.cn}
\affiliation{Purple Mountain Observatory, Chinese Academy of Sciences, 10 Yuanhua Road, Nanjing 210023, China}
\email{dzliu@mpe.mpg.de}

\author[0000-0002-4872-2294]{Georgios~E.~Magdis}
\affiliation{Cosmic Dawn Center (DAWN), Jagtvej 128, DK2200 Copenhagen N, Denmark} \affiliation{DTU-Space, Technical University of Denmark, Elektrovej 327, 2800, Kgs. Lyngby, Denmark}
\email{geoma@space.dtu.dk}

\author[0000-0002-9883-7460]{Jacqueline E. McCleary}
\affiliation{Department of Physics, Northeastern University, 360 Huntington Ave, Boston, MA}
\email{j.mccleary@northeastern.edu}

\author[0000-0002-9489-7765]{Henry~J.~McCracken}
\affiliation{Institut d’Astrophysique de Paris, UMR 7095, CNRS, and Sorbonne Université, 98bis boulevard Arago, 75014 Paris, France}
\email{hjmcc@iap.fr}

\author[0009-0003-9297-6587]{Nicolas McMahon}
\affiliation{Laboratory for Multiwavelength Astrophysics, School of Physics and Astronomy, Rochester Institute of Technology, 84 Lomb Memorial Drive, Rochester, NY 14623, USA}
\email{njm4071@rit.edu}

\author[0000-0002-6149-8178]{Jed McKinney}
\altaffiliation{NASA Hubble Fellow}
\affiliation{Department of Astronomy, The University of Texas at Austin, 2515 Speedway Blvd Stop C1400, Austin, TX 78712, USA}
\email{jed.mckinney@austin.utexas.edu}

\author[0000-0001-8385-3727]{Thomas~Moore}
\affiliation{Space Telescope Science Institute, 3700 San Martin Drive, Baltimore, MD 21218, USA}
\email{tmoore@stsci.edu}

\author[0000-0003-2397-0360]{Louise Paquereau} 
\affiliation{Institut d’Astrophysique de Paris, UMR 7095, CNRS, and Sorbonne Université, 98bis boulevard Arago, 75014 Paris, France}
\email{louise.paquereau@iap.fr}

\author[0000-0002-4485-8549]{Jason Rhodes}
\affiliation{Jet Propulsion Laboratory, California Institute of Technology, 4800 Oak Grove Drive, Pasadena, CA 91001, USA}
\email{jason.d.rhodes@jpl.nasa.gov}

\author[0000-0002-4271-0364]{Brant~E.~Robertson}
\affiliation{Department of Astronomy and Astrophysics, University of California, Santa Cruz, 1156 High Street, Santa Cruz, CA 95064, USA}
\email{brant@ucsc.edu}

\author[0000-0002-1233-9998]{David~B.~Sanders}
\affiliation{Institute for Astronomy, University of Hawai’i at Manoa, 2680 Woodlawn Drive, Honolulu, HI 96822, USA}
\email{sandersd@hawaii.edu}

\author[0009-0009-3048-9090]{Sogol~Sanjaripour}
\affiliation{Department of Physics and Astronomy, University of California Riverside, Riverside, CA 92521, USA}
\email{sogol.sanjaripour@email.ucr.edu}

\author[0000-0002-2798-2943]{Koji~Shukawa}
\affiliation{William H. Miller III Department of Physics \& Astronomy, Johns Hopkins University, 3701 San Martin Dr, Baltimore, MD 21218, USA}
\email{kshukaw1@jh.edu}

\author[0000-0002-7756-4440]{Louis-Gregory~Strolger}
\affiliation{Space Telescope Science Institute, 3700 San Martin Drive, Baltimore, MD 21218, USA}
\email{strolger@stsci.edu}

\author[0000-0003-3631-7176]{Sune~Toft}
\affiliation{Institute for Astronomy, University of Hawai’i at Manoa, 2680 Woodlawn Drive, Honolulu, HI 96822, USA}
\affiliation{Cosmic Dawn Center (DAWN), Jagtvej 128, DK2200 Copenhagen N, Denmark}
\affiliation{Niels Bohr Institute, University of Copenhagen, Jagtvej 128, DK-2200, Copenhagen, Denmark}
\email{sune@nbi.ku.dk}

\author[0000-0001-5233-6989]{Qinan Wang}
\affiliation{Department of Physics and Kavli Institute for Astrophysics and Space Research, Massachusetts Institute of Technology, 77 Massachusetts Avenue, Cambridge, MA 02139, USA}
\email{qnwang@mit.edu}

\author[0000-0002-3742-8460]{Robert~E.~Williams}
\affiliation{Space Telescope Science Institute, 3700 San Martin Drive, Baltimore, MD 21218, USA}
\email{wms@stsci.edu}

\author[0000-0002-0632-8897]{Yossef~Zenati}
\affiliation{William H. Miller III Department of Physics \& Astronomy, Johns Hopkins University, 3701 San Martin Dr, Baltimore, MD 21218, USA}
\affiliation{Astrophysics Research Center of The Open University (ARCO), Ra’anana 4353701, Israel}
\affiliation{Department of Natural Sciences, The Open University of Israel, Ra’anana 4353701, Israel}
\email{yzenati1@jhu.edu}

\begin{abstract}

Transient astronomy in the early Universe ($z \gtrsim 2$) remains largely unexplored, lying beyond the rest-frame optical spectroscopic reach of most current observatories. Yet this regime promises transformative insights, with high-redshift transients providing direct access to the early Universe and enabling studies of how stellar populations and cosmology evolve over cosmic time. JWST is uniquely equipped to probe these redshifts efficiently in the rest-frame optical and near-IR. We present results from an initial pathfinder search, covering an area of $\sim133$ arcmin$^2$ ($\sim$0.037 deg$^2$) independently imaged by the PRIMER and COSMOS-Web (hereafter COSMOS) extragalactic surveys. Although neither program was designed for time-domain astronomy, combining their data results in difference images separated by roughly one year, leading to the discovery of \nsne~supernovae (SNe) with host photometric redshifts reaching $z \lesssim 5$. For most SNe, only a single epoch is available, but the combination of host redshift, classification, color, and magnitude enables us to prioritize candidates for detailed photometric and spectroscopic follow-up. Among the most notable sources are a relatively bright, blue CCSN at $z>3$ (SN 2023aeab) and a young, normal SN Ia at $z>2$ (SN 2023aeax). The sample distribution highlights the increasing likelihood that a wide-area JWST program can uncover younger, bluer, and potentially more extreme explosions. While this pathfinder effort is limited in cadence and number of filters, it demonstrates the strong potential of a dedicated, well-planned time-domain survey with JWST to obtain the sample sizes and rate measurements needed to chart SN populations deep into the early Universe.

\end{abstract}


\section{Introduction} 
\label{sec:intro}

\begin{figure*}[t]
    \plottwo{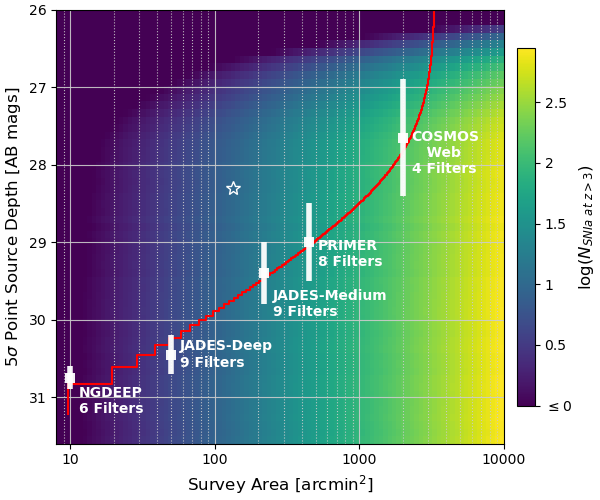}{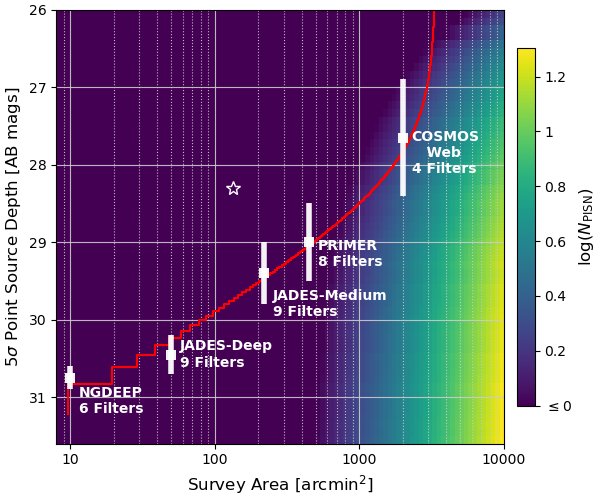}
    \caption{Total expected number of SNe derived from simulations of a corresponding 5-year survey: ({\it left}) Type Ia SNe at redshifts $z>3$; ({\it right}) Pop-III PISNe. For simplicity, the simulated surveys all have a 6-month cadence in four JWST/NIRcam bands (F115W, F150W, F277W, and F444W), all with the same limiting magnitude per epoch. The SN yields correspond to the number of SNe detected in at least one band at any phase. The red line traces the area and magnitude limits possible with 200 hours on JWST (including overheads). Increasing depth has limited benefits below about 28th magnitude, but the survey yields increase proportionally with survey area. The area and depths of several Cycle 1 JWST surveys are marked for comparison, but note these surveys are not uniform and all have different durations, filters, and cadences from the simulations. The open star marks the overlapping area of COSMOS-Web and PRIMER used in this paper (see Figure \ref{fig:cosmosprimer}).}
    \label{fig:depthvsarea}
\end{figure*}

Given their large intrinsic luminosities, supernovae (SNe) offer a powerful alternative to stars as probes of the earlier Universe ($z>1$). These explosions not only enable us to extend current low-$z$~SN investigations to unique redshift ranges and host environments, but they also introduce new opportunities to study the first stars and the most formative years in the evolution of the Universe, including the peak era of star formation and black hole growth in the Universe's history (i.e., Cosmic Noon) and the Epoch of Reionization (EoR). In particular, a statistically significant sample of high-$z$~SNe can uniquely extend SN Ia cosmology to $z>3$ \citep[e.g.,][]{pierel24,pierel25,siebert25}, enable the the discovery and characterization of superluminous (SLSNe) and pair-instability SNe (PISNe) from low-metallicity and/or Pop III stars \citep[e.g.,][]{jeon26,ferrara26}, and track the evolution of Type II/Ib/Ic core-collapse SN (CCSN) explosions and their rates prior Cosmic Noon \citep[e.g.,][]{decoursey25a}. The evolution of the CCSN rates provides an independent test of the evolution of the cosmic star formation rate density (SFRD), independent of conversions from luminosity to SFR and dust corrections. 

Yet transient astronomy at higher redshifts has remained relatively unexplored, given limitations in sensitivity and wavelength ranges of most current ground-based discovery surveys and follow-up facilities. Over the past decade, there has been an increasing effort to discover and follow up high-$z$~SNe. Several ground-based studies have successfully detected a handful of superluminous supernovae (SLSNe) and bright SNe IIn at $2<z<3$~in deep stack survey data \citep[e.g.,][]{cooke09,cooke12,smith18,curtin19,moriya19,galyam19}. These discoveries have set the stage for larger space-based telescopes, such as the {\it James Webb Space Telescope}~(JWST)~and the {\it Nancy Grace Roman Space Telescope}~(hereafter {\it Roman}).

Compared to ground-based telescopes, the distinct sensitivity and wavelength coverage of JWST enables not only detection of high-redshift SNe, but both high signal-to-noise photometric and spectroscopic monitoring throughout most of their evolution during the free-expansion phase. The combination of high angular resolution with stable, low-background imaging facilitates effective host-galaxy subtraction and allows for concurrent characterization of SN environments, including stellar mass, star formation rate (SFR), and metallicity. These measurements are challenging to acquire from ground-based observations at comparable redshifts.

Despite the powerful capabilities of the James Webb Space Telescope, SN discoveries with JWST have so far been relatively limited. Early successes largely came from targeting strongly lensed galaxy clusters through the JWST Prime Extragalactic Areas for Reionization and Lensing Science (PEARLS) program (PID \#1176; PI: R. Windhorst), which identified events such as the triply imaged Type Ia SN H0pe at $z = 1.78$ \citep{frye24,chen24} and two additional SNe at $z\approx$ 0.8 and 2.2, in the MACS J0416.1-2403 field \citep{yan23}. More recently, PEARLS conducted a systematic three-epoch JWST/NIRCam search with a $\sim$6-month cadence over $\sim$14 arcmin$^2$~in the Spitzer IRAC Dark Field (IDF), demonstrating the feasibility of blank-field surveys and yielding 21 transients out to $z\approx$2.6 \citep{yan26}.

Building the necessary statistically significant, well-characterized sample at $z>2$, however,~requires a survey that is both wide and deep with sufficient follow-up opportunities \citep{jha24}. Such surveys demand significant observational time, even with JWST's exquisite sensitivity, because of the telescope's narrow field of view and large overheads. Consequently, these time-domain science observing programs have faced difficulty securing approval. As a result, the time-domain community is increasingly leveraging the existing, serendipitous fields obtained for other science programs (Figure \ref{fig:depthvsarea}). Section \ref{sec:area} provides a more detailed discussion on the tradeoffs between depth and area.

One of the highest-yield discovery programs to date has been the JWST Advanced Deep Extragalactic Survey \citep[JADES;][]{eisenstein23}, which obtained deep, multi-epoch imaging with a primary science goal of exploring high-redshift galaxies. The survey obtained two sets of deep NIRCam images covering an area of $\sim$25 arcmin$^2$ over the GOODS-S field with a separation of 1 year, which has enabled a transient survey of unprecedented depth (down to $\sim$30 AB mag). From these observations, \citet{decoursey25a} identified 79 transients (i.e., $>$1 transient/arcmin$^2$~per epoch) with host redshifts ranging from $0.2 < z \lesssim 5$ \citep{decoursey25b}. A follow-up DD proposal (PID 6541; PI Egami) observed eleven SNe with a single pointing of a NIRSpec/MSA spectroscopy and NIRCam imaging. The results of that spectroscopic program include an broad-lined Type Ic SN (Ic-BL) at $z$=2.83, which indicates this subclass may be more common at high $z$ \citep{siebert24}, an SN Ia at $z$=2.9 that tests an evolving dark energy in a dark-matter dominated Universe \citep{pierel24}, and a UV-bright SN at $z$=3.6 \citep{coulter25}. The follow-up spectroscopy and imaging also realized the first systematic characterization of high-redshift SNe~II \citep{2025PASJ...77..851M}.

While JADES successfully discovered dozens of new SNe, its survey footprint was relatively narrow ($\sim$25~arcmin$^2$). A much larger area is covered by the overlap of JWST Cosmic Evolution Survey (COSMOS-Web; PID 1727) and the Public Release IMaging for Extragalactic Research (PRIMER; PID 1837) Surveys (see Section \ref{sec:obs}). These two Cycle 1 Treasury Programs, when taken together, obtained overlapping images in 4 NIRCam filters (F115W, F150W, F277W, F444W) in an area that is $>5$ times larger than JADES (133 arcmin$^2$), but not as deep ($\sim28$~mag on average). This field also benefits from the fact that the COSMOS data products include extensive photometric and spectroscopic redshift catalogs. 

In this article we present transient discoveries from the overlap of the COSMOS-Web and PRIMER data sets. Section~\ref{sec:obs} presents the observations and reduction techniques. Section~\ref{sec:reduction} outlines the transient/SNe identification criteria, explains how photometry was performed, and provides an analysis of the discovery epoch detection limit. Section \ref{sec:results} presents an analysis of the data, including host galaxy redshifts, single-epoch photometric SN classification, and evaluation of the various SN parameters. Section \ref{sec:followup} describes candidate prioritization, identification, and follow-up photometry and spectroscopy. Section \ref{sec:discussion} provides a discussion of the results, considerations and requirements for an optimal transient survey with JWST, and synergies with {\it Roman}. Finally, in Section~\ref{sec:conclusion}, we summarize the paper and highlight the primary takeaways. Throughout this paper, we use magnitudes in the AB system \citep{oke83} and adopt the standard $\Lambda$CDM cosmological parameters: H$_0$ = 70 km s$^{-1}$ Mpc$^{-1}$, $\Omega _{tot}$ = 1.0, $\Omega_\Lambda$ = 0.7, and $\Omega_m$ = 0.3.

\section{Observations} \label{sec:obs}

\subsection{COSMOS-Web}

COSMOS is a well-known deep, wide area, multi-wavelength survey aimed at measuring the evolution of galaxies on scales from a few kpc to tens of Mpc \citep{scoville07}, centered at (J2000): 10:00:28.600 +02:12:21.00. COSMOS-Web \citep[Program ID $1727$;][]{casey23} is a 255 hour wide-field Cycle 1 JWST treasury program that maps a contiguous $\sim$0.54 deg$^2$ (2000 arcmin$^2$)~area in the center of the COSMOS field that has revolutionized our understanding of reionization's spatial distribution, environments, and drivers at early stages by detecting thousands of galaxies in the EoR ($6<z<11$) on scales large enough to mitigate cosmic variance \citep[e.g.,][]{casey24,akins25}. The mosaiced observations were obtained with NIRCam in four filters: F115W, F150W, F277W, and F444W. Observations were split over four epochs for scheduling purposes: January 2023, April 2023, January 2024, and April 2024. The individual pointings had little to no overlap between the different epochs, although the fourth epoch was primarily focused on previously failed visits. The majority of the mosaic either has an on-sky exposure time of 513\,s or 1030\,s and reaches a depth of roughly 27-28 AB mag.  Additional details regarding the observations and reductions are outlined in the COSMOS-Web overview paper \citep{casey24}.

\subsection{PRIMER}

PRIMER (Public Release IMaging for Extragalactic Research) is another Cycle 1 Program \citep[Program ID $1837$][]{dunlop_primer_2021} designed to provide deep, large-area, homogeneous JWST NIRCam+MIRI imaging survey of the COSMOS and UKIDSS Ultra-Deep Field (UDS). The PRIMER-COSMOS observations only cover $\sim$0.037 deg$^2$ ($\sim$133 arcmin$^2$)~of the COSMOS field, but reach a depth of roughly 28-29 AB mag.  The mosaiced observations were obtained with NIRCam in ten filters (F090W, F115W, F150W, F200W, F277W, F356W, F444W and F410M) and with MIRI in two filters (F770W and F1800W). Observations were split over three epochs for scheduling purposes: December 2022, April 2023, and December 2023. Again, the individual pointings had little to no overlap between epochs except the last epoch that re-observed failed visits. Additional details regarding the observations and reductions are outlined in the PRIMER overview paper \citep{dunlop_primer_2021}.

\begin{figure}[t]
    \centering
    {\includegraphics[width=8cm]{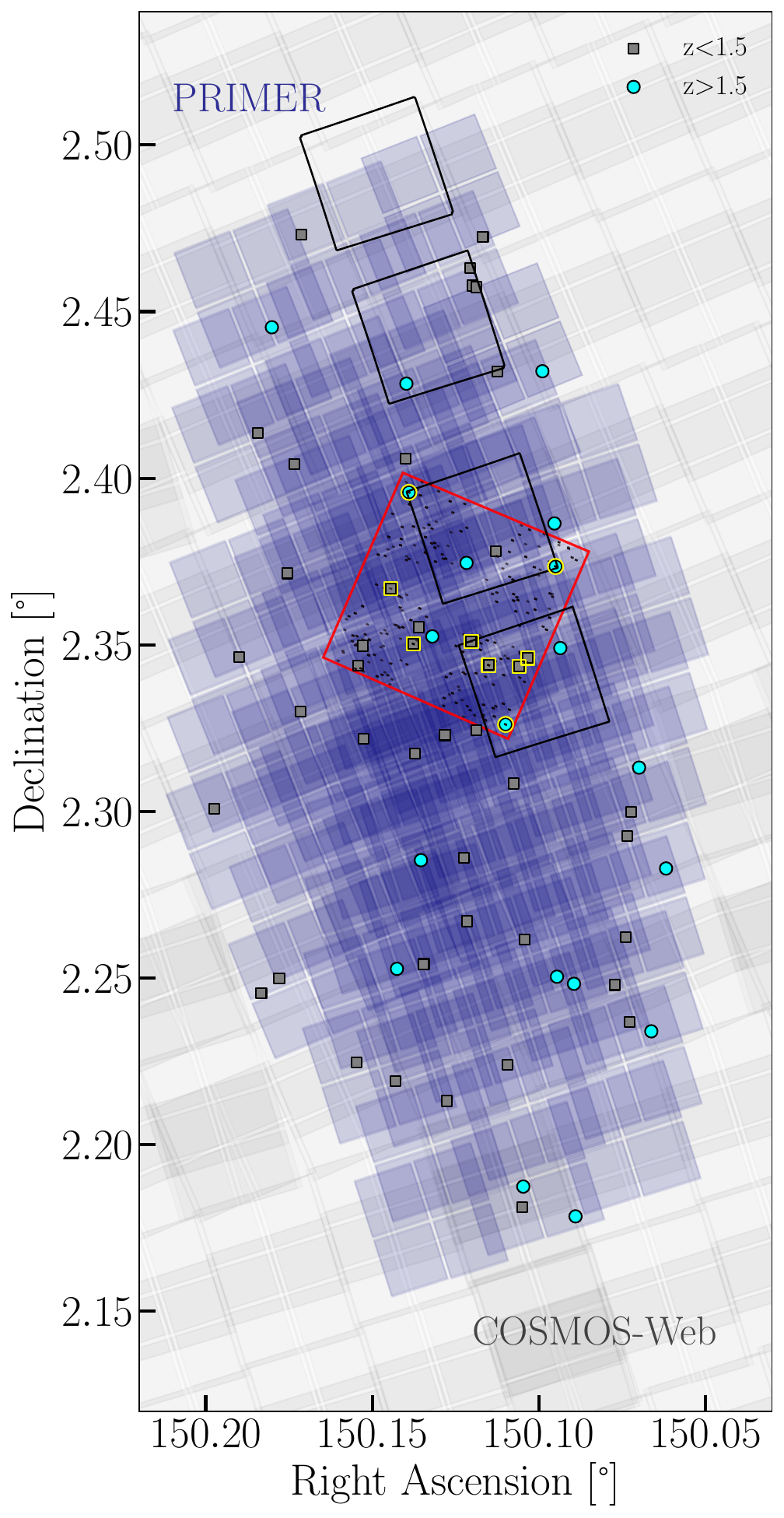}}
    \caption{JWST/NIRCam observations of the COSMOS-Web and PRIMER surveys as gray- and blue-shaded areas, respectively, with roughly $\sim$0.037 deg$^2$~of overlapping area. Grey squares and cyan circles indicate the positions of transients presented in this study with redshifts of $z<1.5$~and $z>1.5$, respectively. Over-plotted are footprints of NIRSpec/MSA (red square) and NIRCam (black square) from follow-up observations obtained via a Director's Discretionary Time Proposal (PID 6585; PI Coulter), described in detail in Section \ref{subsec:follow_up}. NIRSpec MSA slitlets used in that program are shown with small black lines. SNe for which MSA spectra were obtained are outlined in yellow.}
    \label{fig:cosmosprimer}
\end{figure}

Given the overlap of the two surveys and the difference in time of the observations, a number of pointings have two (and sometimes more in rare cases with serendipitous observations from other programs) epochs of observations spaced roughly a year apart. These two epochs offer an ideal pathfinder for a wide-field, relatively deep survey for discovery of transients with JWST in the second epoch, although no dedicated follow-up photometric or spectroscopic observations were available. The overlap between COSMOS and PRIMER is shown in Figure~\ref{fig:cosmosprimer} (see also Figure~15 of \citealt{casey24}). 

\section{Candidate Discovery}  
\label{sec:reduction}

Transient discovery traditionally requires difference imaging by template subtraction (see Section~\ref{sec:differencing}~and~\ref{sec:difference_photometry}). Although this is a well-understood technique for many ground-based telescopes where the point spread function (PSF) is dominated by the ``seeing'' and can be modeled by Gaussian-like functions, space-based telescopes present a whole new level of complication \citep[i.e., non-Gaussian PSF, field rotation, and alignment;][]{berman24}. Furthermore, given the resolution of JWST, optimal differencing requires precise alignment down to sub-pixels (i.e., $\sim0.1$ pixels), but is typically only good to within $\sim$1 pixel using the default pipeline \citep{Bushouse2023,decoursey25a,pierel25}. To make matters even more complicated, optimal subtractions require alignment and differencing to be performed at the exposure level before mosaics are created \citep[e.g.,][]{decoursey25a}, necessitating complex and complicated handling of the many dithers that go into creating the final mosaic. This process entails careful management of numerous individual dithers that collectively contribute to the final integrated image, thereby motivating the development of a specialized, JWST-specific difference-imaging pipeline.

The combined sensitivity of JWST, noise characteristics of the detectors, and cosmic rays in the data yield potentially hundreds of sources that require detailed vetting in each NIRCam pointing \citep[e.g.,][]{decoursey25a}. Filtering out real sources, as well as characterization and prioritization of high-$z$~SNe for dedicated follow-up, is therefore a critical part of the process. In this section, we describe the detailed steps we take to identify SN candidates and apply various cuts to create our final COSMOS-Web/PRIMER SN sample, hereafter referred to as the COSMOS sample.

\subsection{Image reduction and alignment}

\begin{figure*}
    \centering
    {
    \includegraphics[width=3.7in]{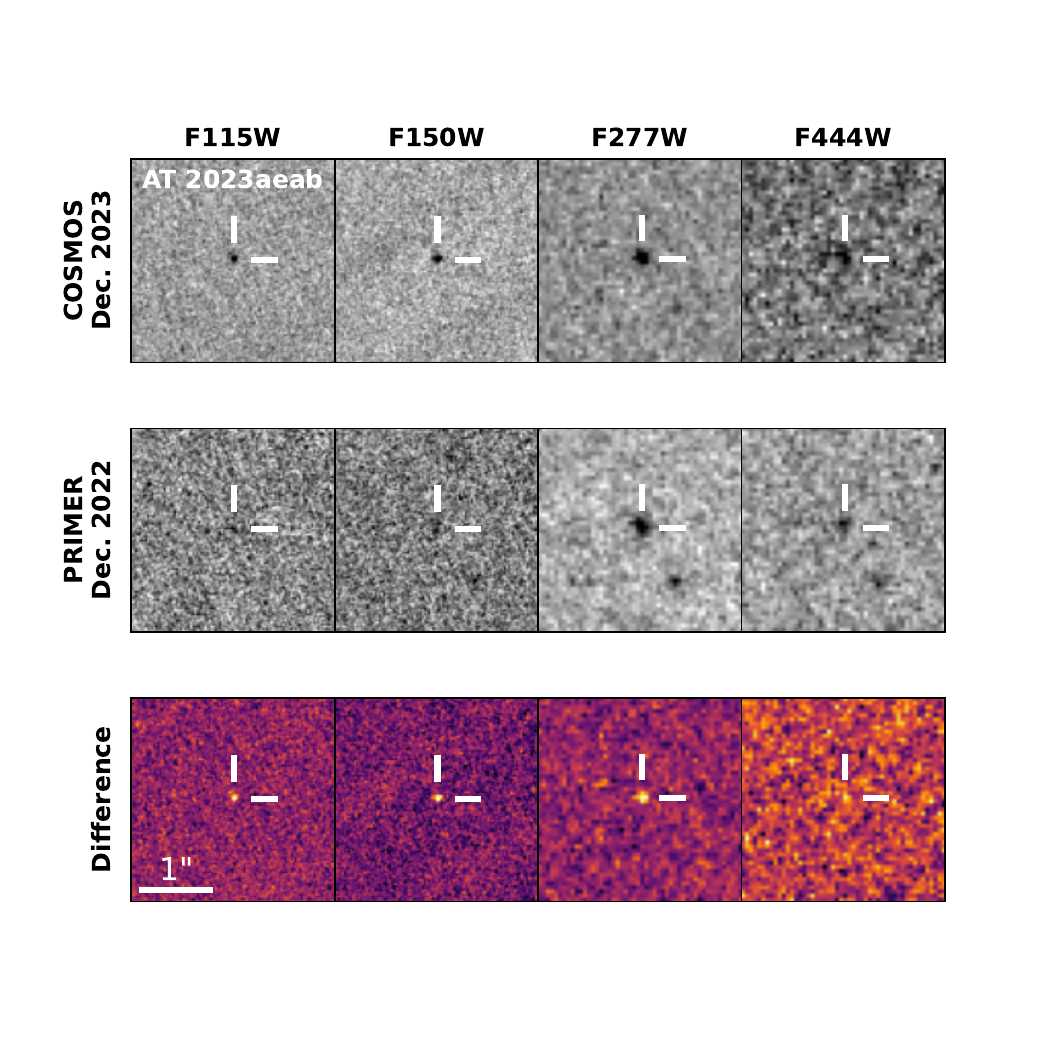}
    \hspace{-0.4in}
    \includegraphics[width=3.7in]{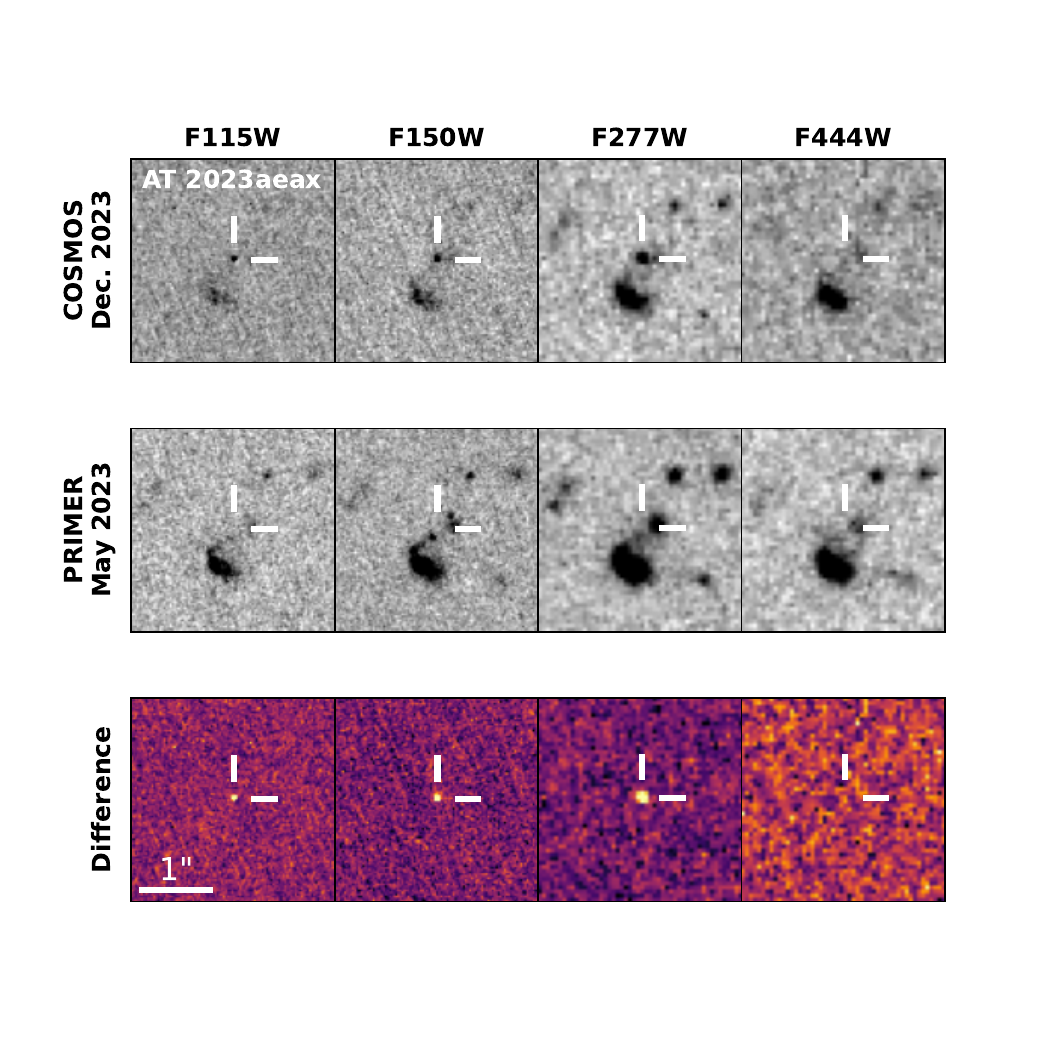}
    \caption{Cutouts for SN 2023aeab and SN 2023aeax (SN\#1 and SN\#4 in Table \ref{tab:targets}, respectively), both discussed throughout the text. Shown are the ({\it top}) discovery epoch from COSMOS-Web, ({\it middle}) template from PRIMER, and ({\it bottom}) difference image.}
    \label{fig:cutout_SN1}
    }
\end{figure*}

All NIRCam observations were processed using the \textit{JWST} Calibration Pipeline \citep{Bushouse2023}, with additional modifications aimed at enhancing image quality and astrometric accuracy. We provide a brief summary here; a detailed account of the data reduction and mosaic creation is provided in \citet{Franco25}. All raw NIRCam exposures used in this paper can be found in the Mikulski Archive for Space Telescopes (MAST)\footnote{\href{https://mast.stsci.edu/}{https://mast.stsci.edu/}} : \dataset[https://doi.org/10.17909/n7kq-ef83]{https://doi.org/10.17909/n7kq-ef83}. The raw NIRCam exposures were processed using pipeline version 1.14.0, incorporating custom improvements, some inspired by  \citet{Bagley2023}. These refinements included suppression of $1/f$ noise, background subtraction, removal of artifacts such as wisps, and detection and masking of defective pixels. Calibration was carried out using the Calibration Reference Data System (CRDS) pmap-1223, corresponding to the NIRCam instrument configuration imap-0285. The final science mosaics for both COSMOS and PRIMER were generated separately with stage 3 of the JWST pipeline at resolutions of $0\farcs03$ per pixel for both short- (SW) and long-wavelength (LW) filters, ensuring optimal spatial detail, and also at the native $0\farcs06$ per pixel for accurate photometric measurements in the LW filters.

Astrometric alignment was performed using the JWST/HST Alignment Tool (JHAT; \citealt{Rest2023}), which registers the individual exposure NIRCam images (i.e., CALs) to a reference catalog constructed from pre-existing HST/ACS F814W mosaics \citep{Koekemoer2007}. The astrometry was tied to Gaia Early Data Release 3 (EDR3; \citealt{gaia21}), yielding a median absolute positional offset of less than $0\farcs005$ and a median absolute deviation (MAD) below $0\farcs012$ across all filters.

\subsection{Difference imaging}
\label{sec:differencing}

We generated difference images by subtracting the PRIMER images from the COSMOS images in each respective overlapping filter (Figure \ref{fig:cutout_SN1}). We use \hotpants\ \citep{Becker_hotpants} to match the spatially-varying PSF kernel within a given filter between the COSMOS and PRIMER data sets, applying the general parameter setup and convolution direction described in \citet{Rest2014} and \citet{Angulo25}. For the LW, we produced two sets of difference images: one set for the transient detection with a pixel scale of 0\farcs03/pixel matching the SW pixel scale (see Section~\ref{sec:difference_photometry}), and the other for forced photometry using a pixel scale of 0\farcs06/pixel nearly identical to the LW native pixel scale in order to avoid correlated noise (see Section~\ref{sec:forced_photometry}).

\subsection{Source detection}
\label{sec:difference_photometry}

To identify transient candidates, source detection is easier when both the SW and LW have the same pixel scale (0\farcs03/pixel), as the pixels are identically registered across bands. For the LW, these mosaics are roughly half the native pixel scale, which is not ideal for the most accurate photometry due to the correlated noise in oversampled images \citep[e.g.,][]{decoursey25a}. We note that for the final forced photometry, all difference images have pixel scales close to the native pixel scale (see section~\ref{sec:forced_photometry}). We use \texttt{Photutils} \citep{bradley25} on the difference images of all filters to identify positive detections \citep{Stetson1987DAOPHOT:Photometry, Bradley2024Astropy/photutils:1.12.0}. We set the \texttt{fwhm} (full-width half-max) to $\sim$2 pixels (0\farcs06) for SW images and $\sim$4 pixels (0\farcs12) for resampled LW images. The \texttt{DAOStarFinder} \texttt{threshold} parameter is set to 3 times the background standard deviation returned from \texttt{astropy} \texttt{sigma\_clipped\_stats} with \texttt{sigma}\,$=$\,3.0.
Following \citet{decoursey25a}, we choose not to constrain the \texttt{sharpness} parameter, as bright transients that are surrounded by subtraction artifacts in the difference images can have a wide variety of \texttt{sharpness} values, and the image border (within $3\times$FWHM from an edge) was excluded from the search.
For all sources with signal-to-noise ratios (S/N) above $3$, aperture photometry is measured using an $r$\,$=$\,0\farcs1 circular aperture with a sky annulus of $r$\,$=$\,0\farcs1--0\farcs2. Photometric errors are based upon the annulus sky noise combined with a Poisson uncertainty is calculated from the measured flux. Using this photometry, measured in all filters, we select transient candidates with \textit{either} S/N\,$\geq$\,5 in at least one band \textit{or} S/N\,$\geq$\,3 in at least two bands.

With these criteria, the sample of potential transients is reduced to a manageable hundreds of sources per difference image. We manually examine individual image and difference image cutouts for all filters, as well as light curve and (apparent) host galaxy information, for all transient candidates. The most common false positives are cosmic rays, diffraction spikes, and difference image artifacts in the cores of bright, point-source-like objects. 

In addition, we must consider contamination from real compact astrophysical sources, including active galactic nuclei (AGN), gamma ray burst (GRB) afterglows and tidal disruption events (TDEs). As discussed in Section 3.3 of \citet{decoursey25a}, we ruled out AGN and TDEs in most cases given their offset from the host nucleus. In a limited number of cases with nuclear variability, we have
cross-matched our transient sources with previously published AGN catalogs \citep{lyu22} and removed any AGN candidates. We also ruled out any nuclear transients where the nucleus showed variability, but did not show clear evidence of a point source appearing (or later disappearing). We have no efficient way to filter GRB afterglows, but we note the rates of GRBs are just a few percent of CCSNe \citep{guetta07}, so we do not consider them as significant contaminants in our sample.

All together, we identify \nsne~transients brightening (Table \ref{tab:cosmos_classifications}). While we don't perform any qualitative or quantitative tests for detection efficiency, we refer to a relevant analysis in \citep{decoursey25a}, which performed artificial star injection tests and found roughly 80\% recovery rates down to $\sim$30 AB mag in the JADES transient sample. COSMOS is shallower by $\sim$2 magnitudes, so we estimate a roughly 80\% detection efficiency at $\sim$28 AB mag and close to 100\% at brighter magnitudes in our COSMOS transient sample. The detection efficiency doesn't directly impact any of our results in this paper, but will be discussed in more detail in future papers that focus on volumetric rates.

\subsection{Forced photometry}
\label{sec:forced_photometry}

For each source detected in Section \ref{sec:difference_photometry}, we perform forced photometry on mosaics from each of the individual epochs with pixel scales of 0$\farcs$03 and 0\farcs06 for the SW and LW channel, respectively. For each candidate, we calculate the average centroid position of all detections in the difference images. At these positions, we measure forced PSF photometry using the \texttt{space\_phot} \citep{pierel24_spacephot} package. We build a drizzed PSF using the \texttt{spike} package \citep{polzin25}, which takes Level 2 PSF models from {\tt webbpsf} v. 1.2.1 \citep{perrin12,perrin14}\footnote{\url{https://webbpsf.readthedocs.io}} that are temporally and spatially dependent and drizzles them together using the same pipeline implementation as the data. We then implement the \texttt{space\_phot} drizzled PSF fitting routine using $5\times5$ pixel cutouts, and fit to the observed flux in all filters. These total fluxes, which are in units of MJy/sr, are converted to AB magnitudes using the pixel scale of each image ($0\farcs03/$pix for SW, $0\farcs06/$pix for LW). A final source of photometric uncertainty is a systematic uncertainty on the zero-points, which is $\lesssim0.01$ mag for all filters \citep{boyer22} and negligible compared to the uncertainties derived by the \texttt{space\_phot} PSF fitting algorithm. Table \ref{tab:cosmos_classifications} lists the photometry for both the brightest and F150W filters, while Table \ref{tab:app:phot} lists the photometry for all filters, which is also provided in machine readable format.

\startlongtable
\begin{deluxetable*}{lccccccrr}
\centering
\tablecaption{COSMOS Classifications Data}
\tablehead{
\colhead{Name} & \colhead{RA} & \colhead{DEC} & \colhead{$z$} & \colhead{Brightest} & \colhead{Brightest} & \colhead{F150W} & \colhead{$\chi^2$~Ia} & \colhead{$\chi^2$~CC}\\
\colhead{(AT)} & \colhead{} & \colhead{} & \colhead{(err)} & \colhead{Mag (err)} & \colhead{Filter} & \colhead{Mag (err)} & \colhead{} & \colhead{} 
}
\label{tab:cosmos_classifications}
\startdata
2023aebb & 10:00:16.783 & +02:18:47.34 & 5.20 (0.14) & 27.80 (0.21) & F277W & -- & -- & -- \\ 
2023aebi & 10:00:21.361 & +02:10:42.55 & 4.22 (0.14) & 27.20 (0.23) & F444W & 29.40 (0.75) & -- & -- \\ 
2023aebl & 10:00:15.901 & +02:14:02.46 & 4.02 (0.14) & 26.90 (0.08) & F150W & 26.90 (0.08) & -- & -- \\ 
2023aeab & 10:00:33.363 & +02:23:44.70 & 3.55 (0.09) & 27.20 (0.09) & F115W & 27.50 (0.07) & -- & -- \\ 
2023aeaf & 10:00:22.800 & +02:22:24.61 & 3.55 (0.12) & 27.50 (0.13) & F150W & 27.50 (0.13) & 3.85 & -- \\ 
2023aeag & 10:00:29.211 & +02:22:28.53 & 3.51 (0.07) & 26.00 (0.12) & F277W & 28.10 (0.16) & -- & -- \\ 
2023adzt & 10:00:22.456 & +02:20:56.46 & 2.35 (0.16) & 27.10 (0.06) & F150W & 27.10 (0.07) & -- & 1.88 \\ 
2023aeax & 10:00:26.401 & +02:19:33.97 & 2.25 (0.10) & 26.00 (0.04) & F150W & 26.00 (0.04) & -- & -- \\ 
2023aeam & 10:00:34.227 & +02:15:09.97 & 2.18 (0.11) & 26.10 (0.04) & F150W & 26.10 (0.04) & -- & -- \\ 
2023aebk & 10:00:25.124 & +02:11:14.74 & 2.14 (0.14) & 27.00 (0.07) & F150W & 27.00 (0.07) & 1.84 & 3.82 \\ 
2023aeal & 10:00:22.698 & +02:15:01.45 & 1.90 (0.14) & 27.90 (0.22) & F150W & 27.90 (0.22) & 2.02 & -- \\ 
2023aeaa & 10:00:22.878 & +02:23:11.13 & 1.89 (0.14) & 27.40 (0.10) & F150W & 27.80 (0.15) & 2.67 & 4.03 \\ 
2023aeak & 10:00:21.473 & +02:14:53.80 & 1.88 (0.10) & 25.60 (0.04) & F115W & 26.10 (0.04) & 2.51 & -- \\ 
2023aebf & 10:00:22.782 & +02:10:31.54 & 1.75 (0.14) & 25.40 (0.11) & F277W & -- & -- & -- \\ 
2023adzw & 10:00:31.677 & +02:21:09.06 & 1.75 (0.14) & 27.30 (0.10) & F115W & 27.50 (0.08) & -- & 2.17 \\ 
2023adze & 10:00:33.547 & +02:25:42.12 & 1.70 (0.13) & 27.00 (0.10) & F115W & 27.10 (0.07) & -- & -- \\ 
2023adzi & 10:00:23.748 & +02:25:55.56 & 1.69 (0.13) & 26.70 (0.10) & F115W & 27.10 (0.07) & -- & -- \\ 
2023adyx & 10:00:43.238 & +02:26:43.05 & 1.64 (0.14) & 27.60 (0.16) & F277W & 27.70 (0.17) & -- & -- \\ 
2023aeaq & 10:00:32.498 & +02:17:07.33 & 1.61 (0.14) & 27.30 (0.08) & F150W & 27.30 (0.08) & -- & 2.50 \\ 
2023adzo & 10:00:39.312 & +02:24:54.30 & 1.52 (0.15) & 27.70 (0.13) & F277W & -- & -- & -- \\ 
2023adyy & 10:00:14.838 & +02:16:58.39 & 1.52 (0.13) & 26.60 (0.09) & F115W & 27.10 (0.10) & -- & -- \\ 
2023adzq & 10:00:25.420 & +02:20:36.49 & 1.49 (0.10) & 25.90 (0.04) & F115W & 26.00 (0.03) & -- & -- \\ 
2023aeav & 10:00:30.774 & +02:19:22.36 & 1.48 (0.10) & 27.20 (0.11) & F115W & 27.70 (0.11) & -- & 1.42 \\ 
2023adzy & 10:00:34.676 & +02:22:00.76 & 1.45 (0.13) & 27.40 (0.14) & F150W & 27.40 (0.14) & -- & -- \\ 
2023aeas & 10:00:25.818 & +02:18:30.19 & 1.43 (0.07) & 25.90 (0.04) & F115W & 25.90 (0.03) & 1.56 & -- \\ 
2023adzp & 10:00:43.731 & +02:20:18.18 & 1.40 (0.13) & 26.80 (0.06) & F115W & 27.10 (0.07) & -- & -- \\ 
2023aebh & 10:00:37.121 & +02:13:28.53 & 1.39 (0.07) & 25.70 (0.04) & F277W & 27.70 (0.15) & -- & -- \\ 
2023adzn & 10:00:45.601 & +02:20:46.62 & 1.38 (0.14) & 28.00 (0.20) & F150W & 28.00 (0.20) & -- & -- \\ 
2023adzx & 10:00:32.664 & +02:21:19.14 & 1.37 (0.10) & 27.00 (0.08) & F115W & 27.10 (0.07) & 1.19 & 3.39 \\ 
2023aean & 10:00:32.292 & +02:15:15.07 & 1.29 (0.10) & 25.70 (0.05) & F115W & 26.00 (0.04) & -- & 1.27 \\ 
2023aebe & 10:00:34.317 & +02:13:08.65 & 1.11 (0.12) & 26.70 (0.12) & F115W & 27.10 (0.13) & -- & -- \\ 
2023aeai & 10:00:44.004 & +02:14:43.65 & 1.10 (0.11) & 25.20 (0.03) & F115W & 25.40 (0.03) & -- & -- \\ 
2023aebj & 10:00:25.184 & +02:10:52.39 & 1.10 (0.14) & 28.50 (0.22) & F277W & 28.70 (0.34) & -- & -- \\ 
2023adzs & 10:00:24.796 & +02:20:45.70 & 1.09 (0.13) & 26.60 (0.08) & F115W & 26.80 (0.07) & 3.53 & 4.91 \\ 
2023aeac & 10:00:33.613 & +02:24:21.12 & 1.01 (0.13) & 26.40 (0.11) & F115W & 26.90 (0.11) & -- & -- \\ 
2023aebc & 10:00:18.539 & +02:14:52.54 & 1.01 (0.10) & 25.80 (0.04) & F115W & 26.30 (0.04) & -- & -- \\ 
2023aebd & 10:00:30.622 & +02:12:47.26 & 1.01 (0.12) & 26.60 (0.10) & F115W & 26.90 (0.09) & -- & -- \\ 
2023aeba & 10:00:17.348 & +02:17:59.73 & 0.93 (0.15) & 27.30 (0.07) & F277W & 27.40 (0.09) & -- & -- \\ 
2023aeap & 10:00:29.165 & +02:16:01.36 & 0.85 (0.12) & 24.00 (0.02) & F115W & 24.30 (0.02) & -- & -- \\ 
2023adzc & 10:00:24.453 & +02:25:34.08 & 0.70 (0.07) & 25.40 (0.06) & F115W & 25.90 (0.06) & -- & -- \\ 
2023adzj & 10:00:42.116 & +02:22:17.16 & 0.68 (0.10) & 27.20 (0.17) & F115W & 28.00 (0.23) & -- & -- \\ 
2023aeat & 10:00:32.912 & +02:19:02.59 & 0.67 (0.15) & 26.70 (0.08) & F150W & 26.80 (0.09) & -- & -- \\ 
2023adyz & 10:00:41.132 & +02:28:23.49 & 0.65 (0.15) & 25.70 (0.04) & F115W & 26.00 (0.05) & -- & -- \\ 
2023adzm & 10:00:44.261 & +02:24:48.99 & 0.60 (0.14) & 25.10 (0.03) & F115W & 25.20 (0.03) & -- & -- \\ 
2023aeaw & 10:00:28.512 & +02:19:27.67 & 0.59 (0.14) & 27.50 (0.26) & F150W & 27.80 (0.33) & -- & -- \\ 
2023aebm & 10:00:17.478 & +02:14:12.79 & 0.59 (0.15) & 27.90 (0.20) & F150W & 27.90 (0.20) & -- & -- \\ 
2023aeao & 10:00:25.046 & +02:15:41.35 & 0.55 (0.15) & 27.40 (0.17) & F115W & 27.80 (0.16) & -- & -- \\ 
2023aeay & 10:00:17.726 & +02:15:44.07 & 0.55 (0.15) & 26.40 (0.07) & F115W & 26.60 (0.05) & -- & -- \\ 
2023adzz & 10:00:27.114 & +02:22:41.22 & 0.54 (0.15) & 26.70 (0.11) & F115W & 27.40 (0.16) & -- & -- \\ 
2023aeah & 10:00:47.380 & +02:18:03.00 & 0.53 (0.15) & 26.00 (0.06) & F115W & 26.20 (0.05) & -- & -- \\ 
2023adzd & 10:00:28.931 & +02:27:46.92 & 0.49 (0.15) & 25.90 (0.06) & F115W & 26.30 (0.05) & -- & -- \\ 
2023aeau & 10:00:36.637 & +02:19:18.31 & 0.48 (0.14) & 26.00 (0.05) & F115W & 26.10 (0.03) & -- & -- \\ 
2023aebg & 10:00:26.253 & +02:13:26.14 & 0.41 (0.14) & 24.30 (0.02) & F115W & 24.80 (0.02) & -- & -- \\ 
2023adzr & 10:00:27.624 & +02:20:38.05 & 0.39 (0.09) & 25.50 (0.01) & F150W & 25.50 (0.02) & -- & -- \\ 
2023adzh & 10:00:26.997 & +02:25:55.38 & 0.30 (0.13) & 26.90 (0.04) & F150W & 26.90 (0.04) & -- & -- \\ 
2023adzg & 10:00:28.671 & +02:27:23.58 & 0.30 (0.09) & 27.30 (0.13) & F115W & -- & -- & -- \\ 
2023adza & 10:00:28.517 & +02:27:26.43 & 0.30 (0.15) & 26.00 (0.08) & F115W & 26.40 (0.08) & -- & -- \\ 
2023adzb & 10:00:28.763 & +02:27:28.14 & 0.26 (0.13) & 25.60 (0.05) & F115W & 26.00 (0.05) & -- & -- \\ 
2023aeaz & 10:00:17.636 & +02:17:33.73 & 0.25 (0.14) & 27.60 (0.10) & F150W & 27.60 (0.10) & -- & -- \\ 
2023adzf & 10:00:28.030 & +02:28:20.46 & 0.17 (0.07) & 27.30 (0.07) & F150W & 27.30 (0.07) & -- & -- \\ 
2023aear & 10:00:29.417 & +02:17:10.03 & 0.11 (0.05) & 26.00 (0.03) & F115W & 26.10 (0.02) & 1.60 & -- \\ 
2023adzu & 10:00:33.048 & +02:21:00.96 & 0.06 (0.04) & 27.40 (0.07) & F150W & 27.40 (0.07) & -- & -- \\ 
2023aeae & 10:00:36.699 & +02:20:58.98 & 0.06 (0.06) & 26.40 (0.06) & F115W & 27.00 (0.06) & -- & -- \\ 
2023aead & 10:00:37.010 & +02:20:37.63 & 0.06 (0.04) & 27.40 (0.08) & F150W & 28.00 (0.13) & -- & -- \\ 
2023adzk & 10:00:41.600 & +02:24:15.36 & 0.03 (0.06) & 27.90 (0.24) & F150W & 27.90 (0.24) & -- & -- \\ 
2023adzl & 10:00:41.177 & +02:19:47.82 & 0.01 (0.06) & 24.40 (0.01) & F150W & 24.40 (0.01) & -- & -- \\ 
2023adzv & 10:00:28.871 & +02:21:03.58 & 0.01 (0.06) & 28.10 (0.25) & F150W & 28.10 (0.25) & -- & -- \\ 
2023aeaj & 10:00:42.706 & +02:14:59.40 & 0.01 (0.06) & 27.60 (0.14) & F150W & 27.60 (0.14) & -- & -- \\\enddata
\tablecomments{
Column (1): Object name (AT designation). 
Column (2): Right ascension (J2000). 
Column (3): Declination (J2000). 
Column (4): Host photo-redshift with uncertainty (Section \ref{sec:photo-z}). 
Column (5): Brightest magnitude (AB) with error. 
Column (6): Filter associated with brightest magnitude. 
Column (7): F150W magnitude with error. 
Column (8): Reduced $\chi^2$~value for Type Ia light-curve fit. Values are not provided when the degrees of freedom are $\leq0$ or for a reduced $\chi^2 > 5$.}
Column (9): Reduced $\chi^2$~value for core–collapse light-curve fit. Values are not provided when the degrees of freedom are $\leq0$ or for a reduced $\chi^2 > 5$.

\end{deluxetable*} 

\section{Analysis}
\label{sec:results}

We have identified a total of \nsne~SNe in the COSMOS sample (Table \ref{tab:cosmos_classifications}). Figure \ref{fig:cosmosprimer} shows each SN position in the COSMOS-Web/PRIMER footprint. With only one epoch of photometry (in most cases), we perform a host-galaxy assignment, host photo-$z$ fit, photometric classification, and color-magnitude diagram evaluation. Taken together, these results provide a useful characterization of each event and enable us to select targets of interest for focused follow-up campaigns.

\subsection{Host galaxies and redshifts}
\label{sec:photo-z}

SN candidates are required to match a nearest-neighbor host galaxies within 4\arcsec\ of the SN because we need a host photo-$z$~to perform our analysis. In other words, there may have been some hostless SNe that we overlooked. In a majority of cases, this technique led to one obvious host candidate. In 5\%\ of cases, the most likely host was more ambiguous, which led to additional visual inspection that incorporated factors such as transient color and the likely lowest redshift host (i.e., for two possible nearby hosts at equal distance from the transient, and all other things being equal, the lower redshift host as assigned).  This technique differs slightly from the ``directional light radius" (DLR) analysis \citep{gupta16} used for the host-galaxy assignment of the JADES transient sample \citep{decoursey25a}. We note, however, that JADES is deeper with higher spatial confusion and therefore requires that more complex approach.

The host galaxy properties are measured using an internally created catalog following the methodology of the COSMOS2025 catalog described in \citet{shuntov25}.  In contrast to the COSMOS2025 catalog, which is based on COSMOS imaging, our internal catalog uses the first epoch of NIRCam imaging from PRIMER to construct a detection image from 8 filters.  PRIMER has the advantage of being deeper than COSMOS and not containing any contaminating light from the SNe discovered in the COSMOS images.

\begin{figure}[t]
    \centering
    {\includegraphics[width=8cm]{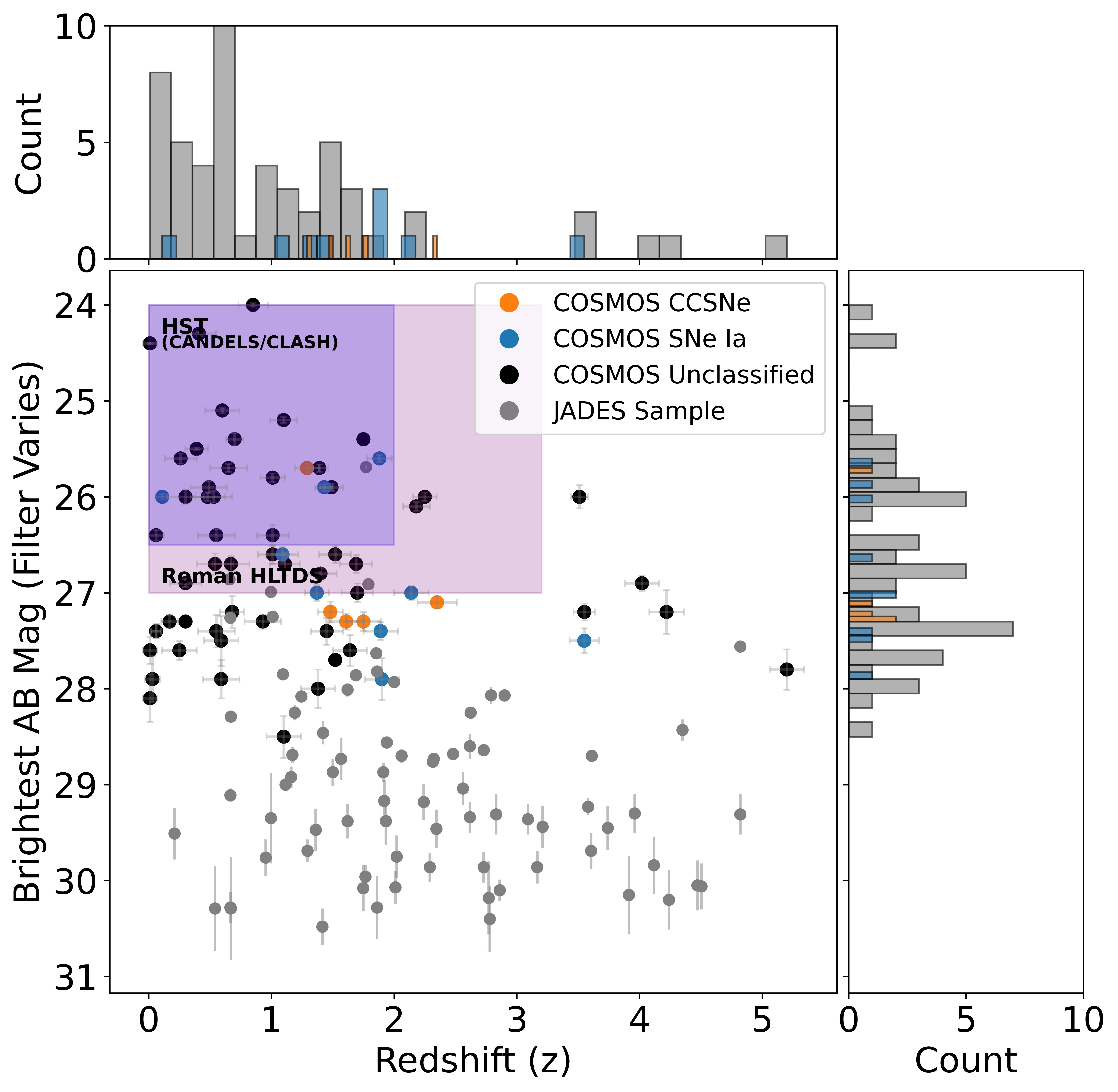}}
    \caption{Magnitude and photometric redshift distribution for the COSMOS sample of SNe (Table \ref{tab:cosmos_classifications}), sorted by classification. Classifications assigned by \texttt{STARDUST2} for targets with at least two filters in a single epoch or two epochs (but see limitations in Section \ref{sec:classification}). The JADES sample \citep{decoursey25a} is plotted for comparison (grey circles). Brightest magnitudes are chosen for plotting purposes because rest-frame photometric bands vary with redshift. However, there is no single standard rest-frame filter that we can use across all redshifts for all SNe. Shaded regions correspond to approximate sensitivity and redshift range for surveys on other telescopes, including HST CANDELS \citep{grogin11,koekemoer11}, CLASH \citep{postman12} and Roman's High-Latitude Time Domain Survey \citep[HLTDS;][]{jha25}. }
    \label{fig:mag_redshift}
\end{figure}

Images are PSF homogenized to F444W, converted to SNR maps, and combined in a positive-truncated $\chi^2$ image for the eight NIRCam filters.  Sources are then extracted with SEP \citep{barbary16}, a Python wrapper for SExtractor \citep{bertin96}, using hot/cold methodology \citep{shuntov25}. Photometry is then performed in all NIRCam, HST/ACS, and WFC3 imaging filters using both fixed circular and Kron apertures.  Uncertainties on photometry from SEP are added in quadrature with Poisson background noise estimates as a function of aperture size.  Photometric redshifts are then fit to all sources using the EAzY SED-fitting tool \citep{brammer08} with a combination of the default Flexible Stellar Population Synthesis templates \citep[][]{conroy10}, specically QSF 12 v3, and bluer templates optimized for selecting less dusty galaxies in the EoR \citep{larson23}, which provide a number of models with variable Ly$\alpha$~escape fractions. Table \ref{tab:cosmos_classifications} lists and Figure \ref{fig:mag_redshift} plots the distribution redshifts for our sample.

We note that our internal PRIMER catalog is roughly a magnitude deeper than the published COSMOS2025 catalog.  Another subtle point is that the COSMOS2025 catalog, based on COSMOS imaging, includes our transient candidates in its photometric measurements and thereby skews any of its resulting photometric redshift estimates.  While most sources (80\%) have consistent photometric redshifts between COSMOS2025 and our internal catalog, the remaining discrepancies are largely due to the greater depth of the PRIMER data and the absence of detected transients from our catalog. 

\subsection{Single-epoch photometric classification}
\label{sec:classification}

Similar to the method implemented for the JADES sample in \citet{decoursey25a}, we attempt to classify each SN candidate using the STARDUST2 Bayesian light curve classification tool \citep{rodney14}, which is built on the underlying \texttt{SNCosmo}
framework and was originally designed for classifying SNe using HST. STARDUST2 uses the SALT3-NIR model to represent Type\,Ia SNe \citep[]{Guy2007SALT2, Kenworthy2021SALT3, pierel22} and a collection of spectrophotometric time series templates to represent CC SNe (27 Type\,II and 15 Type\,Ib/c). The CCSN templates are those developed for the SN analysis software \texttt{SNANA} \citep{Kessler2009SNANA:Analysis}, derived from the SN samples of the Sloan Digital Sky Survey \citep{Frieman2008THESUMMARY,Sako2008THEOBSERVATIONS,DAndrea2010TYPEMETHOD}, Supernova Legacy Survey \citep{Astier2006TheSet}, and Carnegie Supernova Project \citep{Hamuy2006TheSurvey,Stritzinger2009THEWAVELENGTHS,Morrell2012CarnegieSupernovae}. The models produced for \texttt{SNANA} were extended to the NIR by \citet{Pierel2018ExtendingObservations}. Within STARDUST2 a nested sampling algorithm \citep{Skilling2004NestedSampling} measures likelihoods over the SN simulation parameter space, including priors on dust parameters described in \citet{rodney14}. We use all available data for each classification, which is typically a single epoch (and sometimes more in rare cases with serendipitous observations from other programs) and up to 7 wide-band filters. STARDUST2 has been successfully implemented for single-epoch classification in the past \citep{golubchik23, decoursey25a}, and we are able to validate it by comparing our single-epoch photometric classifications to spectroscopic classifications for existing sub-samples with spectra (see DeCoursey et al. in prep for the JADES sample and Siebert et al. in prep for this COSMOS sample). 
 
Table \ref{tab:cosmos_classifications} lists the reduced $\chi^2$ values for both SNe Ia and CCSN fits. We do not include any 
fits where the degrees of freedom are $\leq0$ (i.e., there are fewer photometry points than light-curve parameters). We also do not include any reduced $\chi^2 > 5$~because these fits did not converge with our model library. The majority of the SNe are unclassifiable. Figure \ref{fig:mag_redshift} includes preliminary classifications for illustrative purposes, where a classification is assigned to the lowest reduced $\chi^2$. Even so, single epoch classifications have a number of limitations and should be treated carefully for any future rate studies. DeCoursey et al. (in prep) provide a detailed description of these limitations using the JADES sample. A detailed discussion on the COSMOS rates are beyond the scope of this paper, but will be the focus of future papers on this sample.

\subsection{Color, magnitude, and redshift evaluation}

\begin{figure*}
    \centering
    {
    \includegraphics[width=3.6in]{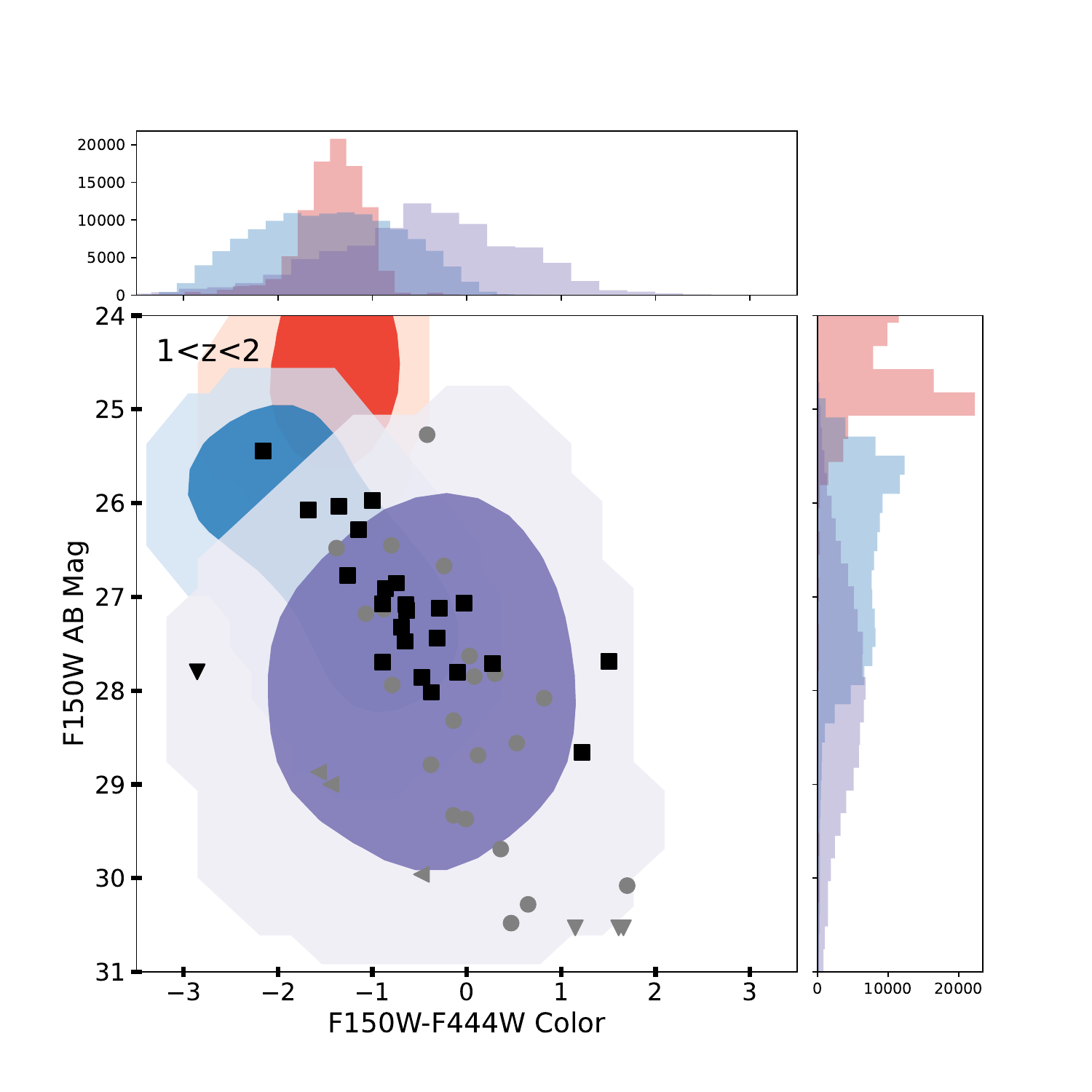}
    \hspace{-0.4in}
    \includegraphics[width=3.6in]{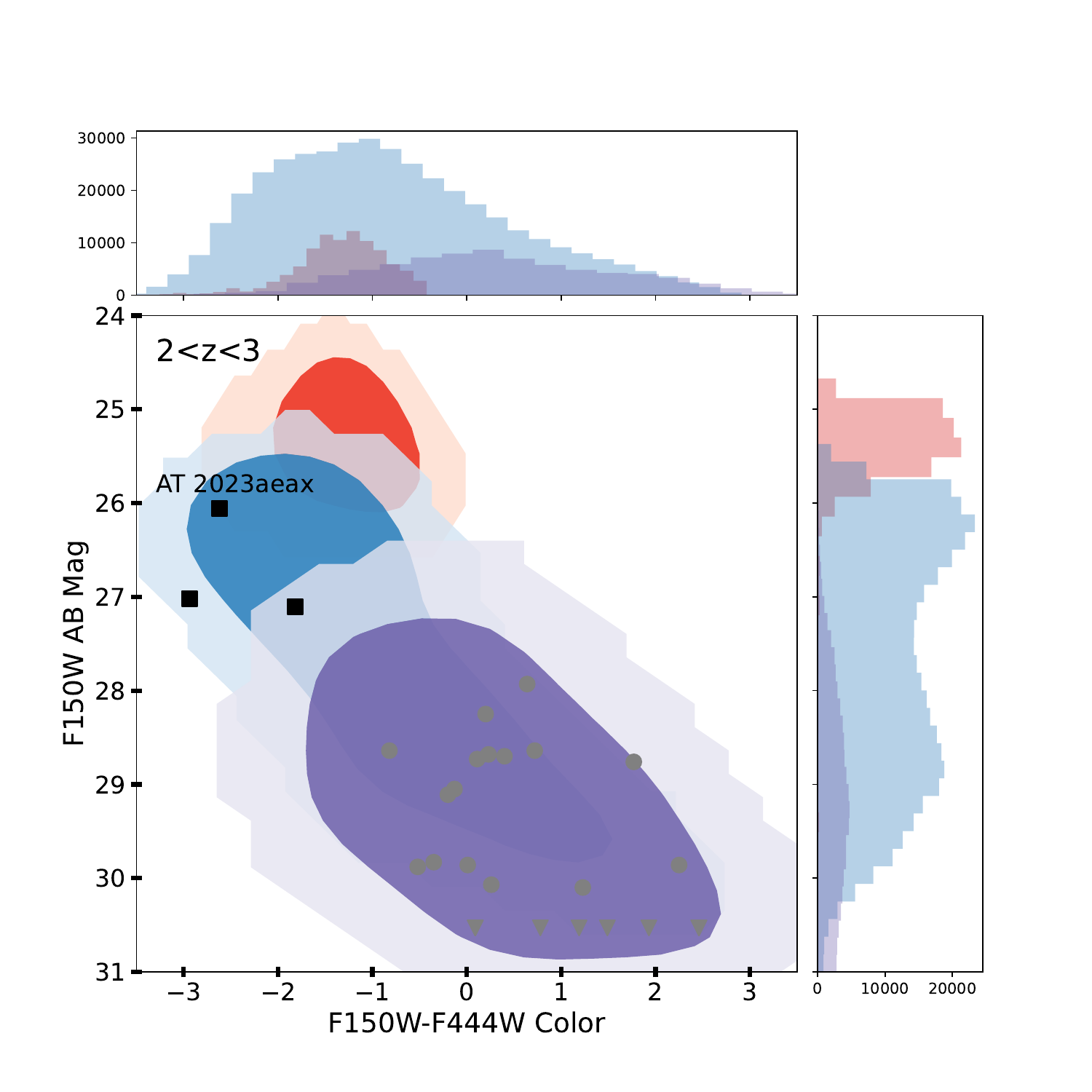}\\
    \vspace{-0.2in}
    \includegraphics[width=3.6in]{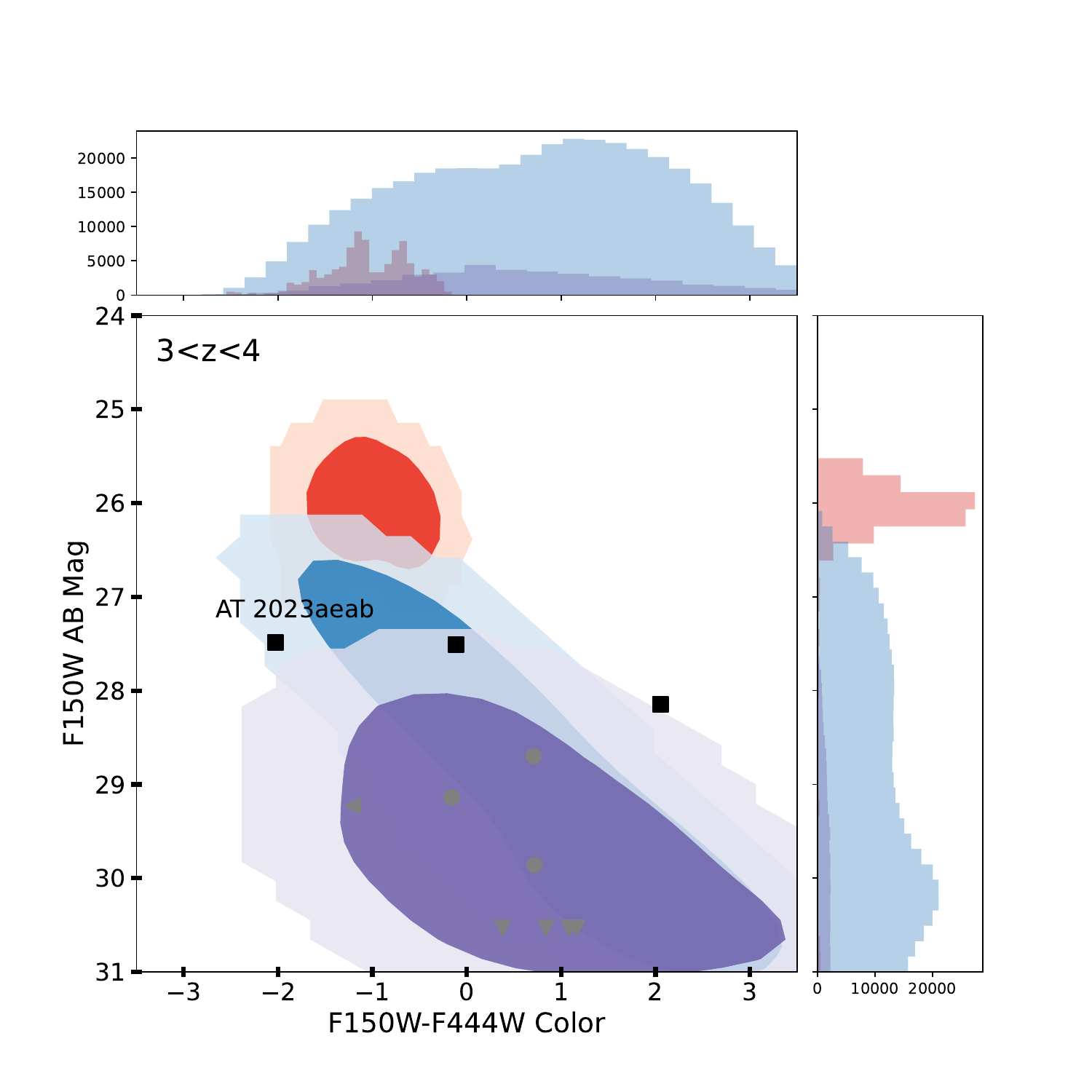}
    \hspace{-0.4in}
    \includegraphics[width=3.6in]{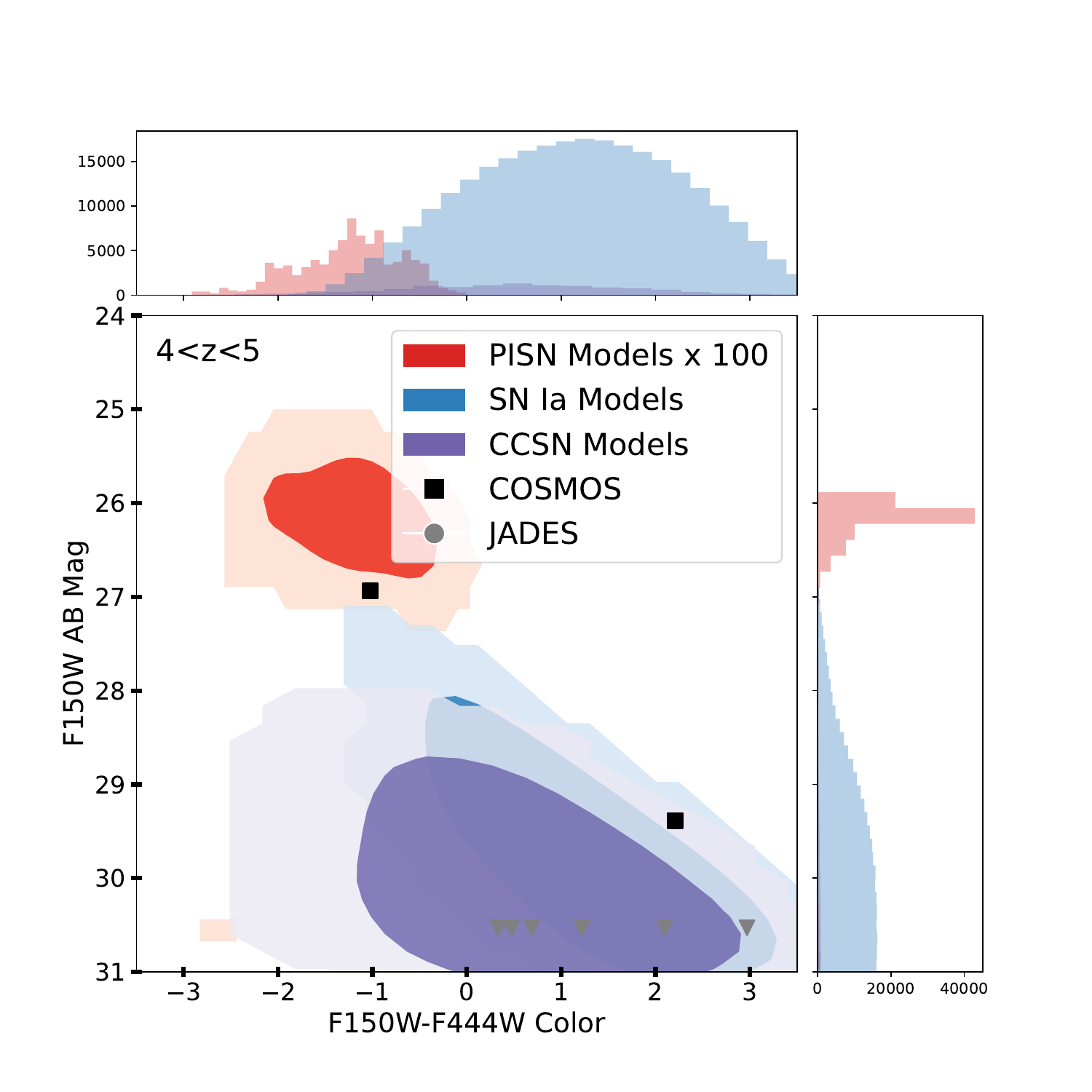}
    \caption{Color-magnitude diagram of theoretical models of SNe\,Ia (blue), SNe II (purple), and PISNe (red) in four redshift bins (where redshifts are photometric redshifts in Table \ref{tab:cosmos_classifications}). Type Ia and CCSN models are described in Section \ref{sec:classification}, while we use the 250 solar mass Red Supergiant (RSG) PISN models taken from \citet{kasen11}. Models range in epoch and redshift, although they are limited to one observer-frame year and no extinction. The shaded tiers for each color correspond to 68\%~and 95\%~of the models for each subclass, but do not provide any information on likelihood of each model. These contours could be loosely interpreted as 1- and 2-$\sigma$~values, although the distributions are not gaussian (see accompanying histograms). Overplotted are empirical results from COSMOS/PRIMER (black squares; this work) and JADES \citep[gray circles and triangles;][]{decoursey25a}. Histograms to the top and right of each contour provide absolute numbers.  Note that the PISN histogram is multiplied by a factor of 100 for visualization purposes.
    }
    \label{fig:cosmos_exotic}
    }
\end{figure*}

Figure \ref{fig:mag_redshift} plots the brightest magnitude (over the range of filters) for each SN against redshift for our COSMOS sample, divided by photometric classification from Section \ref{sec:classification}. Brightest magnitudes are chosen for plotting purposes because rest-frame photometric bands vary with redshift. The scatterplot offers a useful initial visualization of the entire distribution, particularly when comparing it against other JWST samples, such as JADES, or the capabilities of other surveys, such as HST or Roman. However, it is limited in its ability to identify targets for focused follow-up observations. The photometric classifications from Section \ref{sec:classification} are only based on light-curve fits to a single epoch of photometry and have a limited number of models that do not include tails of the CCSNe/SNe Ia distributions or extreme explosions, such as SLSNe or PISNe. Taken all together, the representation of the data in Figure \ref{fig:mag_redshift} alone doesn't have the ability to discern potentially unique or extreme events. 

A color-magnitude diagram (CMD) offers an alternative and convenient visualization of the results and can help to identify exotic SNe with extreme colors and/or luminosities, particularly when only a single epoch of data is available. By also incorporating host-galaxy redshift information, we can filter out low-$z$~contamination events. Figure \ref{fig:cosmos_exotic} plots our observational sample (black filled squares) onto a series of CMDs, each divided into redshift bins. Also included is the observational sample from JADES \citep[grey filled circles;][]{decoursey25a}. Contours show the theoretical color-magnitude loci for different models of SNe~Ia (blue), SNe~II (purple), and PISNe (red). Each SN subclass becomes redder and fainer as it evolves, which explains why each contour generally extends from the top left (bright and blue) to the bottom right (faint and red) of each plot.

The contours include models spanning a range of ages, redshifts, and reddening. The shaded tiers correspond to 68\%~and 95\%~of each distribution of theoretical models. All models are weighted equally. These contours could be loosely interpreted as 1- and 2-$\sigma$~values, although note the distributions are not gaussian (see accompanying histograms) and, furthermore, the models are not weighted by likelihood. In other words, the contours do not communicate the relative frequency of occurrence of each subclass or even the models within each subclass. Histograms to the top and right of each contour provide absolute numbers to get a better set of context and relative probabilities.  Note that the number of PISNe are multiplied by a factor of 100 for visualization purposes in the histogram.

\section{Follow-Up Observations}
\label{sec:followup}

Our initial sample includes only a limited number of epochs and lacks accompanying spectra. Although a broader analysis, such as a rate study, is possible, targeted follow-up of a subsample remains important for several reasons. Detailed observations of individual high-$z$ SNe allow us to probe the early Universe directly. These explosions provide insight into how SNe, their progenitor stars, and their host galaxies evolve with metallicity and redshift. They also offer the opportunity to identify new, extreme, or otherwise exotic events, including those arising from Population III stars or the elusive PISNe. Finally, assembling a well-sampled observational set will enable the construction of training data for machine-learning algorithms, improving our ability to filter and prioritize future candidates without requiring extensive, costly follow-up.

As previously stated, neither the COSMOS nor PRIMER programs were designed as time-domain surveys, so retiling multiple times to build light-curves is not possible. Furthermore, our sample is too large to allow individual follow-up of every source. Given the 0.037 deg$^2$~of overlapping area between the two surveys, any follow-up strategy therefore requires careful prioritization of the candidates and optimized observational techniques.

\subsection{Candidate prioritization and identification}
\label{subsec:follow_up}

\begin{figure}[t]
    \centering
    {
    \includegraphics[width=3.6in]{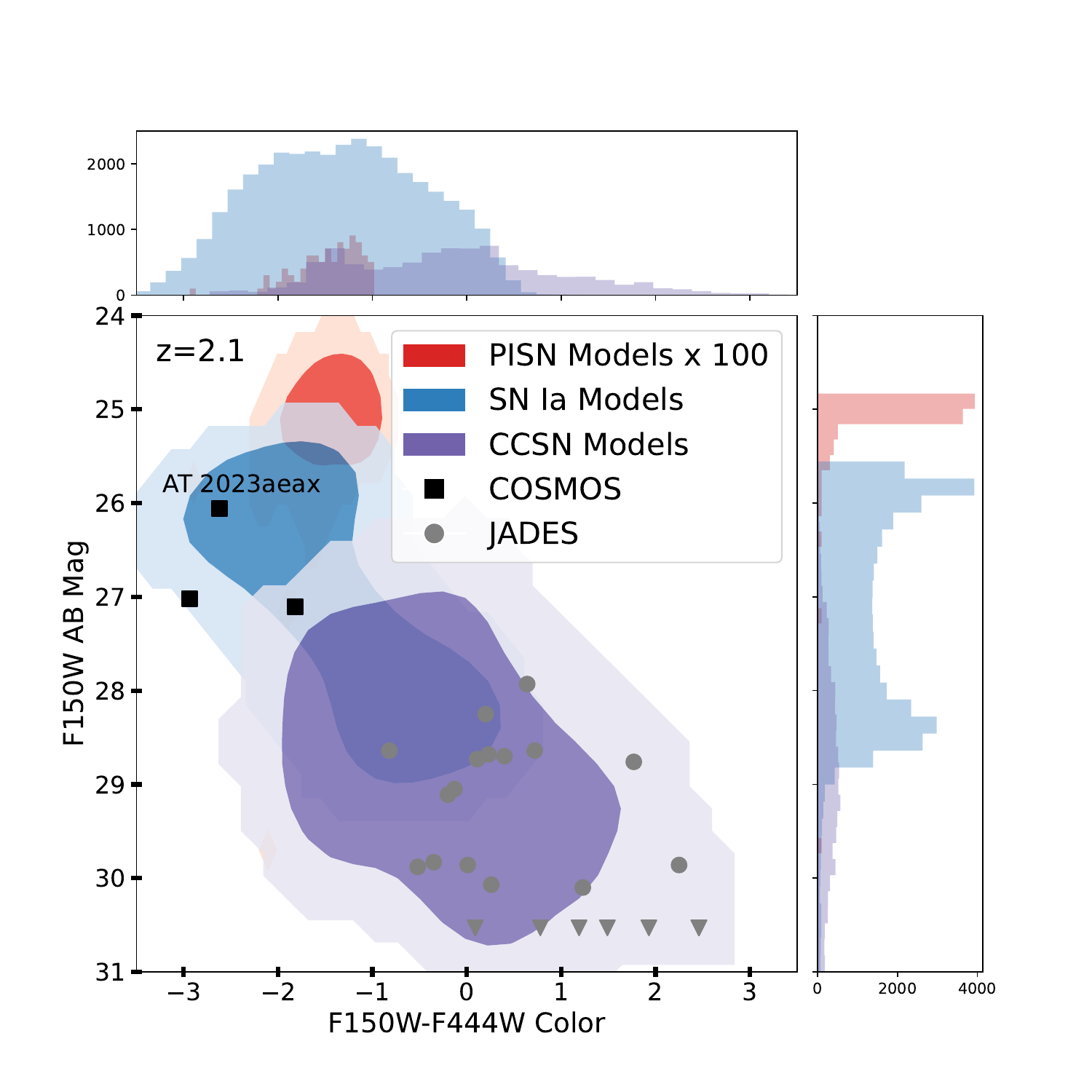}\\
    \vspace{-0.2in}
    \includegraphics[width=3.6in]{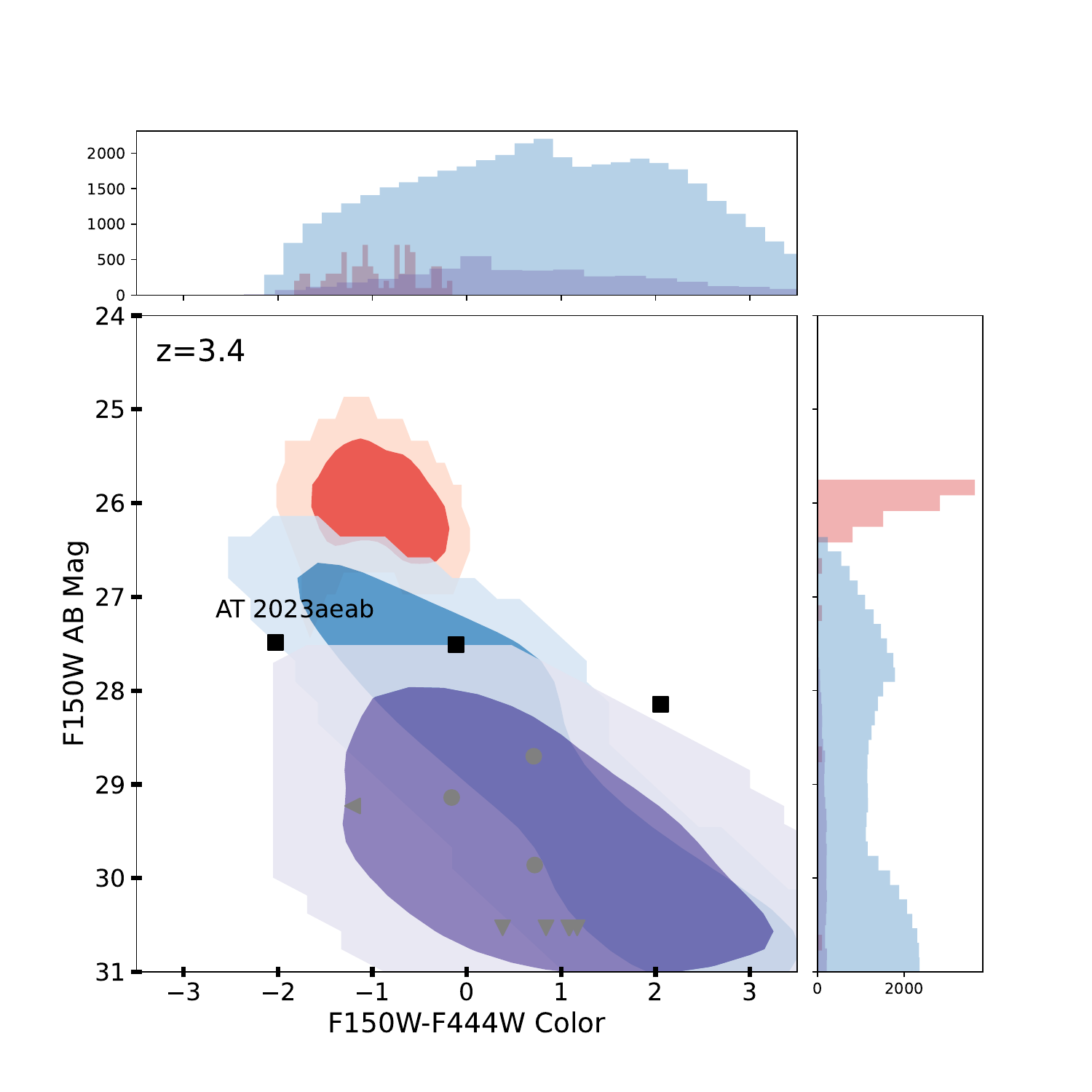}
    \caption{Same as Figure \ref{fig:cosmos_exotic}, but for the specific redshift of both ({\it top}) SN 2023aeax and ({\it bottom}) SN 2023aeab.
    }
    \label{fig:cosmos_exotic_slice}
    }
\end{figure}

Figure \ref{fig:cosmos_exotic} provides a visual tool for identifying SNe that are promising follow-up targets. The position of individual SNe on the CMD can quickly highlight extreme or unusual events that fall outside the typical distribution of SNe Ia and CCSNe at the corresponding redshifts (or within the typical distribution if that is the scientific goal). In the redshift bin $3<z<4$, one source (SN 2023aeab) stood out to us in terms of redshift, brightness, and color. Figure \ref{fig:cosmos_exotic_slice} (bottom) shows the CMD specific to SN 2023aeab's host photometric redshift ($z\approx3.4$; Table \ref{tab:cosmos_classifications}). The other COSMOS and JADES SNe from $3<z<4$~are also included for reference. Selecting targets of interest may vary according to an individual's criteria. SN 2023aeab stood out to us because of its high redshift ($z\approx3.4$), extremely blue color, and the fact that its CMD position was beyond 95\% (i.e., 2$\sigma$)~of the SNe II and Ia models. It is also within the 99\% contour (i.e., 3$\sigma$; not shown) of the PISN distribution. The other COSMOS targets are also potentially interesting given their positions in the CMD, but given the limited follow-up capabilities available to us, we chose SN 2023aeab as our top priority and proceeded to obtain an additional imaging epoch with NIRCam and spectroscopy with NIRSpec as part of the JWST DDT Program (PID 6585; PI Coulter).\\

\begin{table*}[!t]
\begin{center}  
    \small
    \caption{Transients targeted by our followup program (PID 6585; PI Coulter)} \vspace*{-5pt}
\begin{tabular*}{\linewidth}{@{\extracolsep{\stretch{1}}}*{4}{c}}
\hline
ID  & Name & Spectroscopic Redshift & Instrument \\\hline
  1 & AT 2023aeab &  3.368 (0.002) & NIRSpec+ NIRCam \\2 & AT 2023aeaf & 3.195 (0.002) & NIRSpec$+$NIRCam \\3 & AT 2023adzt & $\sim$2.4 (photo-$z$~only) & NIRCam \\4 & AT 2023aeax & 2.13 (0.01) & NIRSpec+NIRCam \\5 & AT 2023adzu & 1.253 (0.002) & NIRSpec+NIRCam \\6 & AT 2023adzy & 1.432 (0.001) & NIRSpec \\7 & AT 2023adzq & $\sim1.5$ (photo-$z$~only) & NIRCam \\9 & AT 2023adzs & $\sim1.1$ (photo-$z$~only) & NIRCam \\10 & AT 2023adzr & 1.26 (0.01) & NIRSpec+NIRCam \\11 & AT 2023adzv & 1.857 (0.001) & NIRSpec+NIRCam \\13 & AT 2023adzz & $z<1$ (photo-$z$~only) & NIRCam \\\end{tabular*}

\vspace*{5pt}
\hspace*{0cm} 
\label{tab:targets}
\end{center}
\normalsize
\end{table*}

\subsection{NIRCam Imaging and NIRSpec Spectroscopy}

\begin{figure}[t]
    \centering
    {\includegraphics[width=8cm]{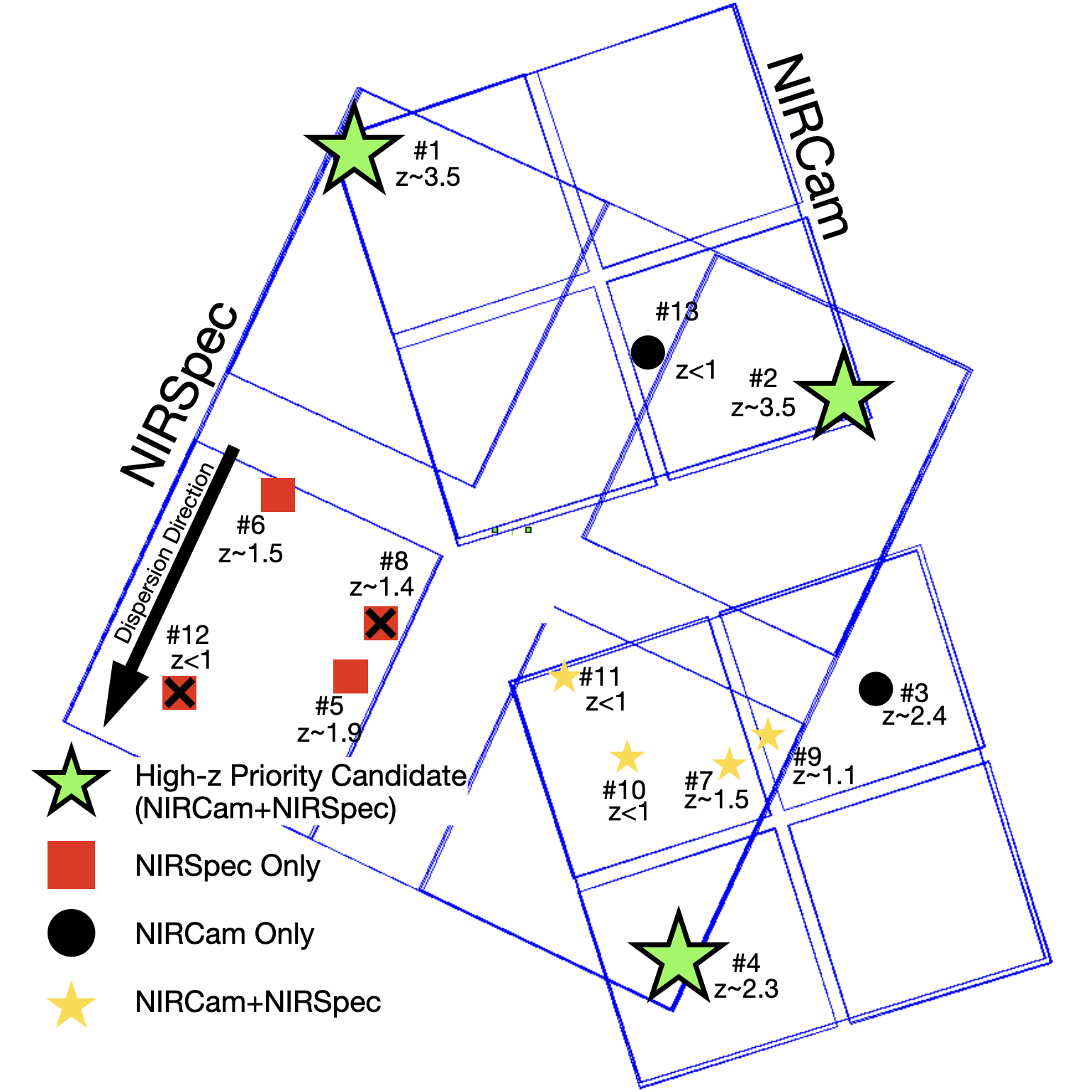}}
    \caption{Design of the follow-up observations to SN 2023aeab (target \#1) from the COSMOS sample (see Table \ref{tab:targets} for target number identification). Dynamic follow-up greatly improves the efficiency, allowing us to capture multiple SNe in both our NIRCam and MSA pointings. NIRspec targets with an `X' were not observed due to overlapping spectra along the dispersion direction.}
    \label{fig:DDfootprint}
\end{figure}

In addition to SN 2023aeab, we designed the spectroscopy using the NIRSpec microshutter array (MSA) and optimized the NIRCam imager rotation to efficiently capture a number of other events in the field-of-view (Figure \ref{fig:DDfootprint}). SN 2023aeax, for example, is another SN that stands out given its blue color, bright magnitude, and high redshift, particularly because it falls into the typical Type Ia distribution at $z>2$, making it an ideal candidate for follow-up (see Figures \ref{fig:cutout_SN1}, \ref{fig:cosmos_exotic}, and  \ref{fig:cosmos_exotic_slice}). SN 2023aeax fits into both our MSA and NIRCam pointings.

\begin{figure}[t]
    \centering
    {\includegraphics[width=8cm]{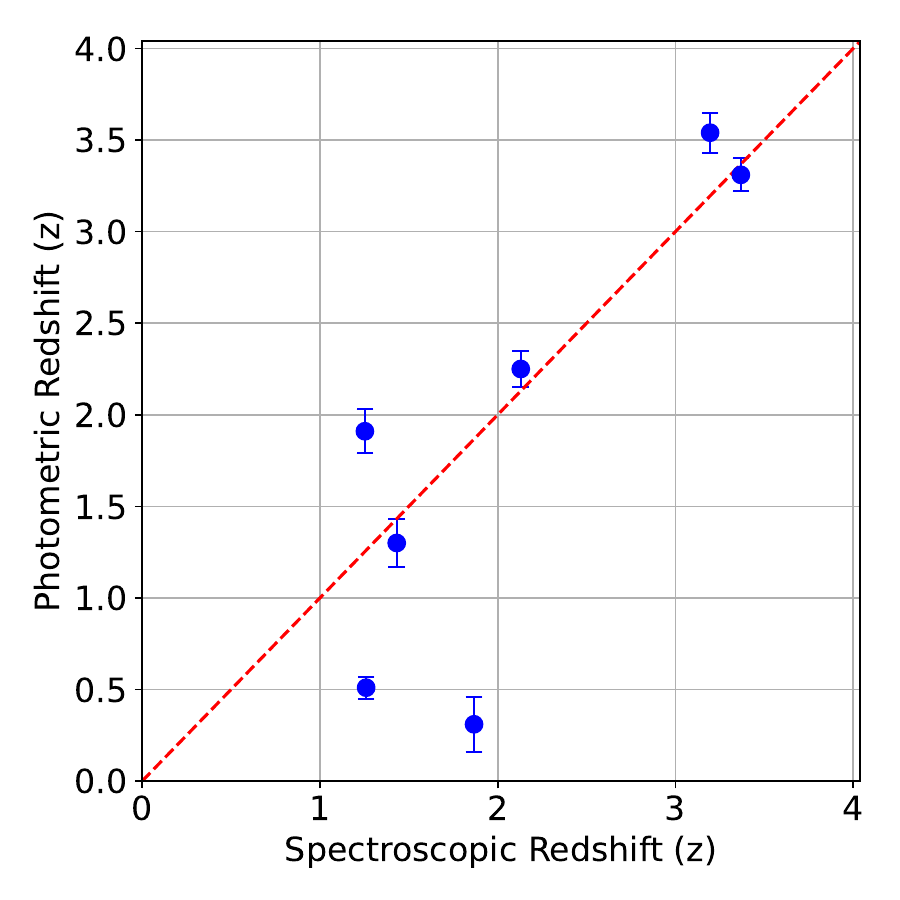}}
    \caption{Plot of redshifts derived spectroscopically (Table \ref{tab:targets}) versus photometrically (Table \ref{tab:cosmos_classifications}).}
    \label{fig:redshifts}
\end{figure}

Such dynamic follow-up dramatically improves the efficiency of follow-up observations. A total of ten transients fit on our MSA pointing, with six covered by NIRSpec+NIRCam, one covered by only NIRSpec, and four covered by only NIRCam (Table \ref{tab:targets}). 

\begin{figure*}[t]
    \centering
    {
    \includegraphics[height=4.5in]{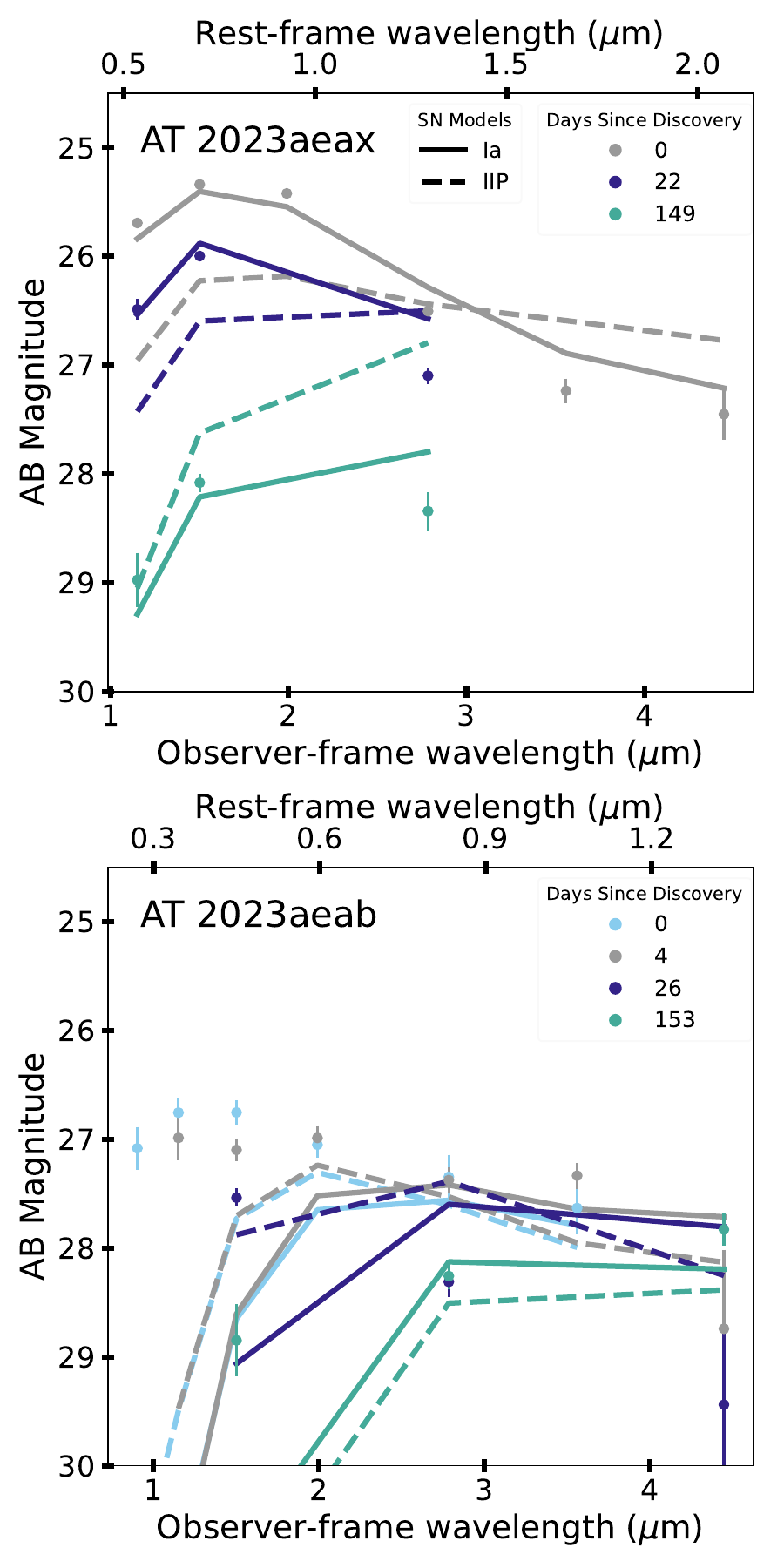}
    \includegraphics[height=4.5in]{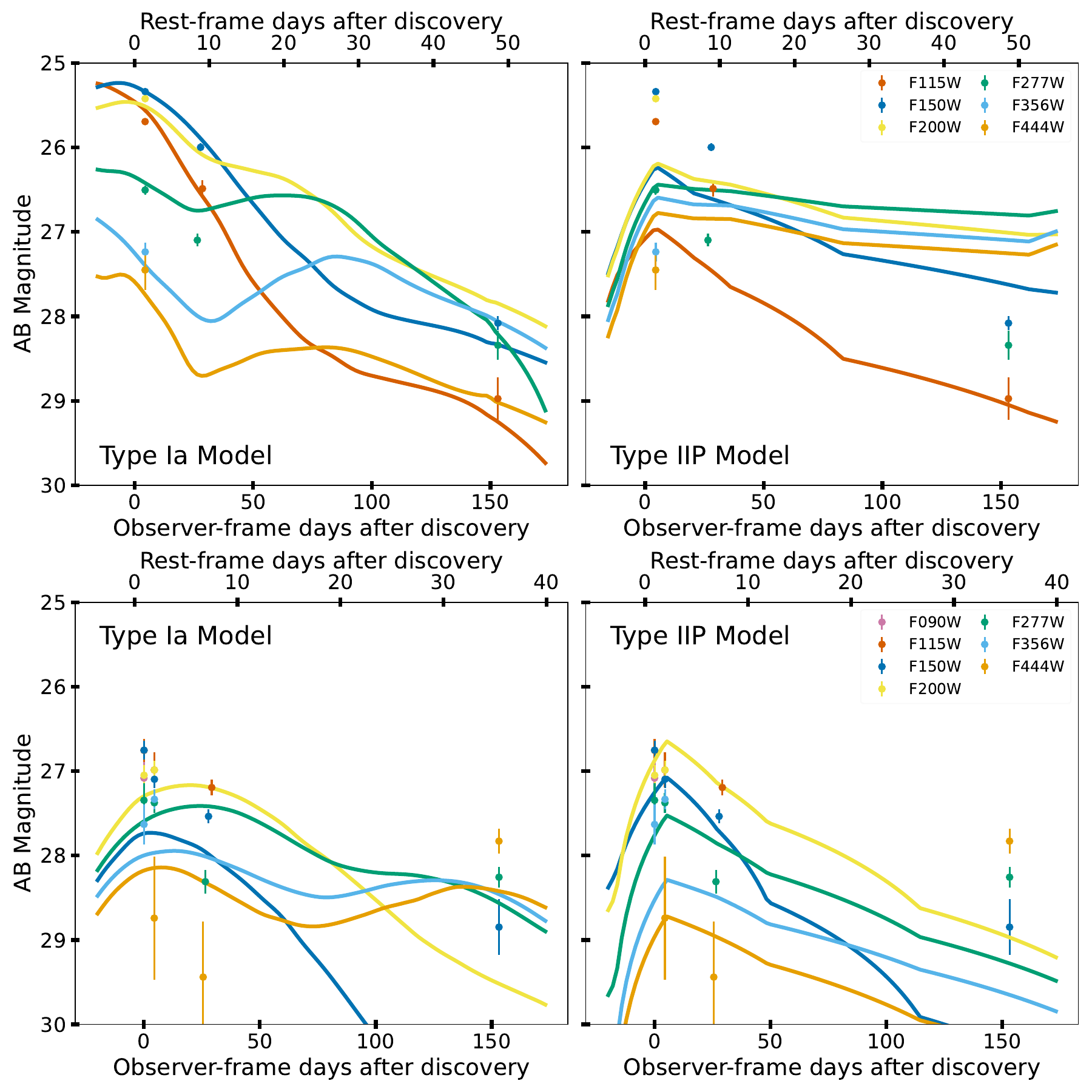}
    \caption{Photometry for ({\it top}) SN 2023aeax and ({\it bottom}) SN 2023aeab (SNe \#4 and \#1, respectively in Table \ref{tab:targets}) The photometry is plotted as (\textit{left}) SEDs and (\textit{right}) light-curves. Best-fitting generic Type Ia and CCSN models (described in Section \ref{sec:classification}) are overplotted. For the SEDs, lines connect synthetic photometry calculated for each model at a particular bandpass. SN 2023aeax illustrates that Type Ia SNe and CCSNe are best differentiated with a minimum of three light-curve points over rest-frame 50 days. The early-time, short-wavelength evolution offers particularly strong leverage for disentangling the two models. For SN 2023aeab, neither SN template adequately reproduces the early-time blue emission, which suggests a more energetic, tailored SN model is needed than typically considered (Coulter et al, in prep).}
    \label{fig:2023aeax}
    }
\end{figure*}
 
We obtained NIRCam images in the filters  F115W, F150W, F277W, and F444W, using the SHALLOW4 readout pattern with 8 groups and 3 standard subpixel dithers, resulting into 1256~sec exposures to a depth of about 28-29~AB mag with a S/N=5. We also obtained NIRSpec MSA observations of a subsample of the transients within the NIRCam field of view, along with their hosts. Observations for both instruments were obtained on $2024$~April~$29$~(MJD $60429$). 

The NIRCam imaging data were reduced and analyzed as described in Sections~\ref{sec:differencing}--\ref{sec:forced_photometry}. The details of the MSA spectroscopic sample observations and reduction will be described in a separate paper (Siebert et al., in prep). However, we report the spectroscopic redshifts in Table \ref{tab:targets}. For comparison, Figure \ref{fig:redshifts} plots the spectroscopic versus host photometric redshifts. Qualitatively, the host photometric redshifts successfully predict the spectroscopic redshift, particularly at $z>1$. At $z<1$, the host photometric redshifts are less reliable, which is expected because of the lack of shorter wavelength data. The Balmer break feature is in $u$/$b$-band for low-$z$~galaxies \citep{crenshaw25}. These results suggest that, assuming similar depth and wavelength coverage in future surveys, we can rely on photo-$z>$1 to trigger for follow-up of high-$z$~targets.h

\subsection{Additional Light-Curve Information}

Even one additional epoch of NIRCam photometry allows us to perform more reliable light-curve fits of the SNe in our follow-up sample using similar methods as described in Section \ref{sec:classification}. These fits also offer a useful sanity check to our initial single epoch results.

SN 2023aeax and 2023aeab (SNe 4 and 1 in Table \ref{tab:targets}, respectively), in particular, are good test cases. SN 1 also has one additional epoch of photometry obtained serendipitously by JWST program PIDs 2514 (PI Williams). Their complete SEDs and multi-filter light-curves are plotted in Figure \ref{fig:2023aeax}. For each light-curve, generic SNe Ia \citep[SALT3;][]{Guy2007SALT2, Kenworthy2021SALT3, pierel22} and IIP \citep[][]{cappellaro97,gilliland99,baron04} templates are overplotted for reference.

For SN 2023aeax, the combined SED and light curve are best fit by a generic Type Ia SN model, particularly at the shortest wavelengths. The brightness of the source in short wavelengths indicates that the SN was observed shortly after explosion. For SN 2023aeab, neither the generic Type Ia nor Type IIP templates provides a strong fit, although the IIP model performs better both qualitatively and quantitatively. The largest discrepancy arises at the shortest wavelengths, where SN 2023aeab also exhibits an apparent blue excess. This is consistent with the CMD in Figure \ref{fig:cosmos_exotic_slice}, which places SN 2023aeab in the 95\% contour ($\sim2\sigma$) of the CCSNe distribution and approaching the PISN regime. Together, these findings reinforce the unusual nature of SN 2023aeab. 

\section{Discussion}
\label{sec:discussion}

\subsection{A New Frontier}
\label{sec:frontier}

This COSMOS sample highlights JWST’s ability to discover high-$z$ SNe. As discussed in Section \ref{sec:classification}, the combined surveys, spanning $\sim\!133$ arcmin$^2$ and reach depths of $\sim\!28$ AB mag, yielded \nsne~SNe ($\sim\!0.5$ SN arcmin$^{-2}$) extending to $z \lesssim 5$. Although the discovery rate is lower than that of the JADES sample \citep[$>1$ SN arcmin$^{-2}$;][]{decoursey25a}, the COSMOS sample include a larger fraction of brighter and bluer SNe (see discussion in Section~\ref{sec:area}). Spectroscopic follow-up observations confirm our initial classifications (Siebert et al. in prep): SN 2023aeab is a unique Type IIP at $z=3.368$, unusually bright and blue for its class, and SN 2023aeax is a Type Ia at $z=2.13$ caught near the peak of the light-curve. These discoveries enable us to study new physics. For example, a detailed analysis of SN 2023aeax has shown a possible redshift-dependent bias/evolution of SNe Ia at $z\gtrsim2$ \citep{pierel25}. Future work on SN 2023aeab will examine the effects of low metallicity on Type IIP explosions in the early Universe (Coulter et al., in prep). The full sample will ultimately enable high-$z$~SN population studies, including volumetric rates (e.g., DeCoursey et al., in prep) and links to cosmic star formation (e.g., Vassallo et al., in prep). Together, these examples illustrate only a small fraction of the high-redshift science now accessible as JWST opens this new observational frontier. 

\subsection{Area vs Depth}
\label{sec:area}

\begin{table*}[!t]
  \begin{center}  
    \small
    \caption{Survey Details and Rates} 
    \label{tab:surveys}
\vspace*{-5pt}
\begin{tabular*}{\linewidth}{@{\extracolsep{\stretch{1}}}*{8}{|c}|}
\hline
Survey$^{\rm a}$    & Area        & Exp. Time$^{\rm a}$  & Exp. Time$^{\rm b}$  & Depth$^{\rm b}$  & \# SNe$^{\rm b}$   & \# SNe$^{\rm b}$   & \# SNe$^{\rm b}$ \\
& deg$^2$  & NIRCam/NIRISS & NIRSpec & F277W/F356W & $1<z<2$ & $2<z<4$ & $z>4$ \\
             &  & (sec) & (hr) &  & & & \\
 \hline
 JADES          & 0.007          &  12000   &   7       & 30.0  & 28 & 33 & 3-7 \\
COSMOS     & 0.037        & 1000     &   4       & 28.3  & 27 & 7 & 3 \\
\hline
\end{tabular*}
\tablecomments{
\noindent $^{\rm a}$JADES and COSMOS taken from \citep{decoursey25a} and \citep{casey23}, respectively; $^{\rm b}$per epoch. 
}
\end{center}
\end{table*}

Table \ref{tab:surveys} compares the COSMOS and JADES surveys and their resulting samples. The COSMOS sample is shallower than the JADES sample by roughly two magnitudes, yet it includes events that are significantly brighter and bluer, even at comparable redshifts (Figures \ref{fig:cosmos_exotic}, and \ref{fig:cosmos_exotic_slice}). 
 This behavior can be understood by considering the detectability of transients such as SNe~Ia and PISNe (see Figure~\ref{fig:depthvsarea}). The depth of COSMOS is sufficient to detect these events near peak brightness even at high redshift, while the additional depth of JADES primarily enables detections at fainter phases of the light curve, typically at later times when the SNe are intrinsically redder, as opposed to detection of higher redshift events. As a result, increasing survey depth does not substantially increase the number of SNe~Ia, CCSNe, and PISNe discovered near maximum light, nor does it preferentially yield more young, luminous events with the highest signal-to-noise. Increasing the survey area, however, directly increases the number of SNe detected near peak and enhances the probability of discovering intrinsically rare transients.

JADES reaches greater depth than COSMOS, but its smaller footprint lends itself to discoveries of transients in the central contours (i.e. 68\% or 1$\sigma$) of color–magnitude space populated by typical CCSNe and SNe Ia. It is more difficult to disentangle young, unique, and/or intrinsically rare SNe in this region of the CMD. The extreme red edge of the CMD remains sparsely populated by either survey simply because PISNe are exceedingly rare, and even the larger area of the COSMOS sample ($\sim$0.037 deg$^2$)~provides little chance for discovery. Taken together, Figures \ref{fig:mag_redshift} and \ref{fig:cosmos_exotic} illustrate that, even in the COSMOS versus JADES comparison, survey area—not depth—is the key factor for building a statistically robust SN sample at $1<z<3$~and for discovering the brightest, rarest events at $z>3$.

In particular, for SNe Ia cosmology, $\sim5-10$~classified, well-sampled events are required to distinguish between a luminosity bias or not at the $5\sigma$~level. For CCSN rate science, existing FUV CSFH measurements out to $z>10$ \cite{bouwens23,harikane23} indicate that $>200$~classified CCSNe per year at $2 < z < 4$ are necessary to obtain $<5\%$ precision for multiple SN subtypes. Both simulations \citep{strolger15} and empirical results \citep{coulter25,decoursey25a} show that $>0.1$ deg$^2$~is the minimum area necessary to build such samples. For PISNe, we don't have empirical results, but models indicate 0.25 deg$^2$~is the minimum area to put the SLSN+PISNe rate in the single digits \citep[][]{moriya22a}.

\subsection{Implications for future high-$z$~surveys}
\label{sec:future}

These JWST discoveries from COSMOS and JADES are spectacular but represent only the tip of the iceberg because neither JADES nor COSMOS were designed for time-domain astronomy. Both data sets were plagued by a minimal number of epochs (see also DeCoursey et al. in prep) and had limited follow-up imaging and spectroscopy. Despite the COSMOS-Web/PRIMER overlap of 0.037 deg$^2$~area, the FOV is still too small to build the most robust statistics for both rate studies and Type Ia cosmology, not to mention achieve the necessary rates to discover the elusive PISNe. Both fields are outside JWST's Continuous Viewing Zone (CVZ), limiting year-round monitoring. A current survey, NEXUS (PID 9263; PI Shen), is located in the CVZ and designed with some SN/time-domain science in mind, but has limited utility for low redshift SN detection because of its slow cadence ($\sim$2 months). Its utility is also limited for high redshift SNe because of its shallow integrations and use of only two filters (F200W+F444W); high-$z$ SN classification is difficult without F115W, which covers the bluest part of the spectrum (see the short-wavelength regime of the SED for SN 2023aeax in Figure \ref{fig:2023aeax}). Note also that new lensing search programs with JWST templates are underway and quite promising (see PID 6882; PI Fujimoto). For example, \citet{coulter26} recently discovered a strongly lensed Type IIP SN at $z = 5.13$. But the strength of these observations is in the detailed follow-up of individual targets, not large samples \citep[e.g,][]{coulter26}.

Nonetheless, the COSMOS sample lays the necessary foundation for many more studies and future work. For example, the COSMOS sample highlights the value of having a wide footprint, as it is necessary to discover the more rare and exotic phenomena. Even if a survey depth reaches only $\sim$28 mag AB, this is more than sufficient to detect SNe Ia and CCSNe at $2<z<4$~and potentially SLSNe and PISNe out to $4<z<6$~(Figures \ref{fig:depthvsarea} and \ref{fig:cosmos_exotic}). This sample can also be used to test the redshift evolution of properties of low-$z$~($z<1$) SN samples (i.e., physics, rates, etc.), perform population studies, and conduct detailed analyses on the more unusual and rare events. Future high-$z$~surveys will benefit from such analyses and can be optimized for specific observational goals. 

A dedicated, well-designed JWST time-domain survey is needed maximize the scientific return and minimize the time spent per epoch collecting data. To achieve the science goals outlined above and properly maximize the unique phase space (primarily redshift) afforded by JWST, a properly designed survey must consider tradeoffs to area, filter choice, integration times, and cadence to allow: (1) high enough number counts for statistically significant conclusions; (2) differentiation of CCSNe from SNe Ia out past cosmic noon ($z\approx4$) for volumetric rate constraints; (3) fit a SN Ia light-curve for cosmology at redshifts higher than {\it Roman} (i.e., $z>3$) ; and (4) discovery and characterization of rare SNe of interest (i.e., SLSNe and PISNe) in the low-metallicity Universe around the Epoch of Reionization (i.e., $z>6$), where Pop III stars are most likely to exist. 

Figure \ref{fig:depthvsarea} puts some of these points into context. We model the number of SNe derived from simulations of a corresponding 5-year survey for both Type Ia SNe at redshifts $z>3$ and Pop-III PISNe. For simplicity, the simulated surveys all have a 6-month cadence in four JWST/NIRcam bands (F115W, F150W, F277W, and F444W), all with the same limiting magnitude per epoch. The SN yields correspond to the number of SNe detected in at least one band at any phase. The red line traces the area and magnitude limits possible with 200 hours on JWST (including overheads; see Appendix~\ref{appendix}). Increasing depth has limited benefits below about 28th magnitude, but the survey yields increase proportionally with survey area. The area and depths of several Cycle 1 JWST surveys are marked for comparison, but note these surveys are not uniform and all have different durations, filters, and cadences from the simulations. A properly designed JWST survey with an area and depth similar to the entire COSMOS-Web field ($\sim$28th AB mag over $\sim$0.5~deg$^2$) would be most ideal for our desired science goals.

\subsection{Synergies with Roman}
\label{sec:roman}

Coordinating such a survey with the Roman High-Latitude Time-Domain Survey \citep[HLTDS;][]{jha25} offers a remarkable opportunity to combine the strengths of both surveys, maximizing JWST’s science return and enhancing HLTDS results. JWST and {\it Roman} could together produce an unprecedented foundation for time-domain science. Their complementary data would enable measurements of SN rates through cosmic noon, Type Ia cosmology out to $z>2.5$, and spectroscopic confirmation of extreme SNe (SLSNe and PISNe) observed by both telescopes. JWST’s deep, long-wavelength imaging and spectroscopy of select transients will aid classification and supply a machine-learning ``truth'' set to identify high-$z$ events across Roman’s wider field. We note, even for such a joint survey, the necessity for shorter wavelength photo-$z$'s (Section \ref{sec:photo-z}). At $z<5$, JWST alone cannot detect the Ly$\alpha$ dropout. Unlike the COSMOS field, there is not shorter wavelength data currently available in all parts of the HLTDS field. However, the HLTDS South will have {\it Roman}+Rubin for unimodal galaxy SEDs, while the North field is partly covered by Subaru HSC, with completion expected soon.

\section{Conclusions} 
\label{sec:conclusion}

Two Cycle 1 treasury programs (COSMOS-Web and PRIMER) obtained overlapping images in 4 NIRCam filters (F115W, F150W, F277W, F444W). The area is $>5$ times larger than JADES but not as deep (133 arcmin$^2$ to $\sim\!28$~mag on average). A total of \nsne~SNe (0.5 SNe/arcmin$^2$) were discovered. Most of the candidates had photo-$z$'s at $1<z<2$, but ten were at $2<z<5$. Compared to JADES, a smaller percentage of the sample is at higher redshifts, but the COSMOS sample tend to be brighter, bluer, and younger than the JADES sample found at comparable redshifts. This result is due to the nature of having a wide-area survey, which increases the statistical likelihood of finding not only rarer events, but also SNe at positions on the light-curve where they spend little time (i.e., near peak). This is best exemplified by SN 2023aeab (Coulter et al. in prep).

Even with a single epoch of photometry, we were able to characterize, filter, and prioritize individual SNe for detailed photometric and spectroscopic follow-up using several methods. In particular, the CMDs in Figures \ref{fig:cosmos_exotic} and \ref{fig:cosmos_exotic_slice} provide a useful method for quickly visualizing the sample and identifying where individual SNe exist relative to populations of various SN subclasses. Depending on follow-up priorities, this technique offers an efficient way to identify targets of interest. Dynamic follow-up allows the NIRCam and NIRSpec orientation and slit arrangement to maximize the number of targets for follow-up in each pointing. Follow-up observations confirmed that for this limited sample, initial host photometric redshifts and classifications are reliable estimates of the actual redshift, particularly identifying intermediate and high-redshift SNe.

Certain scientific investigations are uniquely enabled by JWST. This sample provides a foundational basis for numerous follow-up studies and future investigations, while underscoring the need for a dedicated, long-term JWST SN discovery program. The results also elucidate the relative strengths and limitations of specific survey design choices—including survey area, filter selection, integration time, and cadence—thereby informing the optimization of future, dedicated JWST time-domain surveys for targeted scientific objectives.

A JWST high-$z$~time-domain survey would also have significant synergies with the {\it Nancy Grace Roman Space Telescope}. Expected to launch in late 2026, \textit{Roman} will usher in a new era of high-$z$ discoveries, with tens of thousands of new transients to be uncovered via its predefined HLTDS and High-Latitude Wide-Area Survey \citep[HLWAS;][]{jha25}. The sheer volume of discoveries will present a significant challenge for sorting and prioritization for detailed follow-up. While machine-learning approaches are increasingly being developed to address this problem, their performance ultimately depends on the availability of high-quality, well-characterized training data. Smaller, but well-studied, JWST ``truth'' catalogs will make it possible to train and validate these tools in advance of \textit{Roman}, enabling efficient sorting and prioritization of the most high-value high-$z$~targets. Ultimately, follow-up photometric and spectroscopic of most of these high-$z$~{\it Roman} SNe will still only be possible with JWST. A new era of time domain astronomy is upon us, and these latest JWST results are just the beginning.

\section*{Author Contributions}

{\noindent Ori D. Fox and Armin Rest contributed equally to this work and are considered co-first authors. All authors discussed the results and commented on the manuscript.} 

\begin{acknowledgments}
Some/all of the data presented in this paper were obtained from the Mikulski Archive for Space Telescopes (MAST) at the Space Telescope Science Institute. The specific observations analyzed can be accessed via \dataset[https://doi.org/10.17909/n7kq-ef83]{https://doi.org/10.17909/n7kq-ef83}. STScI is operated by the Association of Universities for Research in Astronomy, Inc., under NASA contract NAS5–26555. Support to MAST for these data is provided by the NASA Office of Space Science via grant NAG5–7584 and by other grants and contracts. The STScI TSST group acknowledges partial support from  JWST-GO-06541, JWST-GO-06585, and JWST-GO-05324. This work was made possible by utilising the CANDIDE cluster at the Institut d’Astrophysique de Paris. The cluster was funded through grants from the PNCG, CNES, DIM-ACAV, the Euclid Consortium, and the Danish National Research Foundation Cosmic Dawn Center (DNRF140). It is maintained by Stephane Rouberol. GEM acknowledges the Villum Fonden research grants 37440 and 13160. The Cosmic Dawn Center (DAWN) is funded by the Danish National Research Foundation under grant DNRF140. This project has received funding from the European Union’s Horizon 2020 research and innovation programme under the Marie Skłodowska-Curie grant agreement No 101148925. The French contingent of the COSMOS team is partly supported by the Centre National d'Etudes Spatiales (CNES). SF acknowledges support from the Dunlap Institute, funded through an endowment established by the David Dunlap family and the University of Toronto. ODF acknowledges the use of ChatGPT in helping with minor phrasing, grammar, and editing, as well as with generating some tables and figures. DBS gratefully acknowledges support from NSF Grant 2407752. MRS is supported by an STScI Postdoctoral Fellowship. JP is supported by NASA through a Einstein Fellowship grant No. HF2-51541.001 awarded by the Space Telescope Science Institute (STScI), which is operated by the Association of Universities for Research in Astronomy, Inc., for NASA, under contract NAS5-26555. Y.Z. Acknowledges support from the MAOF grant 12641898 and visitor support from the Observatories of the Carnegie Institution for Science, Pasadena, CA, where part of this work was completed. N.E.D acknowledge support from NSF grants LEAPS-2532703 and AST-2510993. 

 \end{acknowledgments}

\vspace{5mm}
\facilities{\textit{JWST}, \textit{HST}}

\software{astropy \citep{astropy22}, EAZY and eazy-py \citep{brammer08}, STARDUST2 \citep{rodney14}, JHAT \citep{Rest2023}}

\appendix

\section{Survey time dependence on Depth and Area}
\label{appendix}

Observations with \textit{JWST} have significant overheads due to various factors (e.g., guide star acquisition, filter change, end slew), which makes it difficult to accurately  estimate the total amount of time needed obtain 1 epoch of a mosaic to a given depth. Instead of trying to estimate each different source of overhead in detail, we empirically determine the relations between survey time, depth, and area using representative mosaics calculated with the Astronomer's Proposal Tool (APT)\footnote{https://apt.stsci.edu}. 

For the purpose of this paper, we use two NIRCam exposures (i.e., 4 filters), which we consider best for transient discovery and preliminary classification with light curve fitting to guide follow-up observations. We also use 3 small dither positions, which we find to be the minimum for adequate cosmic ray rejection. Table~\ref{tab:exposuresetup_appendix} shows the NIRCam setup used for a range of possible exposure times.

For each exposure time, we calculate the total charged time $t_{total}$ for a set of 12 mosaics with (x,y) ranging from (1,1) to (50,20), i.e. $N_{pointing}=x \times y$ ranging from 1 to 1000. We choose $x \geq y$ since there is a few percent less demand on the time if more columns than rows are used in the mosaic. For a given exposure time, we fit the straight line of $t_{total}$ versus $N_{pointing}$  \begin{equation}
    t_{total} = m_1(t_{exp})  N_{pointing} + b_1(t_{exp}) \label{eq:ttot}
\end{equation}
The fits are shown in Figure~\ref{fig:depthvsarea} as solid red lines, and the fitted $m_1(t_{exp})$  and $b_1(t_{exp})$ in Table~\ref{tab:m1b1fit}.

\begin{figure*}[t]
     \centering
     {
     \includegraphics[width=7in]{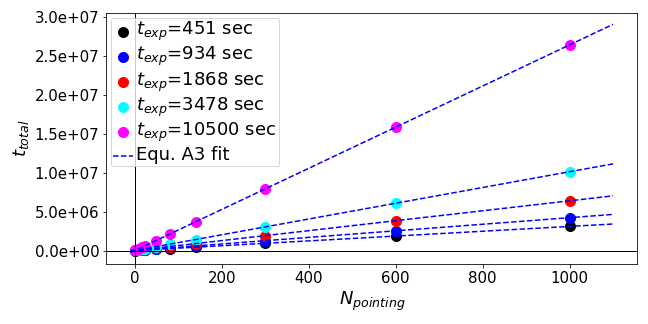}
     \caption{This figure shows the total charged time for a mosaic with $N_{pointing}$ pointings with filled circles for a given exposure time $t_{exp}$, calculated using the APT. The blue dashed line show the calculcated $t_{total}$ using Equ. \ref{eq:ttot_full}}
     \label{fig:t_tot}
     }
\end{figure*}

We then fit a straight line to $m_1(t_{exp})$ versus $t_{exp}$
\begin{equation}
    m_1 = m_2 \times t_{exp} + b_2 \label{eq:m2b2}
\end{equation}
and find an excellent fit (residuals $<<0.1\%$) with $m_2=2.31954$ and $b_2=2050.5014$. We also calculate the average $\overline{b_1}$=1635.51. Using Equ.~\ref{eq:ttot}~and~\ref{eq:m2b2}, we can then express $t_{total}$ as
\begin{equation}
t_{total} = (m_2 \times t_{exp} + b_2) \times N_{pointing}  + \overline{b_1} \label{eq:ttot_full}
\end{equation}
These fits are shown as dashed lines in Figure~\ref{fig:t_tot} for a given exposure time $t_{exp}$. They are excellent, with residuals $\leq 1\%$ for $N_{pointing}>2$. For $N_{pointing}\leq 2$, the residuals are at most on the order of 5\%.

We can solve Equ.~\ref{eq:ttot_full} for $t_{exp}$ and get
\begin{equation}
t_{exp} = \frac{t_{total}-b_2\times N_{pointing}-\overline{b_1}}{m_2 \times N_{pointing}}
\end{equation}
For Figure~\ref{fig:depthvsarea}, we can loop through $N_{pointing}$, and calculate the depth $m_{depth}$ with
\begin{equation}
m_{depth} = m_{0,depth} + 2.5 \log_{10}\left(\sqrt{t_{exp}/t_{0,exp}}\right)
\end{equation}
Here we use as empirical anchor points $m_{0,depth}=28.3$ and $t_{0,exp}=1000$ seconds for F227W and F356W \citep{casey23}.

\begin{table}[!t]
\centering
\small

\begin{minipage}{0.48\columnwidth}
\centering
\caption{NIRCam exposure setup}
\begin{tabular}{cccc}
\hline
$t_{exp}$ [sec] & Readout & $N_{group}$ & $N_{int}$\\
\hline
\phantom{10}451  & SHALLOW4 & 3 & 1 \\
\phantom{10}934  & SHALLOW4 & 6 & 1 \\
\phantom{1}1868  & MEDIUM8  & 6 & 1 \\
\phantom{1}3478  & DEEP8    & 6 & 1 \\
10500            & DEEP8    & 6 & 3 
\end{tabular}
\label{tab:exposuresetup_appendix}
\end{minipage}\hfill
\begin{minipage}{0.48\columnwidth}
\centering
\caption{Straight line fit results of $t_{total}$ versus $N_{pointing}$}
\begin{tabular}{ccc}
\hline
$t_{exp}$ [sec] & $m_1$ & $b_1$\\
\hline
\phantom{10}451	  & \phantom{0}3096.41	& 1636.29 \\
\phantom{10}934	  & \phantom{0}4217.40	& 1635.46\\
\phantom{1}1868	  & \phantom{0}6381.90	& 1633.88\\
\phantom{1}3478	  & 10119.40	        & 1635.63\\
10500	          & 26405.41	        & 1636.29

\end{tabular}
\label{tab:m1b1fit}
\end{minipage}

\normalsize
\end{table}

\section{COSMOS photometry}
\label{appendix:phot}

Table \ref{tab:app:phot} provides the COSMOS photometry of all the sources listed in Table \ref{tab:cosmos_classifications}. 

\begin{deluxetable}{lllrr}
\tablenum{B1}
\tablecaption{COSMOS photometry \label{tab:app:phot}}
\tablehead{
\colhead{IAU ID} & \colhead{MJD} & \colhead{Filter} &
\colhead{mag} & \colhead{$\sigma_{mag}$}
}
\startdata
AT2023adyx&60301.66 &F444W&27.44&0.2 \\
AT2023adyx&60302.6 &F277W&27.6&0.16 \\
AT2023adyx&60303.97 &F150W&27.71&0.17 \\
AT2023adyy&60302.6 &F277W&28.65&0.26 \\
AT2023adyy&60301.66 &F444W&27.78&0.23 \\
AT2023adyy&60303.97 &F150W&27.14&0.09 \\
AT2023adyy&60304.76 &F115W&26.61&0.09 \\
AT2023adyz&60303.97 &F150W&25.99&0.05 \\
AT2023adyz&60301.66 &F444W&27.07&0.16 \\
AT2023adyz&60305.36 &F115W&25.75&0.04 \\
AT2023adyz&60302.6 &F277W&26.67&0.07 \\
\enddata
\tablecomments{Table \ref{tab:app:phot} is published in its entirety in the electronic edition of the {\it Astrophysical Journal}.  A portion is shown here for guidance regarding its form and content.}
\end{deluxetable}

\bibliography{references,new_bibs}{}

@ARTICLE{frye24,
       author = {{Frye}, Brenda L. and {Pascale}, Massimo and {Pierel}, Justin and {Chen}, Wenlei and {Foo}, Nicholas and {Leimbach}, Reagen and {Garuda}, Nikhil and {Cohen}, Seth H. and {Kamieneski}, Patrick S. and {Windhorst}, Rogier A. and {Koekemoer}, Anton M. and {Kelly}, Pat and {Summers}, Jake and {Engesser}, Michael and {Liu}, Daizhong and {Furtak}, Lukas J. and {Polletta}, Maria del Carmen and {Harrington}, Kevin C. and {Willner}, S.~P. and {Diego}, Jose M. and {Jansen}, Rolf A. and {Coe}, Dan and {Conselice}, Christopher J. and {Dai}, Liang and {Dole}, Herv{\'e} and {D'Silva}, Jordan C.~J. and {Driver}, Simon P. and {Grogin}, Norman A. and {Marshall}, Madeline A. and {Meena}, Ashish K. and {Nonino}, Mario and {Ortiz}, Rafael and {Pirzkal}, Nor and {Robotham}, Aaron and {Ryan}, Russell E. and {Strolger}, Lou and {Tompkins}, Scott and {Willmer}, Christopher N.~A. and {Yan}, Haojing and {Yun}, Min S. and {Zitrin}, Adi},
        title = "{The JWST Discovery of the Triply Imaged Type Ia ``Supernova H0pe'' and Observations of the Galaxy Cluster PLCK G165.7+67.0}",
      journal = {\apj},
         year = 2024,
        month = feb,
       volume = {961},
       number = {2},
          eid = {171},
        pages = {171},
          doi = {10.3847/1538-4357/ad1034},
archivePrefix = {arXiv},
       eprint = {2309.07326},
 primaryClass = {astro-ph.GA},
       adsurl = {https://ui.adsabs.harvard.edu/abs/2024ApJ...961..171F}
}

@ARTICLE{yan26,
       author = {{Yan}, Haojing and {Sun}, Bangzheng and {Ma}, Zhiyuan and {Wang}, Lifan and {Willmer}, Christopher N.~A. and {Chen}, Wenlei and {Grogin}, Norman A. and {Beacom}, John F. and {Willner}, S.~P. and {Cohen}, Seth H. and {Windhorst}, Rogier A. and {Jansen}, Rolf A. and {Cheng}, Cheng and {Huang}, Jia-Sheng and {Yun}, Min and {Gim}, Hansung B. and {Hammel}, Heidi B. and {Milam}, Stefanie N. and {Koekemoer}, Anton M. and {Hu}, Lei and {Diego}, Jos{\'e} M. and {Summers}, Jake and {D'Silva}, Jordan C.~J. and {Coe}, Dan and {Conselice}, Christopher J. and {Driver}, Simon P. and {Frye}, Brenda and {Marshall}, Madeline A. and {Ortiz}, III, Rafael and {Pirzkal}, Nor and {Robotham}, Aaron and {Ryan}, Jr., Russell E. and {Honor}, Rachel and {O'Brien}, Rosalia and {Fazio}, Giovanni G. and {Adams}, Nathan J. and {Ricotti}, Massimo and {Saikia}, Payaswini and {Hathi}, Nimish P. and {Smith}, Brent and {Holwerda}, Benne W. and {Kelly}, Patrick},
        title = "{PEARLS: 21 Transients Found in the Three-epoch NIRCam Observations in the Continuous Viewing Zone of the James Webb Space Telescope}",
      journal = {\apj},
         year = 2026,
        month = feb,
       volume = {998},
       number = {1},
          eid = {115},
        pages = {115},
          doi = {10.3847/1538-4357/ae2c5d},
archivePrefix = {arXiv},
       eprint = {2506.12175},
 primaryClass = {astro-ph.GA},
       adsurl = {https://ui.adsabs.harvard.edu/abs/2026ApJ...998..115Y}
}

@ARTICLE{siebert25,
       author = {{Siebert}, M.~R. and {Pierel}, J.~D.~R. and {Engesser}, M. and {Coulter}, D.~A. and {Decoursey}, C. and {Fox}, O.~D. and {Rest}, A. and {Chen}, W. and {Derkacy}, J.~M. and {Egami}, E. and {Foley}, R.~J. and {Jones}, D.~O. and {Koekemoer}, A.~M. and {Larison}, C. and {Leonard}, D.~C. and {Moriya}, T.~J. and {Quimby}, R.~M. and {Shukawa}, K. and {Strolger}, L.~G. and {Zenati}, Yossef},
        title = "{SN 2025ogs: A Spectroscopically-Normal Type Ia Supernova at z = 2 as a Benchmark for Redshift Evolution}",
      journal = {arXiv e-prints},
         year = 2025,
        month = dec,
          eid = {arXiv:2512.19783},
        pages = {arXiv:2512.19783},
          doi = {10.48550/arXiv.2512.19783},
archivePrefix = {arXiv},
       eprint = {2512.19783},
 primaryClass = {astro-ph.CO},
       adsurl = {https://ui.adsabs.harvard.edu/abs/2025arXiv251219783S}
}

@ARTICLE{jeon26,
       author = {{Jeon}, Junehyoung and {Bromm}, Volker and {Venditti}, Alessandra and {Finkelstein}, Steven L. and {Hsiao}, Tiger Yu-Yang},
        title = "{Hunting for the First Explosions at the High-redshift Frontier}",
      journal = {\apj},
         year = 2026,
        month = apr,
       volume = {1001},
       number = {1},
          eid = {3},
        pages = {3},
          doi = {10.3847/1538-4357/ae517d},
archivePrefix = {arXiv},
       eprint = {2601.02469},
 primaryClass = {astro-ph.GA},
       adsurl = {https://ui.adsabs.harvard.edu/abs/2026ApJ..1001....3J}
}

@ARTICLE{coulter26,
       author = {{Coulter}, David A. and {Larison}, Conor and {Pierel}, Justin D.~R. and {Fujimoto}, Seiji and {Kokorev}, Vasily and {Allingham}, Joseph F.~V. and {Moriya}, Takashi J. and {Siebert}, Matthew and {Asada}, Yoshihisa and {Bezanson}, Rachel and {Brada{\v{c}}}, Maru{\v{s}}a and {Brammer}, Gabriel and {Chisholm}, John and {Coe}, Dan and {Dayal}, Pratika and {Engesser}, Michael and {Finkelstein}, Steven L. and {Fox}, Ori D. and {Furtak}, Lukas J. and {Koekemoer}, Anton M. and {Moore}, Thomas and {Nakane}, Minami and {Ouchi}, Masami and {Pan}, Richard and {Quimby}, Robert and {Rest}, Armin and {Richard}, Johan and {Robbins}, Luke and {Strolger}, Louis-Gregory and {Sun}, Fengwu and {Treu}, Tommaso and {Yanagisawa}, Hiroto and {Abdurro'uf} and {Agrawal}, Aadya and {Amor{\'\i}n}, Ricardo and {Anderson}, Joseph P. and {Angulo}, Rodrigo and {Atek}, Hakim and {Bauer}, Franz E. and {Bradley}, Larry D. and {Bromm}, Volker and {Bronikowski}, Mateusz and {Conselice}, Christopher J. and {DeCoursey}, Christa and {DerKacy}, James M. and {Desprez}, Guillaume and {Dhawan}, Suhail and {Diego}, Jose M. and {Egami}, Eiichi and {Faisst}, Andreas and {Frye}, Brenda and {Gomez}, Sebastian and {Gonz{\'a}lez-Otero}, Mauro and {Griggio}, Massimo and {Harikane}, Yuichi and {Inayoshi}, Kohei and {Jha}, Saurabh W. and {Jim{\'e}nez-Teja}, Yolanda and {Kartaltepe}, Jeyhan S. and {Kelly}, Patrick L. and {Kwok}, Lindsey A. and {Lane}, Zachary G. and {Li}, Xiaolong and {Lobbe}, Ivo and {Lucas}, Ray A. and {Magdis}, Georgios E. and {Martis}, Nicholas S. and {Matthee}, Jorryt and {Meena}, Ashish K. and {Naidu}, Rohan P. and {Noirot}, Ga{\"e}l and {Oguri}, Masamune and {Padilla Gonzalez}, Estefania and {Pascale}, Massimo and {Petrushevska}, Tanja and {Ricotti}, Massimo and {Schaerer}, Daniel and {Schuldt}, Stefan and {Shahbandeh}, Melissa and {Sheu}, William and {Shukawa}, Koji and {Tsujita}, Akiyoshi and {Vanzella}, Eros and {Wang}, Qinan and {Weaver}, John and {Windhorst}, Rogier and {Xu}, Yi and {Zenati}, Yossef and {Zitrin}, Adi},
        title = "{A spectroscopically confirmed, strongly lensed, metal-poor Type II supernova at z = 5.13}",
      journal = {arXiv e-prints},
         year = 2026,
        month = jan,
          eid = {arXiv:2601.04156},
        pages = {arXiv:2601.04156},
          doi = {10.48550/arXiv.2601.04156},
archivePrefix = {arXiv},
       eprint = {2601.04156},
 primaryClass = {astro-ph.HE},
       adsurl = {https://ui.adsabs.harvard.edu/abs/2026arXiv260104156C}
}

@ARTICLE{ferrara26,
       author = {{Ferrara}, Andrea and {Carniani}, Stefano and {Morishita}, Takahiro and {Stiavelli}, Massimo},
        title = "{Possible evidence for a pair-instability supernova nature of ultra-early JWST sources}",
      journal = {arXiv e-prints},
         year = 2026,
        month = jan,
          eid = {arXiv:2601.07374},
        pages = {arXiv:2601.07374},
archivePrefix = {arXiv},
       eprint = {2601.07374},
 primaryClass = {astro-ph.GA},
       adsurl = {https://ui.adsabs.harvard.edu/abs/2026arXiv260107374F}
}

@ARTICLE{gilliland99,
       author = {{Gilliland}, Ronald L. and {Nugent}, Peter E. and {Phillips}, M.~M.},
        title = "{High-Redshift Supernovae in the Hubble Deep Field}",
      journal = {\apj},
         year = 1999,
        month = aug,
       volume = {521},
       number = {1},
        pages = {30-49},
          doi = {10.1086/307549},
archivePrefix = {arXiv},
       eprint = {astro-ph/9903229},
 primaryClass = {astro-ph},
       adsurl = {https://ui.adsabs.harvard.edu/abs/1999ApJ...521...30G}
}

@ARTICLE{cappellaro97,
       author = {{Cappellaro}, E. and {Turatto}, M. and {Tsvetkov}, D. Yu. and {Bartunov}, O.~S. and {Pollas}, C. and {Evans}, R. and {Hamuy}, M.},
        title = "{The rate of supernovae from the combined sample of five searches.}",
      journal = {\aap},
         year = 1997,
        month = jun,
       volume = {322},
        pages = {431-441},
          doi = {10.48550/arXiv.astro-ph/9611191},
archivePrefix = {arXiv},
       eprint = {astro-ph/9611191},
 primaryClass = {astro-ph},
       adsurl = {https://ui.adsabs.harvard.edu/abs/1997A&A...322..431C}
}

@ARTICLE{baron04,
       author = {{Baron}, E. and {Nugent}, Peter E. and {Branch}, David and {Hauschildt}, Peter H.},
        title = "{Type IIP Supernovae as Cosmological Probes: A Spectral-fitting Expanding Atmosphere Model Distance to SN 1999em}",
      journal = {\apjl},
         year = 2004,
        month = dec,
       volume = {616},
       number = {2},
        pages = {L91-L94},
          doi = {10.1086/426506},
archivePrefix = {arXiv},
       eprint = {astro-ph/0410153},
 primaryClass = {astro-ph},
       adsurl = {https://ui.adsabs.harvard.edu/abs/2004ApJ...616L..91B}
}

@ARTICLE{crenshaw25,
       author = {{Crenshaw}, John Franklin and {Leistedt}, Boris and {Graham}, Melissa Lynn and {Payerne}, Constantin and {Connolly}, Andrew J. and {Gawiser}, Eric and {Karim}, Tanveer and {Malz}, Alex I. and {Newman}, Jeffrey A. and {Ricci}, Marina and {The LSST Dark Energy Science Collaboration}},
        title = "{Quantifying the Impact of LSST u-band Survey Strategy on Photometric Redshift Estimation and the Detection of Lyman-break Galaxies}",
      journal = {\apjs},
         year = 2025,
        month = dec,
       volume = {281},
       number = {2},
          eid = {54},
        pages = {54},
          doi = {10.3847/1538-4365/ae14f0},
archivePrefix = {arXiv},
       eprint = {2503.06016},
 primaryClass = {astro-ph.CO},
       adsurl = {https://ui.adsabs.harvard.edu/abs/2025ApJS..281...54C}
}

@INPROCEEDINGS{perrin14,
       author = {{Perrin}, Marshall D. and {Sivaramakrishnan}, Anand and {Lajoie}, Charles-Philippe and {Elliott}, Erin and {Pueyo}, Laurent and {Ravindranath}, Swara and {Albert}, Lo{\"\i}c.},
        title = "{Updated point spread function simulations for JWST with WebbPSF}",
    booktitle = {Space Telescopes and Instrumentation 2014: Optical, Infrared, and Millimeter Wave},
         year = 2014,
       editor = {{Oschmann}, Jr., Jacobus M. and {Clampin}, Mark and {Fazio}, Giovanni G. and {MacEwen}, Howard A.},
       series = {Society of Photo-Optical Instrumentation Engineers (SPIE) Conference Series},
       volume = {9143},
        month = aug,
          eid = {91433X},
        pages = {91433X},
          doi = {10.1117/12.2056689},
       adsurl = {https://ui.adsabs.harvard.edu/abs/2014SPIE.9143E..3XP}
}

@INPROCEEDINGS{perrin12,
       author = {{Perrin}, Marshall D. and {Soummer}, R{\'e}mi and {Elliott}, Erin M. and {Lallo}, Matthew D. and {Sivaramakrishnan}, Anand},
        title = "{Simulating point spread functions for the James Webb Space Telescope with WebbPSF}",
    booktitle = {Space Telescopes and Instrumentation 2012: Optical, Infrared, and Millimeter Wave},
         year = 2012,
       editor = {{Clampin}, Mark C. and {Fazio}, Giovanni G. and {MacEwen}, Howard A. and {Oschmann}, Jr., Jacobus M.},
       series = {Society of Photo-Optical Instrumentation Engineers (SPIE) Conference Series},
       volume = {8442},
        month = sep,
          eid = {84423D},
        pages = {84423D},
          doi = {10.1117/12.925230},
       adsurl = {https://ui.adsabs.harvard.edu/abs/2012SPIE.8442E..3DP}
}

@software{bradley25,
  author       = {Larry Bradley and
                  Brigitta Sip{\H o}cz and
                  Thomas Robitaille and
                  Erik Tollerud and
                  Z\`e Vin{\'{\i}}cius and
                  Christoph Deil and
                  Kyle Barbary and
                  Tom J Wilson and
                  Ivo Busko and
                  Axel Donath and
                  Hans Moritz G{\"u}nther and
                  Mihai Cara and
                  P. L. Lim and
                  Sebastian Me{\ss}linger and
                  Zach Burnett and
                  Simon Conseil and
                  Michael Droettboom and
                  Azalee Bostroem and
                  E. M. Bray and
                  Lars Andersen Bratholm and
                  William Jamieson and
                  Adam Ginsburg and
                  Geert Barentsen and
                  Matt Craig and
                  Sergio Pascual and
                  Shivangee Rathi and
                  Marshall Perrin and
                  Brett M. Morris},
  title        = {astropy/photutils: 2.2.0},
  month        = feb,
  year         = 2025,
  publisher    = {Zenodo},
  version      = {2.2.0},
  doi          = {10.5281/zenodo.14889440},
  url          = {https://doi.org/10.5281/zenodo.14889440},
  swhid        = {swh:1:dir:11159107f27a28985192ed1118b1f2055709d093
                   ;origin=https://doi.org/10.5281/zenodo.596036;visi
                   t=swh:1:snp:ae8c4a55d349d43e53cfe9ce92e678fcfe840f
                   3b;anchor=swh:1:rel:0117f67e8888adcdfc85308287dd9c
                   854b466389;path=astropy-photutils-ffb96c5
                  },
}

@ARTICLE{larson23,
       author = {{Larson}, Rebecca L. and {Hutchison}, Taylor A. and {Bagley}, Micaela and {Finkelstein}, Steven L. and {Yung}, L.~Y. Aaron and {Somerville}, Rachel S. and {Hirschmann}, Michaela and {Brammer}, Gabriel and {Holwerda}, Benne W. and {Papovich}, Casey and {Morales}, Alexa M. and {Wilkins}, Stephen M.},
        title = "{Spectral Templates Optimal for Selecting Galaxies at z > 8 with the JWST}",
      journal = {\apj},
         year = 2023,
        month = dec,
       volume = {958},
       number = {2},
          eid = {141},
        pages = {141},
          doi = {10.3847/1538-4357/acfed4},
archivePrefix = {arXiv},
       eprint = {2211.10035},
 primaryClass = {astro-ph.GA},
       adsurl = {https://ui.adsabs.harvard.edu/abs/2023ApJ...958..141L}
}

@software{conroy10,
       author = {{Conroy}, Charlie and {Gunn}, James E.},
        title = "{FSPS: Flexible Stellar Population Synthesis}",
 howpublished = {Astrophysics Source Code Library, record ascl:1010.043},
         year = 2010,
        month = oct,
          eid = {ascl:1010.043},
archivePrefix = {ascl},
       eprint = {1010.043},
       adsurl = {https://ui.adsabs.harvard.edu/abs/2010ascl.soft10043C}
}

@ARTICLE{strolger15,
       author = {{Strolger}, Louis-Gregory and {Dahlen}, Tomas and {Rodney}, Steven A. and {Graur}, Or and {Riess}, Adam G. and {McCully}, Curtis and {Ravindranath}, Swara and {Mobasher}, Bahram and {Shahady}, A. Kristin},
        title = "{The Rate of Core Collapse Supernovae to Redshift 2.5 from the CANDELS and CLASH Supernova Surveys}",
      journal = {\apj},
         year = 2015,
        month = nov,
       volume = {813},
       number = {2},
          eid = {93},
        pages = {93},
          doi = {10.1088/0004-637X/813/2/93},
archivePrefix = {arXiv},
       eprint = {1509.06574},
 primaryClass = {astro-ph.GA},
       adsurl = {https://ui.adsabs.harvard.edu/abs/2015ApJ...813...93S}
}

@ARTICLE{moriya22a,
       author = {{Moriya}, T.~J. and {Inserra}, C. and {Tanaka}, M. and {Cappellaro}, E. and {Della Valle}, M. and {Hook}, I. and {Kotak}, R. and {Longo}, G. and {Mannucci}, F. and {Mattila}, S. and {Tao}, C. and {Altieri}, B. and {Amara}, A. and {Auricchio}, N. and {Bonino}, D. and {Branchini}, E. and {Brescia}, M. and {Brinchmann}, J. and {Camera}, S. and {Capobianco}, V. and {Carbone}, C. and {Carretero}, J. and {Castellano}, M. and {Cavuoti}, S. and {Cimatti}, A. and {Cledassou}, R. and {Congedo}, G. and {Conselice}, C.~J. and {Conversi}, L. and {Copin}, Y. and {Corcione}, L. and {Courbin}, F. and {Cropper}, M. and {Da Silva}, A. and {Degaudenzi}, H. and {Douspis}, M. and {Dubath}, F. and {Duncan}, C.~A.~J. and {Dupac}, X. and {Dusini}, S. and {Ealet}, A. and {Farrens}, S. and {Ferriol}, S. and {Frailis}, M. and {Franceschi}, E. and {Fumana}, M. and {Garilli}, B. and {Gillard}, W. and {Gillis}, B. and {Giocoli}, C. and {Grazian}, A. and {Grupp}, F. and {Haugan}, S.~V.~H. and {Holmes}, W. and {Hormuth}, F. and {Hornstrup}, A. and {Jahnke}, K. and {Kermiche}, S. and {Kiessling}, A. and {Kilbinger}, M. and {Kitching}, T. and {Kurki-Suonio}, H. and {Ligori}, S. and {Lilje}, P.~B. and {Lloro}, I. and {Maiorano}, E. and {Mansutti}, O. and {Marggraf}, O. and {Markovic}, K. and {Marulli}, F. and {Massey}, R. and {McCracken}, H.~J. and {Melchior}, M. and {Meneghetti}, M. and {Meylan}, G. and {Moresco}, M. and {Moscardini}, L. and {Munari}, E. and {Niemi}, S.~M. and {Padilla}, C. and {Paltani}, S. and {Pasian}, F. and {Pedersen}, K. and {Pettorino}, V. and {Poncet}, M. and {Popa}, L. and {Raison}, F. and {Rhodes}, J. and {Riccio}, G. and {Rossetti}, E. and {Saglia}, R. and {Sartoris}, B. and {Schneider}, P. and {Secroun}, A. and {Seidel}, G. and {Sirignano}, C. and {Sirri}, G. and {Stanco}, L. and {Tallada-Cresp{\'\i}}, P. and {Taylor}, A.~N. and {Tereno}, I. and {Toledo-Moreo}, R. and {Torradeflot}, F. and {Wang}, Y. and {Zamorani}, G. and {Zoubian}, J. and {Andreon}, S. and {Scottez}, V. and {Morris}, P.~W.},
        title = "{Euclid: Searching for pair-instability supernovae with the Deep Survey}",
      journal = {\aap},
         year = 2022,
        month = oct,
       volume = {666},
          eid = {A157},
        pages = {A157},
          doi = {10.1051/0004-6361/202243810},
archivePrefix = {arXiv},
       eprint = {2204.08727},
 primaryClass = {astro-ph.HE},
       adsurl = {https://ui.adsabs.harvard.edu/abs/2022A&A...666A.157M}
}

@ARTICLE{bouwens23,
       author = {{Bouwens}, Rychard J. and {Stefanon}, Mauro and {Brammer}, Gabriel and {Oesch}, Pascal A. and {Herard-Demanche}, Thomas and {Illingworth}, Garth D. and {Matthee}, Jorryt and {Naidu}, Rohan P. and {van Dokkum}, Pieter G. and {van Leeuwen}, Ivana F.},
        title = "{Evolution of the UV LF from z   15 to z   8 using new JWST NIRCam medium-band observations over the HUDF/XDF}",
      journal = {\mnras},
         year = 2023,
        month = jul,
       volume = {523},
       number = {1},
        pages = {1036-1055},
          doi = {10.1093/mnras/stad1145},
archivePrefix = {arXiv},
       eprint = {2211.02607},
 primaryClass = {astro-ph.GA},
       adsurl = {https://ui.adsabs.harvard.edu/abs/2023MNRAS.523.1036B}
}

@ARTICLE{harikane23,
       author = {{Harikane}, Yuichi and {Zhang}, Yechi and {Nakajima}, Kimihiko and {Ouchi}, Masami and {Isobe}, Yuki and {Ono}, Yoshiaki and {Hatano}, Shun and {Xu}, Yi and {Umeda}, Hiroya},
        title = "{A JWST/NIRSpec First Census of Broad-line AGNs at z = 4-7: Detection of 10 Faint AGNs with M $_{BH}$ {}10$^{6}$-{}10$^{8}$ M $_{{\ensuremath{\odot}}}$ and Their Host Galaxy Properties}",
      journal = {\apj},
         year = 2023,
        month = dec,
       volume = {959},
       number = {1},
          eid = {39},
        pages = {39},
          doi = {10.3847/1538-4357/ad029e},
archivePrefix = {arXiv},
       eprint = {2303.11946},
 primaryClass = {astro-ph.GA},
       adsurl = {https://ui.adsabs.harvard.edu/abs/2023ApJ...959...39H}
}

@ARTICLE{bertin96,
       author = {{Bertin}, E. and {Arnouts}, S.},
        title = "{SExtractor: Software for source extraction.}",
      journal = {\aaps},
         year = 1996,
        month = jun,
       volume = {117},
        pages = {393-404},
          doi = {10.1051/aas:1996164},
       adsurl = {https://ui.adsabs.harvard.edu/abs/1996A&AS..117..393B}
}

@article{barbary16, 
    doi = {10.21105/joss.00058}, 
    url = {https://doi.org/10.21105/joss.00058}, 
    year = {2016}, 
    publisher = {The Open Journal}, 
    volume = {1}, 
    number = {6}, 
    pages = {58}, 
    author = {Barbary, Kyle}, 
    title = {SEP: Source Extractor as a library}, 
    journal = {Journal of Open Source Software} 
}

@ARTICLE{guetta07,
       author = {{Guetta}, Dafne and {Della Valle}, Massimo},
        title = "{On the Rates of Gamma-Ray Bursts and Type Ib/c Supernovae}",
      journal = {\apjl},
         year = 2007,
        month = mar,
       volume = {657},
       number = {2},
        pages = {L73-L76},
          doi = {10.1086/511417},
archivePrefix = {arXiv},
       eprint = {astro-ph/0612194},
 primaryClass = {astro-ph},
       adsurl = {https://ui.adsabs.harvard.edu/abs/2007ApJ...657L..73G}
}

@ARTICLE{lyu22,
       author = {{Lyu}, Jianwei and {Alberts}, Stacey and {Rieke}, George H. and {Rujopakarn}, Wiphu},
        title = "{AGN Selection and Demographics in GOODS-S/HUDF from X-Ray to Radio}",
      journal = {\apj},
         year = 2022,
        month = dec,
       volume = {941},
       number = {2},
          eid = {191},
        pages = {191},
          doi = {10.3847/1538-4357/ac9e5d},
archivePrefix = {arXiv},
       eprint = {2209.06219},
 primaryClass = {astro-ph.GA},
       adsurl = {https://ui.adsabs.harvard.edu/abs/2022ApJ...941..191L}
}

@ARTICLE{berman24,
       author = {{Berman}, Edward M. and {McCleary}, Jacqueline E. and {Koekemoer}, Anton M. and {Franco}, Maximilien and {Drakos}, Nicole E. and {Liu}, Daizhong and {Nightingale}, James W. and {Shuntov}, Marko and {Scognamiglio}, Diana and {Massey}, Richard and {Mahler}, Guillaume and {McCracken}, Henry Joy and {Robertson}, Brant E. and {Faisst}, Andreas L. and {Casey}, Caitlin M. and {Kartaltepe}, Jeyhan S. and {Cosmos-Web: The Jwst Cosmic Origins Survey}},
        title = "{Efficient Point-spread Function Modeling with ShOpt.jl: A Point-spread Function Benchmarking Study with JWST NIRCam Imaging}",
      journal = {\aj},
         year = 2024,
        month = oct,
       volume = {168},
       number = {4},
          eid = {174},
        pages = {174},
          doi = {10.3847/1538-3881/ad6a0f},
archivePrefix = {arXiv},
       eprint = {2401.11625},
 primaryClass = {astro-ph.IM},
       adsurl = {https://ui.adsabs.harvard.edu/abs/2024AJ....168..174B}
}

@ARTICLE{akins25,
       author = {{Akins}, Hollis B. and {Casey}, Caitlin M. and {Lambrides}, Erini and {Allen}, Natalie and {Andika}, Irham T. and {Brinch}, Malte and {Champagne}, Jaclyn B. and {Cooper}, Olivia and {Ding}, Xuheng and {Drakos}, Nicole E. and {Faisst}, Andreas and {Finkelstein}, Steven L. and {Franco}, Maximilien and {Fujimoto}, Seiji and {Gentile}, Fabrizio and {Gillman}, Steven and {Gozaliasl}, Ghassem and {Harish}, Santosh and {Hayward}, Christopher C. and {Hirschmann}, Michaela and {Ilbert}, Olivier and {Kartaltepe}, Jeyhan S. and {Kocevski}, Dale D. and {Koekemoer}, Anton M. and {Kokorev}, Vasily and {Liu}, Daizhong and {Long}, Arianna S. and {McCracken}, Henry Joy and {McKinney}, Jed and {Onoue}, Masafusa and {Paquereau}, Louise and {Renzini}, Alvio and {Rhodes}, Jason and {Robertson}, Brant E. and {Shuntov}, Marko and {Silverman}, John D. and {Tanaka}, Takumi S. and {Toft}, Sune and {Trakhtenbrot}, Benny and {Valentino}, Francesco and {Zavala}, Jorge},
        title = "{COSMOS-Web: The Overabundance and Physical Nature of ``Little Red Dots''{\textemdash}Implications for Early Galaxy and SMBH Assembly}",
      journal = {\apj},
         year = 2025,
        month = sep,
       volume = {991},
       number = {1},
          eid = {37},
        pages = {37},
          doi = {10.3847/1538-4357/ade984},
archivePrefix = {arXiv},
       eprint = {2406.10341},
 primaryClass = {astro-ph.GA},
       adsurl = {https://ui.adsabs.harvard.edu/abs/2025ApJ...991...37A}
}

@ARTICLE{eisenstein23,
       author = {{Eisenstein}, Daniel J. and {Willott}, Chris and {Alberts}, Stacey and {Arribas}, Santiago and {Bonaventura}, Nina and {Bunker}, Andrew J. and {Cameron}, Alex J. and {Carniani}, Stefano and {Charlot}, Stephane and {Curtis-Lake}, Emma and {D'Eugenio}, Francesco and {Ferruit}, Pierre and {Giardino}, Giovanna and {Hainline}, Kevin and {Hausen}, Ryan and {Jakobsen}, Peter and {Johnson}, Benjamin D. and {Maiolino}, Roberto and {Rauscher}, Bernard J. and {Rieke}, Marcia and {Rieke}, George and {Rix}, Hans-Walter and {Robertson}, Brant and {Stark}, Daniel P. and {Tacchella}, Sandro and {Williams}, Christina C. and {Willmer}, Christopher N.~A. and {Baker}, William M. and {Baum}, Stefi and {Bhatawdekar}, Rachana and {Boyett}, Kristan and {Chen}, Zuyi and {Chevallard}, Jacopo and {Circosta}, Chiara and {Curti}, Mirko and {Danhaive}, A. Lola and {DeCoursey}, Christa and {Endsley}, Ryan and {de Graaff}, Anna and {Dressler}, Alan and {Egami}, Eiichi and {Helton}, Jakob M. and {Hviding}, Raphael E. and {Ji}, Zhiyuan and {Jones}, Gareth C. and {Kumari}, Nimisha and {L{\"u}tzgendorf}, Nora and {Laseter}, Isaac and {Looser}, Tobias J. and {Lyu}, Jianwei and {Maseda}, Michael V. and {Nelson}, Erica and {Parlanti}, Eleonora and {Perna}, Michele and {Pusk{\'a}s}, D{\'a}vid and {Rawle}, Tim and {Rodr{\'\i}guez Del Pino}, Bruno and {Rujopakarn}, Wiphu and {Sandles}, Lester and {Saxena}, Aayush and {Scholtz}, Jan and {Sharpe}, Katherine and {Shivaei}, Irene and {Silcock}, Maddie S. and {Simmonds}, Charlotte and {Skarbinski}, Maya and {Smit}, Renske and {Stone}, Meredith and {Suess}, Katherine A. and {Sun}, Fengwu and {Tang}, Mengtao and {Topping}, Michael W. and {{\"U}bler}, Hannah and {Villanueva}, Natalia C. and {Wallace}, Imaan E.~B. and {Whitler}, Lily and {Witstok}, Joris and {Woodrum}, Charity},
        title = "{Overview of the JWST Advanced Deep Extragalactic Survey (JADES)}",
      journal = {arXiv e-prints},
         year = 2023,
        month = jun,
          eid = {arXiv:2306.02465},
        pages = {arXiv:2306.02465},
          doi = {10.48550/arXiv.2306.02465},
archivePrefix = {arXiv},
       eprint = {2306.02465},
 primaryClass = {astro-ph.GA},
       adsurl = {https://ui.adsabs.harvard.edu/abs/2023arXiv230602465E}
}

@ARTICLE{2025PASJ...77..851M,
       author = {{Moriya}, Takashi J. and {Coulter}, David A. and {DeCoursey}, Christa and {Pierel}, Justin D.~R. and {Hainline}, Kevin and {Siebert}, Matthew R. and {Rest}, Armin and {Egami}, Eiichi and {Gomez}, Sebastian and {Quimby}, Robert M. and {Fox}, Ori D. and {Engesser}, Michael and {Sun}, Fengwu and {Chen}, Wenlei and {Zenati}, Yossef and {Gezari}, Suvi and {Joshi}, Bhavin A. and {Shahbandeh}, Melissa and {Strolger}, Louis-Gregory and {Wang}, Qinan and {Alberts}, Stacey and {Bhatawdekar}, Rachana and {Bunker}, Andrew J. and {Rinaldi}, Pierluigi and {Robertson}, Brant E. and {Tacchella}, Sandro},
        title = "{Properties of high-redshift Type II supernovae discovered by the JADES transient survey}",
      journal = {\pasj},
         year = 2025,
        month = aug,
       volume = {77},
       number = {4},
        pages = {851-862},
          doi = {10.1093/pasj/psaf052},
archivePrefix = {arXiv},
       eprint = {2501.08969},
 primaryClass = {astro-ph.HE},
       adsurl = {https://ui.adsabs.harvard.edu/abs/2025PASJ...77..851M}
}

@ARTICLE{pierel25,
       author = {{Pierel}, J.~D.~R. and {Coulter}, D.~A. and {Siebert}, M.~R. and {Akins}, H.~B. and {Engesser}, M. and {Fox}, O.~D. and {Franco}, M. and {Rest}, A. and {Agrawal}, A. and {Ajay}, Y. and {Allen}, N. and {Casey}, C.~M. and {DeCoursey}, C. and {Drakos}, N.~E. and {Egami}, E. and {Faisst}, A.~L. and {Gezari}, S. and {Gozaliasl}, G. and {Ilbert}, O. and {Jones}, D.~O. and {Karmen}, M. and {Kartaltepe}, J.~S. and {Koekemoer}, A.~M. and {Lane}, Z.~G. and {Larson}, R.~L. and {Li}, T. and {Liu}, D. and {Moriya}, T.~J. and {McCracken}, H.~J. and {Paquereau}, L. and {Quimby}, R.~M. and {Rich}, R.~M. and {Rhodes}, J. and {Robertson}, B.~E. and {Sanders}, D.~B. and {Shahbandeh}, M. and {Shuntov}, M. and {Silverman}, J.~D. and {Strolger}, L.~G. and {Toft}, S. and {Zenati}, Y.},
        title = "{Testing for Intrinsic Type Ia Supernova Luminosity Evolution at z > 2 with JWST}",
      journal = {\apjl},
         year = 2025,
        month = mar,
       volume = {981},
       number = {1},
          eid = {L9},
        pages = {L9},
          doi = {10.3847/2041-8213/adb1d9},
archivePrefix = {arXiv},
       eprint = {2411.11953},
 primaryClass = {astro-ph.CO},
       adsurl = {https://ui.adsabs.harvard.edu/abs/2025ApJ...981L...9P}
}

@ARTICLE{casey23,
       author = {{Casey}, Caitlin M. and {Kartaltepe}, Jeyhan S. and {Drakos}, Nicole E. and {Franco}, Maximilien and {Harish}, Santosh and {Paquereau}, Louise and {Ilbert}, Olivier and {Rose}, Caitlin and {Cox}, Isabella G. and {Nightingale}, James W. and {Robertson}, Brant E. and {Silverman}, John D. and {Koekemoer}, Anton M. and {Massey}, Richard and {McCracken}, Henry Joy and {Rhodes}, Jason and {Akins}, Hollis B. and {Allen}, Natalie and {Amvrosiadis}, Aristeidis and {Arango-Toro}, Rafael C. and {Bagley}, Micaela B. and {Bongiorno}, Angela and {Capak}, Peter L. and {Champagne}, Jaclyn B. and {Chartab}, Nima and {Ch{\'a}vez Ortiz}, {\'O}scar A. and {Chworowsky}, Katherine and {Cooke}, Kevin C. and {Cooper}, Olivia R. and {Darvish}, Behnam and {Ding}, Xuheng and {Faisst}, Andreas L. and {Finkelstein}, Steven L. and {Fujimoto}, Seiji and {Gentile}, Fabrizio and {Gillman}, Steven and {Gould}, Katriona M.~L. and {Gozaliasl}, Ghassem and {Hayward}, Christopher C. and {He}, Qiuhan and {Hemmati}, Shoubaneh and {Hirschmann}, Michaela and {Jahnke}, Knud and {Jin}, Shuowen and {Khostovan}, Ali Ahmad and {Kokorev}, Vasily and {Lambrides}, Erini and {Laigle}, Clotilde and {Larson}, Rebecca L. and {Leung}, Gene C.~K. and {Liu}, Daizhong and {Liaudat}, Tobias and {Long}, Arianna S. and {Magdis}, Georgios and {Mahler}, Guillaume and {Mainieri}, Vincenzo and {Manning}, Sinclaire M. and {Maraston}, Claudia and {Martin}, Crystal L. and {McCleary}, Jacqueline E. and {McKinney}, Jed and {McPartland}, Conor J.~R. and {Mobasher}, Bahram and {Pattnaik}, Rohan and {Renzini}, Alvio and {Rich}, R. Michael and {Sanders}, David B. and {Sattari}, Zahra and {Scognamiglio}, Diana and {Scoville}, Nick and {Sheth}, Kartik and {Shuntov}, Marko and {Sparre}, Martin and {Suzuki}, Tomoko L. and {Talia}, Margherita and {Toft}, Sune and {Trakhtenbrot}, Benny and {Urry}, C. Megan and {Valentino}, Francesco and {Vanderhoof}, Brittany N. and {Vardoulaki}, Eleni and {Weaver}, John R. and {Whitaker}, Katherine E. and {Wilkins}, Stephen M. and {Yang}, Lilan and {Zavala}, Jorge A.},
        title = "{COSMOS-Web: An Overview of the JWST Cosmic Origins Survey}",
      journal = {\apj},
         year = 2023,
        month = sep,
       volume = {954},
       number = {1},
          eid = {31},
        pages = {31},
          doi = {10.3847/1538-4357/acc2bc},
archivePrefix = {arXiv},
       eprint = {2211.07865},
 primaryClass = {astro-ph.GA},
       adsurl = {https://ui.adsabs.harvard.edu/abs/2023ApJ...954...31C}
}

@ARTICLE{shuntov25,
       author = {{Shuntov}, Marko and {Akins}, Hollis B. and {Paquereau}, Louise and {Casey}, Caitlin M. and {Ilbert}, Olivier and {Arango-Toro}, Rafael C. and {McCracken}, Henry Joy and {Franco}, Maximilien and {Harish}, Santosh and {Kartaltepe}, Jeyhan S. and {Koekemoer}, Anton M. and {Yang}, Lilan and {Huertas-Company}, Marc and {Berman}, Edward M. and {McCleary}, Jacqueline E. and {Toft}, Sune and {Gavazzi}, Rapha{\"e}l and {Achenbach}, Mark J. and {Bertin}, Emmanuel and {Brinch}, Malte and {Champagne}, Jackie and {Chartab}, Nima and {Drakos}, Nicole E. and {Egami}, Eiichi and {Endsley}, Ryan and {Faisst}, Andreas L. and {Fan}, Xiaohui and {Flayhart}, Carter and {Hartley}, William G. and {Hatamnia}, Hossein and {Gozaliasl}, Ghassem and {Gentile}, Fabrizio and {Jermann}, Iris and {Jin}, Shuowen and {Kakiichi}, Koki and {Khostovan}, Ali Ahmad and {K{\"u}mmel}, Martin and {Laigle}, Clotilde and {Laishram}, Ronaldo and {Lambrides}, Erini and {Liu}, Daizhong and {Lyu}, Jianwei and {Magdis}, Georgios and {Mobasher}, Bahram and {Moutard}, Thibaud and {Renzini}, Alvio and {Rich}, R. Michael and {Sanders}, David B. and {Sattari}, Zahra and {Robertson}, Brant E. and {Schefer}, Marc and {Scognamiglio}, Diana and {Scoville}, Nick and {Silverman}, John D. and {Taamoli}, Sina and {Trakhtenbrot}, Benny and {Valentino}, Francesco and {Wang}, Feige and {Weaver}, John R. and {Yang}, Jinyi},
        title = "{COSMOS2025: The COSMOS-Web galaxy catalog of photometry, morphology, redshifts, and physical parameters from JWST, HST, and ground-based imaging}",
      journal = {\aap},
         year = 2025,
        month = dec,
       volume = {704},
          eid = {A339},
        pages = {A339},
          doi = {10.1051/0004-6361/202555799},
archivePrefix = {arXiv},
       eprint = {2506.03243},
 primaryClass = {astro-ph.GA},
       adsurl = {https://ui.adsabs.harvard.edu/abs/2025A&A...704A.339S}
}

@ARTICLE{astropy22,
       author = {{Astropy Collaboration} and {Price-Whelan}, Adrian M. and {Lim}, Pey Lian and {Earl}, Nicholas and {Starkman}, Nathaniel and {Bradley}, Larry and {Shupe}, David L. and {Patil}, Aarya A. and {Corrales}, Lia and {Brasseur}, C.~E. and {N{\"o}the}, Maximilian and {Donath}, Axel and {Tollerud}, Erik and {Morris}, Brett M. and {Ginsburg}, Adam and {Vaher}, Eero and {Weaver}, Benjamin A. and {Tocknell}, James and {Jamieson}, William and {van Kerkwijk}, Marten H. and {Robitaille}, Thomas P. and {Merry}, Bruce and {Bachetti}, Matteo and {G{\"u}nther}, H. Moritz and {Aldcroft}, Thomas L. and {Alvarado-Montes}, Jaime A. and {Archibald}, Anne M. and {B{\'o}di}, Attila and {Bapat}, Shreyas and {Barentsen}, Geert and {Baz{\'a}n}, Juanjo and {Biswas}, Manish and {Boquien}, M{\'e}d{\'e}ric and {Burke}, D.~J. and {Cara}, Daria and {Cara}, Mihai and {Conroy}, Kyle E. and {Conseil}, Simon and {Craig}, Matthew W. and {Cross}, Robert M. and {Cruz}, Kelle L. and {D'Eugenio}, Francesco and {Dencheva}, Nadia and {Devillepoix}, Hadrien A.~R. and {Dietrich}, J{\"o}rg P. and {Eigenbrot}, Arthur Davis and {Erben}, Thomas and {Ferreira}, Leonardo and {Foreman-Mackey}, Daniel and {Fox}, Ryan and {Freij}, Nabil and {Garg}, Suyog and {Geda}, Robel and {Glattly}, Lauren and {Gondhalekar}, Yash and {Gordon}, Karl D. and {Grant}, David and {Greenfield}, Perry and {Groener}, Austen M. and {Guest}, Steve and {Gurovich}, Sebastian and {Handberg}, Rasmus and {Hart}, Akeem and {Hatfield-Dodds}, Zac and {Homeier}, Derek and {Hosseinzadeh}, Griffin and {Jenness}, Tim and {Jones}, Craig K. and {Joseph}, Prajwel and {Kalmbach}, J. Bryce and {Karamehmetoglu}, Emir and {Ka{\l}uszy{\'n}ski}, Miko{\l}aj and {Kelley}, Michael S.~P. and {Kern}, Nicholas and {Kerzendorf}, Wolfgang E. and {Koch}, Eric W. and {Kulumani}, Shankar and {Lee}, Antony and {Ly}, Chun and {Ma}, Zhiyuan and {MacBride}, Conor and {Maljaars}, Jakob M. and {Muna}, Demitri and {Murphy}, N.~A. and {Norman}, Henrik and {O'Steen}, Richard and {Oman}, Kyle A. and {Pacifici}, Camilla and {Pascual}, Sergio and {Pascual-Granado}, J. and {Patil}, Rohit R. and {Perren}, Gabriel I. and {Pickering}, Timothy E. and {Rastogi}, Tanuj and {Roulston}, Benjamin R. and {Ryan}, Daniel F. and {Rykoff}, Eli S. and {Sabater}, Jose and {Sakurikar}, Parikshit and {Salgado}, Jes{\'u}s and {Sanghi}, Aniket and {Saunders}, Nicholas and {Savchenko}, Volodymyr and {Schwardt}, Ludwig and {Seifert-Eckert}, Michael and {Shih}, Albert Y. and {Jain}, Anany Shrey and {Shukla}, Gyanendra and {Sick}, Jonathan and {Simpson}, Chris and {Singanamalla}, Sudheesh and {Singer}, Leo P. and {Singhal}, Jaladh and {Sinha}, Manodeep and {Sip{\H{o}}cz}, Brigitta M. and {Spitler}, Lee R. and {Stansby}, David and {Streicher}, Ole and {{\v{S}}umak}, Jani and {Swinbank}, John D. and {Taranu}, Dan S. and {Tewary}, Nikita and {Tremblay}, Grant R. and {de Val-Borro}, Miguel and {Van Kooten}, Samuel J. and {Vasovi{\'c}}, Zlatan and {Verma}, Shresth and {de Miranda Cardoso}, Jos{\'e} Vin{\'\i}cius and {Williams}, Peter K.~G. and {Wilson}, Tom J. and {Winkel}, Benjamin and {Wood-Vasey}, W.~M. and {Xue}, Rui and {Yoachim}, Peter and {Zhang}, Chen and {Zonca}, Andrea and {Astropy Project Contributors}},
        title = "{The Astropy Project: Sustaining and Growing a Community-oriented Open-source Project and the Latest Major Release (v5.0) of the Core Package}",
      journal = {\apj},
         year = 2022,
        month = aug,
       volume = {935},
       number = {2},
          eid = {167},
        pages = {167},
          doi = {10.3847/1538-4357/ac7c74},
archivePrefix = {arXiv},
       eprint = {2206.14220},
 primaryClass = {astro-ph.IM},
       adsurl = {https://ui.adsabs.harvard.edu/abs/2022ApJ...935..167A}
}

@ARTICLE{brammer08,
       author = {{Brammer}, Gabriel B. and {van Dokkum}, Pieter G. and {Coppi}, Paolo},
        title = "{EAZY: A Fast, Public Photometric Redshift Code}",
      journal = {\apj},
         year = 2008,
        month = oct,
       volume = {686},
       number = {2},
        pages = {1503-1513},
          doi = {10.1086/591786},
archivePrefix = {arXiv},
       eprint = {0807.1533},
 primaryClass = {astro-ph},
       adsurl = {https://ui.adsabs.harvard.edu/abs/2008ApJ...686.1503B}
}

@ARTICLE{golubchik23,
       author = {{Golubchik}, Miriam and {Zitrin}, Adi and {Pierel}, Justin and {Furtak}, Lukas J. and {Meena}, Ashish K. and {Graur}, Or and {Kelly}, Patrick L. and {Coe}, Dan and {Andrade-Santos}, Felipe and {Asif}, Maor and {Bradley}, Larry D. and {Chen}, Wenlei and {Frye}, Brenda L. and {Gomez}, Sebastian and {Jha}, Saurabh and {Mahler}, Guillaume and {Nonino}, Mario and {Strolger}, Louis-Gregory and {Su}, Yuanyuan},
        title = "{A search for transients in the Reionization Lensing Cluster Survey (RELICS): three new supernovae}",
      journal = {\mnras},
         year = 2023,
        month = jul,
       volume = {522},
       number = {3},
        pages = {4718-4727},
          doi = {10.1093/mnras/stad1238},
archivePrefix = {arXiv},
       eprint = {2302.11158},
 primaryClass = {astro-ph.GA},
       adsurl = {https://ui.adsabs.harvard.edu/abs/2023MNRAS.522.4718G}
}

@ARTICLE{gupta16,
       author = {{Gupta}, Ravi R. and {Kuhlmann}, Steve and {Kovacs}, Eve and {Spinka}, Harold and {Kessler}, Richard and {Goldstein}, Daniel A. and {Liotine}, Camille and {Pomian}, Katarzyna and {D'Andrea}, Chris B. and {Sullivan}, Mark and {Carretero}, Jorge and {Castander}, Francisco J. and {Nichol}, Robert C. and {Finley}, David A. and {Fischer}, John A. and {Foley}, Ryan J. and {Kim}, Alex G. and {Papadopoulos}, Andreas and {Sako}, Masao and {Scolnic}, Daniel M. and {Smith}, Mathew and {Tucker}, Brad E. and {Uddin}, Syed and {Wolf}, Rachel C. and {Yuan}, Fang and {Abbott}, Tim M.~C. and {Abdalla}, Filipe B. and {Benoit-L{\'e}vy}, Aur{\'e}lien and {Bertin}, Emmanuel and {Brooks}, David and {Carnero Rosell}, Aurelio and {Carrasco Kind}, Matias and {Cunha}, Carlos E. and {da Costa}, Luiz N. and {Desai}, Shantanu and {Doel}, Peter and {Eifler}, Tim F. and {Evrard}, August E. and {Flaugher}, Brenna and {Fosalba}, Pablo and {Gazta{\~n}aga}, Enrique and {Gruen}, Daniel and {Gruendl}, Robert and {James}, David J. and {Kuehn}, Kyler and {Kuropatkin}, Nikolay and {Maia}, Marcio A.~G. and {Marshall}, Jennifer L. and {Miquel}, Ramon and {Plazas}, Andr{\'e}s A. and {Romer}, A. Kathy and {S{\'a}nchez}, Eusebio and {Schubnell}, Michael and {Sevilla-Noarbe}, Ignacio and {Sobreira}, Fl{\'a}via and {Suchyta}, Eric and {Swanson}, Molly E.~C. and {Tarle}, Gregory and {Walker}, Alistair R. and {Wester}, William},
        title = "{Host Galaxy Identification for Supernova Surveys}",
      journal = {\aj},
         year = 2016,
        month = dec,
       volume = {152},
       number = {6},
          eid = {154},
        pages = {154},
          doi = {10.3847/0004-6256/152/6/154},
archivePrefix = {arXiv},
       eprint = {1604.06138},
 primaryClass = {astro-ph.CO},
       adsurl = {https://ui.adsabs.harvard.edu/abs/2016AJ....152..154G}
}

@ARTICLE{oke83,
       author = {{Oke}, J.~B. and {Gunn}, J.~E.},
        title = "{Secondary standard stars for absolute spectrophotometry.}",
      journal = {\apj},
         year = 1983,
        month = mar,
       volume = {266},
        pages = {713-717},
          doi = {10.1086/160817},
       adsurl = {https://ui.adsabs.harvard.edu/abs/1983ApJ...266..713O}
}

@ARTICLE{pierel22,
       author = {{Pierel}, J.~D.~R. and {Jones}, D.~O. and {Kenworthy}, W.~D. and {Dai}, M. and {Kessler}, R. and {Ashall}, C. and {Do}, A. and {Peterson}, E.~R. and {Shappee}, B.~J. and {Siebert}, M.~R. and {Barna}, T. and {Brink}, T.~G. and {Burke}, J. and {Calamida}, A. and {Camacho-Neves}, Y. and {de Jaeger}, T. and {Filippenko}, A.~V. and {Foley}, R.~J. and {Galbany}, L. and {Fox}, O.~D. and {Gomez}, S. and {Hiramatsu}, D. and {Hounsell}, R. and {Howell}, D.~A. and {Jha}, S.~W. and {Kwok}, L.~A. and {P{\'e}rez-Fournon}, I. and {Poidevin}, F. and {Rest}, A. and {Rubin}, D. and {Scolnic}, D.~M. and {Shirley}, R. and {Strolger}, L.~G. and {Tinyanont}, S. and {Wang}, Q.},
        title = "{SALT3-NIR: Taking the Open-source Type Ia Supernova Model to Longer Wavelengths for Next-generation Cosmological Measurements}",
      journal = {\apj},
         year = 2022,
        month = nov,
       volume = {939},
       number = {1},
          eid = {11},
        pages = {11},
          doi = {10.3847/1538-4357/ac93f9},
archivePrefix = {arXiv},
       eprint = {2209.05594},
 primaryClass = {astro-ph.CO},
       adsurl = {https://ui.adsabs.harvard.edu/abs/2022ApJ...939...11P}
}

@ARTICLE{rodney14,
       author = {{Rodney}, Steven A. and {Riess}, Adam G. and {Strolger}, Louis-Gregory and {Dahlen}, Tomas and {Graur}, Or and {Casertano}, Stefano and {Dickinson}, Mark E. and {Ferguson}, Henry C. and {Garnavich}, Peter and {Hayden}, Brian and {Jha}, Saurabh W. and {Jones}, David O. and {Kirshner}, Robert P. and {Koekemoer}, Anton M. and {McCully}, Curtis and {Mobasher}, Bahram and {Patel}, Brandon and {Weiner}, Benjamin J. and {Cenko}, S. Bradley and {Clubb}, Kelsey I. and {Cooper}, Michael and {Filippenko}, Alexei V. and {Frederiksen}, Teddy F. and {Hjorth}, Jens and {Leibundgut}, Bruno and {Matheson}, Thomas and {Nayyeri}, Hooshang and {Penner}, Kyle and {Trump}, Jonathan and {Silverman}, Jeffrey M. and {U}, Vivian and {Azalee Bostroem}, K. and {Challis}, Peter and {Rajan}, Abhijith and {Wolff}, Schuyler and {Faber}, S.~M. and {Grogin}, Norman A. and {Kocevski}, Dale},
        title = "{Type Ia Supernova Rate Measurements to Redshift 2.5 from CANDELS: Searching for Prompt Explosions in the Early Universe}",
      journal = {\aj},
         year = 2014,
        month = jul,
       volume = {148},
       number = {1},
          eid = {13},
        pages = {13},
          doi = {10.1088/0004-6256/148/1/13},
archivePrefix = {arXiv},
       eprint = {1401.7978},
 primaryClass = {astro-ph.CO},
       adsurl = {https://ui.adsabs.harvard.edu/abs/2014AJ....148...13R}
}

@ARTICLE{grogin11,
       author = {{Grogin}, Norman A. and {Kocevski}, Dale D. and {Faber}, S.~M. and {Ferguson}, Henry C. and {Koekemoer}, Anton M. and {Riess}, Adam G. and {Acquaviva}, Viviana and {Alexander}, David M. and {Almaini}, Omar and {Ashby}, Matthew L.~N. and {Barden}, Marco and {Bell}, Eric F. and {Bournaud}, Fr{\'e}d{\'e}ric and {Brown}, Thomas M. and {Caputi}, Karina I. and {Casertano}, Stefano and {Cassata}, Paolo and {Castellano}, Marco and {Challis}, Peter and {Chary}, Ranga-Ram and {Cheung}, Edmond and {Cirasuolo}, Michele and {Conselice}, Christopher J. and {Roshan Cooray}, Asantha and {Croton}, Darren J. and {Daddi}, Emanuele and {Dahlen}, Tomas and {Dav{\'e}}, Romeel and {de Mello}, Du{\'\i}lia F. and {Dekel}, Avishai and {Dickinson}, Mark and {Dolch}, Timothy and {Donley}, Jennifer L. and {Dunlop}, James S. and {Dutton}, Aaron A. and {Elbaz}, David and {Fazio}, Giovanni G. and {Filippenko}, Alexei V. and {Finkelstein}, Steven L. and {Fontana}, Adriano and {Gardner}, Jonathan P. and {Garnavich}, Peter M. and {Gawiser}, Eric and {Giavalisco}, Mauro and {Grazian}, Andrea and {Guo}, Yicheng and {Hathi}, Nimish P. and {H{\"a}ussler}, Boris and {Hopkins}, Philip F. and {Huang}, Jia-Sheng and {Huang}, Kuang-Han and {Jha}, Saurabh W. and {Kartaltepe}, Jeyhan S. and {Kirshner}, Robert P. and {Koo}, David C. and {Lai}, Kamson and {Lee}, Kyoung-Soo and {Li}, Weidong and {Lotz}, Jennifer M. and {Lucas}, Ray A. and {Madau}, Piero and {McCarthy}, Patrick J. and {McGrath}, Elizabeth J. and {McIntosh}, Daniel H. and {McLure}, Ross J. and {Mobasher}, Bahram and {Moustakas}, Leonidas A. and {Mozena}, Mark and {Nandra}, Kirpal and {Newman}, Jeffrey A. and {Niemi}, Sami-Matias and {Noeske}, Kai G. and {Papovich}, Casey J. and {Pentericci}, Laura and {Pope}, Alexandra and {Primack}, Joel R. and {Rajan}, Abhijith and {Ravindranath}, Swara and {Reddy}, Naveen A. and {Renzini}, Alvio and {Rix}, Hans-Walter and {Robaina}, Aday R. and {Rodney}, Steven A. and {Rosario}, David J. and {Rosati}, Piero and {Salimbeni}, Sara and {Scarlata}, Claudia and {Siana}, Brian and {Simard}, Luc and {Smidt}, Joseph and {Somerville}, Rachel S. and {Spinrad}, Hyron and {Straughn}, Amber N. and {Strolger}, Louis-Gregory and {Telford}, Olivia and {Teplitz}, Harry I. and {Trump}, Jonathan R. and {van der Wel}, Arjen and {Villforth}, Carolin and {Wechsler}, Risa H. and {Weiner}, Benjamin J. and {Wiklind}, Tommy and {Wild}, Vivienne and {Wilson}, Grant and {Wuyts}, Stijn and {Yan}, Hao-Jing and {Yun}, Min S.},
        title = "{CANDELS: The Cosmic Assembly Near-infrared Deep Extragalactic Legacy Survey}",
      journal = {\apjs},
         year = 2011,
        month = dec,
       volume = {197},
       number = {2},
          eid = {35},
        pages = {35},
          doi = {10.1088/0067-0049/197/2/35},
archivePrefix = {arXiv},
       eprint = {1105.3753},
 primaryClass = {astro-ph.CO},
       adsurl = {https://ui.adsabs.harvard.edu/abs/2011ApJS..197...35G}
}

@ARTICLE{postman12,
       author = {{Postman}, Marc and {Coe}, Dan and {Ben{\'\i}tez}, Narciso and {Bradley}, Larry and {Broadhurst}, Tom and {Donahue}, Megan and {Ford}, Holland and {Graur}, Or and {Graves}, Genevieve and {Jouvel}, Stephanie and {Koekemoer}, Anton and {Lemze}, Doron and {Medezinski}, Elinor and {Molino}, Alberto and {Moustakas}, Leonidas and {Ogaz}, Sara and {Riess}, Adam and {Rodney}, Steve and {Rosati}, Piero and {Umetsu}, Keiichi and {Zheng}, Wei and {Zitrin}, Adi and {Bartelmann}, Matthias and {Bouwens}, Rychard and {Czakon}, Nicole and {Golwala}, Sunil and {Host}, Ole and {Infante}, Leopoldo and {Jha}, Saurabh and {Jimenez-Teja}, Yolanda and {Kelson}, Daniel and {Lahav}, Ofer and {Lazkoz}, Ruth and {Maoz}, Dani and {McCully}, Curtis and {Melchior}, Peter and {Meneghetti}, Massimo and {Merten}, Julian and {Moustakas}, John and {Nonino}, Mario and {Patel}, Brandon and {Reg{\"o}s}, Enik{\"o} and {Sayers}, Jack and {Seitz}, Stella and {Van der Wel}, Arjen},
        title = "{The Cluster Lensing and Supernova Survey with Hubble: An Overview}",
      journal = {\apjs},
         year = 2012,
        month = apr,
       volume = {199},
       number = {2},
          eid = {25},
        pages = {25},
          doi = {10.1088/0067-0049/199/2/25},
archivePrefix = {arXiv},
       eprint = {1106.3328},
 primaryClass = {astro-ph.CO},
       adsurl = {https://ui.adsabs.harvard.edu/abs/2012ApJS..199...25P}
}

@ARTICLE{jha25,
       author = {{Roman Observations Time Allocation Committee} and {Core Community Survey Definition Committees}},
        title = "{Roman Observations Time Allocation Committee: Final Report and Recommendations}",
      journal = {arXiv e-prints},
         year = 2025,
        month = may,
          eid = {arXiv:2505.10574},
        pages = {arXiv:2505.10574},
          doi = {10.48550/arXiv.2505.10574},
archivePrefix = {arXiv},
       eprint = {2505.10574},
 primaryClass = {astro-ph.IM},
       adsurl = {https://ui.adsabs.harvard.edu/abs/2025arXiv250510574O}
}

@misc{pierel24_spacephot,
	title = {Space-{Phot}: {Simple} {Python}-{Based} {Photometry} for {Space} {Telescopes}},
	url = {https://doi.org/10.5281/zenodo.12100100},
	publisher = {Zenodo},
	author = {Pierel, Justin},
	month = jun,
	year = {2024},
	doi = {10.5281/zenodo.12100100},
}

@ARTICLE{polzin25,
       author = {{Polzin}, Ava},
        title = "{spike: A Tool to Drizzle HST, JWST, and Roman PSFs for Improved Analyses}",
      journal = {The Journal of Open Source Software},
         year = 2025,
        month = jul,
       volume = {10},
       number = {111},
          eid = {8200},
        pages = {8200},
          doi = {10.21105/joss.08200},
archivePrefix = {arXiv},
       eprint = {2503.02288},
 primaryClass = {astro-ph.IM},
       adsurl = {https://ui.adsabs.harvard.edu/abs/2025JOSS...10.8200P}
}

@ARTICLE{boyer22,
       author = {{Boyer}, Martha L. and {Anderson}, Jay and {Gennaro}, Mario and {Geha}, Marla and {Wingfield McQuinn}, Kristen B. and {Tollerud}, Erik and {Correnti}, Matteo and {Brenner Newman}, Max J. and {Cohen}, Roger E. and {Kallivayalil}, Nitya and {Beaton}, Rachel and {Cole}, Andrew A. and {Dolphin}, Andrew and {Kalirai}, Jason S. and {Sandstrom}, Karin M. and {Savino}, Alessandro and {Skillman}, Evan D. and {Weisz}, Daniel R. and {Williams}, Benjamin F.},
        title = "{The JWST Resolved Stellar Populations Early Release Science Program. I. NIRCam Flux Calibration}",
      journal = {Research Notes of the American Astronomical Society},
         year = 2022,
        month = sep,
       volume = {6},
       number = {9},
          eid = {191},
        pages = {191},
          doi = {10.3847/2515-5172/ac923a},
archivePrefix = {arXiv},
       eprint = {2209.03348},
 primaryClass = {astro-ph.IM},
       adsurl = {https://ui.adsabs.harvard.edu/abs/2022RNAAS...6..191B}
}

@ARTICLE{jha24,
       author = {{Jha}, Saurabh W. and {Casetti-Dinescu}, Dana I. and {Bernstein}, Gary M. and {Hayes}, Matthew J. and {Oskinova}, Lidia M. and {Pace}, Andrew B. and {Quimby}, Robert M. and {Reiter}, Megan and {Rest}, Armin and {Riess}, Adam G. and {Sand}, David J. and {Weisz}, Daniel R.},
        title = "{HST/JWST Long-Term Monitoring Working Group Final Report}",
      journal = {arXiv e-prints},
         year = 2024,
        month = may,
          eid = {arXiv:2405.12297},
        pages = {arXiv:2405.12297},
          doi = {10.48550/arXiv.2405.12297},
archivePrefix = {arXiv},
       eprint = {2405.12297},
 primaryClass = {astro-ph.IM},
       adsurl = {https://ui.adsabs.harvard.edu/abs/2024arXiv240512297J}
}

@ARTICLE{kasen11,
       author = {{Kasen}, Daniel and {Woosley}, S.~E. and {Heger}, Alexander},
        title = "{Pair Instability Supernovae: Light Curves, Spectra, and Shock Breakout}",
      journal = {\apj},
         year = 2011,
        month = jun,
       volume = {734},
       number = {2},
          eid = {102},
        pages = {102},
          doi = {10.1088/0004-637X/734/2/102},
archivePrefix = {arXiv},
       eprint = {1101.3336},
 primaryClass = {astro-ph.HE},
       adsurl = {https://ui.adsabs.harvard.edu/abs/2011ApJ...734..102K}
}

@ARTICLE{Kenworthy2021SALT3,
       author = {{Kenworthy}, W.~D. and {Jones}, D.~O. and {Dai}, M. and {Kessler}, R. and {Scolnic}, D. and {Brout}, D. and {Siebert}, M.~R. and {Pierel}, J.~D.~R. and {Dettman}, K.~G. and {Dimitriadis}, G. and {Foley}, R.~J. and {Jha}, S.~W. and {Pan}, Y. -C. and {Riess}, A. and {Rodney}, S. and {Rojas-Bravo}, C.},
        title = "{SALT3: An Improved Type Ia Supernova Model for Measuring Cosmic Distances}",
      journal = {\apj},
         year = 2021,
        month = dec,
       volume = {923},
       number = {2},
          eid = {265},
        pages = {265},
          doi = {10.3847/1538-4357/ac30d8},
archivePrefix = {arXiv},
       eprint = {2104.07795},
 primaryClass = {astro-ph.CO},
       adsurl = {https://ui.adsabs.harvard.edu/abs/2021ApJ...923..265K}
}

@ARTICLE{Guy2007SALT2,
       author = {{Guy}, J. and {Astier}, P. and {Baumont}, S. and {Hardin}, D. and {Pain}, R. and {Regnault}, N. and {Basa}, S. and {Carlberg}, R.~G. and {Conley}, A. and {Fabbro}, S. and {Fouchez}, D. and {Hook}, I.~M. and {Howell}, D.~A. and {Perrett}, K. and {Pritchet}, C.~J. and {Rich}, J. and {Sullivan}, M. and {Antilogus}, P. and {Aubourg}, E. and {Bazin}, G. and {Bronder}, J. and {Filiol}, M. and {Palanque-Delabrouille}, N. and {Ripoche}, P. and {Ruhlmann-Kleider}, V.},
        title = "{SALT2: using distant supernovae to improve the use of type Ia supernovae as distance indicators}",
      journal = {\aap},
         year = 2007,
        month = apr,
       volume = {466},
       number = {1},
        pages = {11-21},
          doi = {10.1051/0004-6361:20066930},
archivePrefix = {arXiv},
       eprint = {astro-ph/0701828},
 primaryClass = {astro-ph},
       adsurl = {https://ui.adsabs.harvard.edu/abs/2007A&A...466...11G}
}

@ARTICLE{scoville07,
       author = {{Scoville}, N. and {Aussel}, H. and {Brusa}, M. and {Capak}, P. and {Carollo}, C.~M. and {Elvis}, M. and {Giavalisco}, M. and {Guzzo}, L. and {Hasinger}, G. and {Impey}, C. and {Kneib}, J. -P. and {LeFevre}, O. and {Lilly}, S.~J. and {Mobasher}, B. and {Renzini}, A. and {Rich}, R.~M. and {Sanders}, D.~B. and {Schinnerer}, E. and {Schminovich}, D. and {Shopbell}, P. and {Taniguchi}, Y. and {Tyson}, N.~D.},
        title = "{The Cosmic Evolution Survey (COSMOS): Overview}",
      journal = {\apjs},
         year = 2007,
        month = sep,
       volume = {172},
       number = {1},
        pages = {1-8},
          doi = {10.1086/516585},
archivePrefix = {arXiv},
       eprint = {astro-ph/0612305},
 primaryClass = {astro-ph},
       adsurl = {https://ui.adsabs.harvard.edu/abs/2007ApJS..172....1S}
}

@ARTICLE{casey24,
       author = {{Casey}, Caitlin M. and {Akins}, Hollis B. and {Shuntov}, Marko and {Ilbert}, Olivier and {Paquereau}, Louise and {Franco}, Maximilien and {Hayward}, Christopher C. and {Finkelstein}, Steven L. and {Boylan-Kolchin}, Michael and {Robertson}, Brant E. and {Allen}, Natalie and {Brinch}, Malte and {Cooper}, Olivia R. and {Ding}, Xuheng and {Drakos}, Nicole E. and {Faisst}, Andreas L. and {Fujimoto}, Seiji and {Gillman}, Steven and {Harish}, Santosh and {Hirschmann}, Michaela and {Jin}, Shuowen and {Kartaltepe}, Jeyhan S. and {Koekemoer}, Anton M. and {Kokorev}, Vasily and {Liu}, Daizhong and {Long}, Arianna S. and {Magdis}, Georgios and {Maraston}, Claudia and {Martin}, Crystal L. and {McCracken}, Henry Joy and {McKinney}, Jed and {Mobasher}, Bahram and {Rhodes}, Jason and {Rich}, R. Michael and {Sanders}, David B. and {Silverman}, John D. and {Toft}, Sune and {Vijayan}, Aswin P. and {Weaver}, John R. and {Wilkins}, Stephen M. and {Yang}, Lilan and {Zavala}, Jorge A.},
        title = "{COSMOS-Web: Intrinsically Luminous z {\ensuremath{\gtrsim}} 10 Galaxy Candidates Test Early Stellar Mass Assembly}",
      journal = {\apj},
         year = 2024,
        month = apr,
       volume = {965},
       number = {1},
          eid = {98},
        pages = {98},
          doi = {10.3847/1538-4357/ad2075},
archivePrefix = {arXiv},
       eprint = {2308.10932},
 primaryClass = {astro-ph.GA},
       adsurl = {https://ui.adsabs.harvard.edu/abs/2024ApJ...965...98C}
}

@ARTICLE{yan23,
       author = {{Yan}, Haojing and {Ma}, Zhiyuan and {Sun}, Bangzheng and {Wang}, Lifan and {Kelly}, Patrick and {Diego}, Jos{\'e} M. and {Cohen}, Seth H. and {Windhorst}, Rogier A. and {Jansen}, Rolf A. and {Grogin}, Norman A. and {Beacom}, John F. and {Conselice}, Christopher J. and {Driver}, Simon P. and {Frye}, Brenda and {Coe}, Dan and {Marshall}, Madeline A. and {Koekemoer}, Anton and {Willmer}, Christopher N.~A. and {Robotham}, Aaron and {D'Silva}, Jordan C.~J. and {Summers}, Jake and {Nonino}, Mario and {Pirzkal}, Nor and {Ryan}, Russell E. and {Ortiz}, Rafael and {Tompkins}, Scott and {Bhatawdekar}, Rachana A. and {Cheng}, Cheng and {Zitrin}, Adi and {Willner}, S.~P.},
        title = "{JWST's PEARLS: Transients in the MACS J0416.1-2403 Field}",
      journal = {\apjs},
         year = 2023,
        month = dec,
       volume = {269},
       number = {2},
          eid = {43},
        pages = {43},
          doi = {10.3847/1538-4365/ad0298},
archivePrefix = {arXiv},
       eprint = {2307.07579},
 primaryClass = {astro-ph.GA},
       adsurl = {https://ui.adsabs.harvard.edu/abs/2023ApJS..269...43Y}
}

@ARTICLE{chen24,
       author = {{Chen}, Wenlei and {Kelly}, Patrick L. and {Frye}, Brenda L. and {Pierel}, Justin and {Willner}, S.~P. and {Pascale}, Massimo and {Cohen}, Seth H. and {Conselice}, Christopher J. and {Engesser}, Michael and {Furtak}, Lukas J. and {Gilman}, Daniel and {Grogin}, Norman A. and {Huber}, Simon and {Jha}, Saurabh W. and {Johansson}, Joel and {Koekemoer}, Anton M. and {Larison}, Conor and {Meena}, Ashish K. and {Siebert}, Matthew R. and {Windhorst}, Rogier A. and {Yan}, Haojing and {Zitrin}, Adi},
        title = "{JWST Spectroscopy of SN H0pe: Classification and Time Delays of a Triply Imaged Type Ia Supernova at z = 1.78}",
      journal = {\apj},
         year = 2024,
        month = aug,
       volume = {970},
       number = {2},
          eid = {102},
        pages = {102},
          doi = {10.3847/1538-4357/ad50a5},
archivePrefix = {arXiv},
       eprint = {2403.19029},
 primaryClass = {astro-ph.GA},
       adsurl = {https://ui.adsabs.harvard.edu/abs/2024ApJ...970..102C}
}

@ARTICLE{Rest2014,
    author = {{Rest}, A. and {Scolnic}, D. and {Foley}, R.~J. and {Huber}, M.~E. and {Chornock}, R. and {Narayan}, G. and {Tonry}, J.~L. and {Berger}, E. and {Soderberg}, A.~M. and {Stubbs}, C.~W. and {Riess}, A. and {Kirshner}, R.~P. and {Smartt}, S.~J. and {Schlafly}, E. and {Rodney}, S. and {Botticella}, M.~T. and {Brout}, D. and {Challis}, P. and {Czekala}, I. and {Drout}, M. and {Hudson}, M.~J. and {Kotak}, R. and {Leibler}, C. and {Lunnan}, R. and {Marion}, G.~H. and {McCrum}, M. and {Milisavljevic}, D. and {Pastorello}, A. and {Sanders}, N.~E. and {Smith}, K. and {Stafford}, E. and {Thilker}, D. and {Valenti}, S. and {Wood-Vasey}, W.~M. and {Zheng}, Z. and {Burgett}, W.~S. and {Chambers}, K.~C. and {Denneau}, L. and {Draper}, P.~W. and {Flewelling}, H. and {Hodapp}, K.~W. and {Kaiser}, N. and {Kudritzki}, R.-P. and {Magnier}, E.~A. and {Metcalfe}, N. and {Price}, P.~A. and {Sweeney}, W. and {Wainscoat}, R. and {Waters}, C.},
    title = "{Cosmological Constraints from Measurements of Type Ia Supernovae Discovered during the First 1.5 yr of the Pan-STARRS1 Survey}",
    journal = {\apj},
    archivePrefix = "arXiv",
    eprint = {1310.3828},
    year = 2014,
    month = nov,
    volume = 795,
    eid = {44},
    pages = {44},
    doi = {10.1088/0004-637X/795/1/44},
    adsurl = {http://adsabs.harvard.edu/abs/2014ApJ...795...44R}
}

@MISC{Becker_hotpants,
       author = {{Becker}, Andrew},
        title = "{HOTPANTS: High Order Transform of PSF ANd Template Subtraction}",
 howpublished = {Astrophysics Source Code Library, record ascl:1504.004},
         year = 2015,
        month = apr,
          eid = {ascl:1504.004},
        pages = {ascl:1504.004},
archivePrefix = {ascl},
       eprint = {1504.004},
       adsurl = {https://ui.adsabs.harvard.edu/abs/2015ascl.soft04004B}
}

@ARTICLE{decoursey25b,
       author = {{DeCoursey}, Christa and {Egami}, Eiichi and {Sun}, Fengwu and {Akhtarkavan}, Arshia and {Bhatawdekar}, Rachana and {Bunker}, Andrew J. and {Coulter}, David A. and {Engesser}, Michael and {Fox}, Ori D. and {Gomez}, Sebastian and {Inayoshi}, Kohei and {Johnson}, Benjamin D. and {Karmen}, Mitchell and {Larison}, Conor and {Lin}, Xiaojing and {Lyu}, Jianwei and {Mattila}, Seppo and {Moriya}, Takashi J. and {Pierel}, Justin D.~R. and {Pusk{\'a}s}, D{\'a}vid and {Rest}, Armin and {Rieke}, George H. and {Robertson}, Brant and {Salamat}, Sepehr and {Strolger}, Louis-Gregory and {Tacchella}, Sandro and {Vassallo}, Christian and {Williams}, Christina C. and {Zenati}, Yossef and {Zhang}, Junyu},
        title = "{The First Photometric Evidence of a Transient/Variable Source at z>5 with JWST}",
      journal = {arXiv e-prints},
         year = 2025,
        month = apr,
          eid = {arXiv:2504.17007},
        pages = {arXiv:2504.17007},
          doi = {10.48550/arXiv.2504.17007},
archivePrefix = {arXiv},
       eprint = {2504.17007},
 primaryClass = {astro-ph.HE},
       adsurl = {https://ui.adsabs.harvard.edu/abs/2025arXiv250417007D}
}

@ARTICLE{pierel24,
       author = {{Pierel}, J.~D.~R. and {Engesser}, M. and {Coulter}, D.~A. and {DeCoursey}, C. and {Siebert}, M.~R. and {Rest}, A. and {Egami}, E. and {Chen}, W. and {Fox}, O.~D. and {Jones}, D.~O. and {Joshi}, B.~A. and {Moriya}, T.~J. and {Zenati}, Y. and {Bunker}, A.~J. and {Cargile}, P.~A. and {Curti}, M. and {Eisenstein}, D.~J. and {Gezari}, S. and {Gomez}, S. and {Guolo}, M. and {Johnson}, B.~D. and {Karmen}, M. and {Maiolino}, R. and {Quimby}, R.~M. and {Robertson}, B. and {Shahbandeh}, M. and {Strolger}, L.~G. and {Sun}, F. and {Wang}, Q. and {Wevers}, T.},
        title = "{Discovery of an Apparent Red, High-velocity Type Ia Supernova at z = 2.9 with JWST}",
      journal = {\apjl},
         year = 2024,
        month = aug,
       volume = {971},
       number = {2},
          eid = {L32},
        pages = {L32},
          doi = {10.3847/2041-8213/ad6908},
archivePrefix = {arXiv},
       eprint = {2406.05089},
 primaryClass = {astro-ph.GA},
       adsurl = {https://ui.adsabs.harvard.edu/abs/2024ApJ...971L..32P}
}

@ARTICLE{decoursey25a,
       author = {{DeCoursey}, Christa and {Egami}, Eiichi and {Pierel}, Justin D.~R. and {Sun}, Fengwu and {Rest}, Armin and {Coulter}, David A. and {Engesser}, Michael and {Siebert}, Matthew R. and {Hainline}, Kevin N. and {Johnson}, Benjamin D. and {Bunker}, Andrew J. and {Cargile}, Phillip A. and {Charlot}, Stephane and {Chen}, Wenlei and {Curti}, Mirko and {DeFour-Remy}, Shea and {Eisenstein}, Daniel J. and {Fox}, Ori D. and {Gezari}, Suvi and {Gomez}, Sebastian and {Jencson}, Jacob and {Joshi}, Bhavin A. and {Khairnar}, Sanvi and {Lyu}, Jianwei and {Maiolino}, Roberto and {Moriya}, Takashi J. and {Quimby}, Robert M. and {Rieke}, George H. and {Rieke}, Marcia J. and {Robertson}, Brant and {Shahbandeh}, Melissa and {Strolger}, Louis-Gregory and {Tacchella}, Sandro and {Wang}, Qinan and {Williams}, Christina C. and {Willmer}, Christopher N.~A. and {Willott}, Chris and {Zenati}, Yossef},
        title = "{The JADES Transient Survey: Discovery and Classification of Supernovae in the JADES Deep Field}",
      journal = {\apj},
         year = 2025,
        month = feb,
       volume = {979},
       number = {2},
          eid = {250},
        pages = {250},
          doi = {10.3847/1538-4357/ad8fab},
archivePrefix = {arXiv},
       eprint = {2406.05060},
 primaryClass = {astro-ph.HE},
       adsurl = {https://ui.adsabs.harvard.edu/abs/2025ApJ...979..250D}
}

@ARTICLE{siebert24,
       author = {{Siebert}, M.~R. and {DeCoursey}, C. and {Coulter}, D.~A. and {Engesser}, M. and {Pierel}, J.~D.~R. and {Rest}, A. and {Egami}, E. and {Shahbandeh}, M. and {Chen}, W. and {Fox}, O.~D. and {Zenati}, Y. and {Moriya}, T.~J. and {Bunker}, A.~J. and {Cargile}, P.~A. and {Curti}, M. and {Eisenstein}, D.~J. and {Gezari}, S. and {Gomez}, S. and {Guolo}, M. and {Johnson}, B.~D. and {Joshi}, B.~A. and {Karmen}, M. and {Maiolino}, R. and {Quimby}, R.~M. and {Robertson}, B. and {Strolger}, L.~G. and {Sun}, F. and {Wang}, Q. and {Wevers}, T.},
        title = "{Discovery of a Relativistic Stripped-envelope Type Ic-BL Supernova at z = 2.83 with JWST}",
      journal = {\apjl},
         year = 2024,
        month = sep,
       volume = {972},
       number = {1},
          eid = {L13},
        pages = {L13},
          doi = {10.3847/2041-8213/ad6c32},
archivePrefix = {arXiv},
       eprint = {2406.05076},
 primaryClass = {astro-ph.HE},
       adsurl = {https://ui.adsabs.harvard.edu/abs/2024ApJ...972L..13S}
}

@ARTICLE{coulter25,
       author = {{Coulter}, D.~A. and {Pierel}, J.~D.~R. and {DeCoursey}, C. and {Moriya}, T.~J. and {Siebert}, M.~R. and {Joshi}, B.~A. and {Engesser}, M. and {Rest}, A. and {Egami}, E. and {Shahbandeh}, M. and {Chen}, W. and {Fox}, O.~D. and {Strolger}, L.~G. and {Zenati}, Y. and {Bunker}, A.~J. and {Cargile}, P.~A. and {Curti}, M. and {Eisenstein}, D.~J. and {Gezari}, S. and {Gomez}, S. and {Guolo}, M. and {Hainline}, K. and {Jencson}, J. and {Johnson}, B.~D. and {Karmen}, M. and {Maiolino}, R. and {Quimby}, R.~M. and {Rinaldi}, P. and {Robertson}, B. and {Tacchella}, S. and {Sun}, F. and {Wang}, Q. and {Wevers}, T.},
        title = "{Discovery of a likely Type II SN at $z$=3.6 with JWST}",
      journal = {arXiv e-prints},
         year = 2025,
        month = jan,
          eid = {arXiv:2501.05513},
        pages = {arXiv:2501.05513},
          doi = {10.48550/arXiv.2501.05513},
archivePrefix = {arXiv},
       eprint = {2501.05513},
 primaryClass = {astro-ph.HE},
       adsurl = {https://ui.adsabs.harvard.edu/abs/2025arXiv250105513C}
}

@ARTICLE{smith18,
       author = {{Smith}, M. and {Sullivan}, M. and {Nichol}, R.~C. and {Galbany}, L. and {D'Andrea}, C.~B. and {Inserra}, C. and {Lidman}, C. and {Rest}, A. and {Schirmer}, M. and {Filippenko}, A.~V. and {Zheng}, W. and {Cenko}, S. Bradley and {Angus}, C.~R. and {Brown}, P.~J. and {Davis}, T.~M. and {Finley}, D.~A. and {Foley}, R.~J. and {Gonz{\'a}lez-Gait{\'a}n}, S. and {Guti{\'e}rrez}, C.~P. and {Kessler}, R. and {Kuhlmann}, S. and {Marriner}, J. and {M{\"o}ller}, A. and {Nugent}, P.~E. and {Prajs}, S. and {Thomas}, R. and {Wolf}, R. and {Zenteno}, A. and {Abbott}, T.~M.~C. and {Abdalla}, F.~B. and {Allam}, S. and {Annis}, J. and {Bechtol}, K. and {Benoit-L{\'e}vy}, A. and {Bertin}, E. and {Brooks}, D. and {Burke}, D.~L. and {Carnero Rosell}, A. and {Carrasco Kind}, M. and {Carretero}, J. and {Castander}, F.~J. and {Crocce}, M. and {Cunha}, C.~E. and {da Costa}, L.~N. and {Davis}, C. and {Desai}, S. and {Diehl}, H.~T. and {Doel}, P. and {Eifler}, T.~F. and {Flaugher}, B. and {Fosalba}, P. and {Frieman}, J. and {Garc{\'\i}a-Bellido}, J. and {Gaztanaga}, E. and {Gerdes}, D.~W. and {Goldstein}, D.~A. and {Gruen}, D. and {Gruendl}, R.~A. and {Gschwend}, J. and {Gutierrez}, G. and {Honscheid}, K. and {James}, D.~J. and {Johnson}, M.~W.~G. and {Kuehn}, K. and {Kuropatkin}, N. and {Li}, T.~S. and {Lima}, M. and {Maia}, M.~A.~G. and {Marshall}, J.~L. and {Martini}, P. and {Menanteau}, F. and {Miller}, C.~J. and {Miquel}, R. and {Ogando}, R.~L.~C. and {Petravick}, D. and {Plazas}, A.~A. and {Romer}, A.~K. and {Rykoff}, E.~S. and {Sako}, M. and {Sanchez}, E. and {Scarpine}, V. and {Schindler}, R. and {Schubnell}, M. and {Sevilla-Noarbe}, I. and {Smith}, R.~C. and {Soares-Santos}, M. and {Sobreira}, F. and {Suchyta}, E. and {Swanson}, M.~E.~C. and {Tarle}, G. and {Walker}, A.~R. and {DES Collaboration}},
        title = "{Studying the Ultraviolet Spectrum of the First Spectroscopically Confirmed Supernova at Redshift Two}",
      journal = {\apj},
         year = 2018,
        month = feb,
       volume = {854},
       number = {1},
          eid = {37},
        pages = {37},
          doi = {10.3847/1538-4357/aaa126},
archivePrefix = {arXiv},
       eprint = {1712.04535},
 primaryClass = {astro-ph.HE},
       adsurl = {https://ui.adsabs.harvard.edu/abs/2018ApJ...854...37S}
}

@ARTICLE{curtin19,
       author = {{Curtin}, Chris and {Cooke}, Jeff and {Moriya}, Takashi J. and {Tanaka}, Masayuki and {Quimby}, Robert M. and {Bernard}, Stephanie R. and {Galbany}, Llu{\'\i}s and {Jiang}, Ji-an and {Lee}, Chien-Hsiu and {Maeda}, Keiichi and {Morokuma}, Tomoki and {Nomoto}, Ken'ichi and {Pignata}, Giuliano and {Pritchard}, Tyler and {Suzuki}, Nao and {Takahashi}, Ichiro and {Tanaka}, Masaomi and {Tominaga}, Nozomu and {Yamaguchi}, Masaki and {Yasuda}, Naoki},
        title = "{First Release of High-redshift Superluminous Supernovae from the Subaru HIgh-Z SUpernova CAmpaign (SHIZUCA). II. Spectroscopic Properties}",
      journal = {\apjs},
         year = 2019,
        month = apr,
       volume = {241},
       number = {2},
          eid = {17},
        pages = {17},
          doi = {10.3847/1538-4365/ab07c8},
archivePrefix = {arXiv},
       eprint = {1801.08241},
 primaryClass = {astro-ph.HE},
       adsurl = {https://ui.adsabs.harvard.edu/abs/2019ApJS..241...17C}
}

@ARTICLE{moriya19,
       author = {{Moriya}, Takashi J. and {Tanaka}, Masaomi and {Yasuda}, Naoki and {Jiang}, Ji-an and {Lee}, Chien-Hsiu and {Maeda}, Keiichi and {Morokuma}, Tomoki and {Nomoto}, Ken'ichi and {Quimby}, Robert M. and {Suzuki}, Nao and {Takahashi}, Ichiro and {Tanaka}, Masayuki and {Tominaga}, Nozomu and {Yamaguchi}, Masaki and {Bernard}, Stephanie R. and {Cooke}, Jeff and {Curtin}, Chris and {Galbany}, Llu{\'\i}s and {Gonz{\'a}lez-Gait{\'a}n}, Santiago and {Pignata}, Giuliano and {Pritchard}, Tyler and {Komiyama}, Yutaka and {Lupton}, Robert H.},
        title = "{First Release of High-Redshift Superluminous Supernovae from the Subaru HIgh-Z SUpernova CAmpaign (SHIZUCA). I. Photometric Properties}",
      journal = {\apjs},
         year = 2019,
        month = apr,
       volume = {241},
       number = {2},
          eid = {16},
        pages = {16},
          doi = {10.3847/1538-4365/ab07c5},
archivePrefix = {arXiv},
       eprint = {1801.08240},
 primaryClass = {astro-ph.HE},
       adsurl = {https://ui.adsabs.harvard.edu/abs/2019ApJS..241...16M}
}

@ARTICLE{galyam19,
       author = {{Gal-Yam}, Avishay},
        title = "{The Most Luminous Supernovae}",
      journal = {\araa},
         year = 2019,
        month = aug,
       volume = {57},
        pages = {305-333},
          doi = {10.1146/annurev-astro-081817-051819},
archivePrefix = {arXiv},
       eprint = {1812.01428},
 primaryClass = {astro-ph.HE},
       adsurl = {https://ui.adsabs.harvard.edu/abs/2019ARA&A..57..305G}
}

@ARTICLE{cooke09,
       author = {{Cooke}, Jeff and {Sullivan}, Mark and {Barton}, Elizabeth J. and {Bullock}, James S. and {Carlberg}, Ray G. and {Gal-Yam}, Avishay and {Tollerud}, Erik},
        title = "{Type IIn supernovae at redshift z\raisebox{-0.5ex}\textasciitilde2 from archival data}",
      journal = {\nat},
         year = 2009,
        month = jul,
       volume = {460},
       number = {7252},
        pages = {237-239},
          doi = {10.1038/nature08082},
archivePrefix = {arXiv},
       eprint = {0907.1928},
 primaryClass = {astro-ph.CO},
       adsurl = {https://ui.adsabs.harvard.edu/abs/2009Natur.460..237C}
}

@ARTICLE{cooke12,
       author = {{Cooke}, Jeff and {Sullivan}, Mark and {Gal-Yam}, Avishay and {Barton}, Elizabeth J. and {Carlberg}, Raymond G. and {Ryan-Weber}, Emma V. and {Horst}, Chuck and {Omori}, Yuuki and {D{\'\i}az}, C. Gonzalo},
        title = "{Superluminous supernovae at redshifts of 2.05 and 3.90}",
      journal = {\nat},
         year = 2012,
        month = nov,
       volume = {491},
       number = {7423},
        pages = {228-231},
          doi = {10.1038/nature11521},
archivePrefix = {arXiv},
       eprint = {1211.2003},
 primaryClass = {astro-ph.CO},
       adsurl = {https://ui.adsabs.harvard.edu/abs/2012Natur.491..228C}
}

@misc{dunlop_primer_2021,
	title = {{PRIMER}: {Public} {Release} {IMaging} for {Extragalactic} {Research}},
	author = {Dunlop, James S. and Abraham, Roberto G. and Ashby, Matthew L. N. and Bagley, Micaela and Best, Philip N. and Bongiorno, Angela and Bouwens, Rychard and Bowler, Rebecca A. A. and Brammer, Gabriel and Bremer, Malcolm and Calabro', Antonello and Carnall, Adam and Castellano, Marco and Cirasuolo, Michele and Conselice, Christopher and Cullen, Fergus and Dave, Romeel and Dayal, Pratika and Dekel, Avishai and Dickinson, Mark and Duncan, Kenneth James and Elbaz, David and Ellis, Richard S. and Ferguson, Harry C. and Ferrara, Andrea and Finkelstein, Steven L. and Fontana, Adriano and Furlanetto, Steven and Fynbo, Johan P. U. and Gallerani, Simona and Gardner, Jonathan P. and Giavalisco, Mauro and Grazian, Andrea and Grogin, Norman and Harikane, Yuichi and Hopkins, Philip F. and Ilbert, Olivier and Illingworth, Garth D. and Juneau, Stephanie and Jung, Intae and Kartaltepe, Jeyhan and Kassin, Susan and Kauffmann, Olivier Benjamin and Khochfar, Sadegh and Kirkpatrick, Allison and Kocevski, Dale D. and Koekemoer, Anton M. and Labbe, Ivo and Laporte, Nicolas and Larson, Rebecca L. and Lucas, Ray A. and Magee, Daniel K. and Mason, Charlotte and McCracken, Henry Joy and McLeod, Derek and McLure, Ross and Merlin, Emiliano and Mesinger, Andrei and Milvang-Jensen, Bo and Newman, Jeffrey Allen and Oesch, Pascal and Ouchi, Masami and Pacifici, Camilla and Papovich, Casey and Peacock, John and Peeples, Molly and Pentericci, Laura and Perez-Gonzalez, Pablo G. and Pirzkal, Norbert and Pope, Alexandra and Pye, John P. and Reddy, Naveen A. and Robertson, Brant and Salvato, Mara and Santini, Paola and Schaerer, Daniel and Shapley, Alice E. and Simons, Raymond and Smit, Renske and Smith, Britton D. and Snyder, Greg and Somerville, Rachel S. and Stanway, Elizabeth R. and Stefanon, Mauro and Tasca, Lidia and Tikkanen, Tuomo and Tresse, Laurence and Trump, Jonathan R. and Whitaker, Katherine E. and Wilkins, Stephen Matthew and Wright, Gillian and Wyithe, J. Stuart B. and van Dokkum, Pieter and van der Werf, Paul},
	month = mar,
	year = {2021},
	note = {Pages: 1837
Published: JWST Proposal. Cycle 1, ID. \#1837},
}

@software{Rest2023,
       author = {{Rest}, Armin and {Pierel}, Justin and {Correnti}, Matteo and {Canipe}, Alicia and {Hilbert}, Bryan and {Engesser}, Mike and {Sunnquist}, Ben and {Fox}, Ori},
        title = "{arminrest/jhat: The JWST HST Alignment Tool (JHAT)}",
         year = 2023,
        month = may,
          eid = {10.5281/zenodo.7892935},
          doi = {10.5281/zenodo.7892935},
      version = {v2},
    publisher = {Zenodo},
       adsurl = {https://ui.adsabs.harvard.edu/abs/2023zndo...7892935R}
}

@software{Bushouse2023,
       author = {{Bushouse}, Howard and {Eisenhamer}, Jonathan and {Dencheva}, Nadia and {Davies}, James and {Greenfield}, Perry and {Morrison}, Jane and {Hodge}, Phil and {Simon}, Bernie and {Grumm}, David and {Droettboom}, Michael and {Slavich}, Edward and {Sosey}, Megan and {Pauly}, Tyler and {Miller}, Todd and {Jedrzejewski}, Robert and {Hack}, Warren and {Davis}, David and {Crawford}, Steven and {Law}, David and {Gordon}, Karl and {Regan}, Michael and {Cara}, Mihai and {MacDonald}, Ken and {Bradley}, Larry and {Shanahan}, Clare and {Jamieson}, William and {Teodoro}, Mairan and {Williams}, Thomas},
        title = "{JWST Calibration Pipeline}",
         year = 2023,
        month = jan,
          eid = {10.5281/zenodo.7577320},
          doi = {10.5281/zenodo.7577320},
      version = {1.9.4},
    publisher = {Zenodo},
       adsurl = {https://ui.adsabs.harvard.edu/abs/2023zndo...7577320B}
}

@ARTICLE{Bagley2023,
       author = {{Bagley}, Micaela B. and {Finkelstein}, Steven L. and {Koekemoer}, Anton M. and {Ferguson}, Henry C. and {Arrabal Haro}, Pablo and {Dickinson}, Mark and {Kartaltepe}, Jeyhan S. and {Papovich}, Casey and {P{\'e}rez-Gonz{\'a}lez}, Pablo G. and {Pirzkal}, Nor and {Somerville}, Rachel S. and {Willmer}, Christopher N.~A. and {Yang}, Guang and {Yung}, L.~Y. Aaron and {Fontana}, Adriano and {Grazian}, Andrea and {Grogin}, Norman A. and {Hirschmann}, Michaela and {Kewley}, Lisa J. and {Kirkpatrick}, Allison and {Kocevski}, Dale D. and {Lotz}, Jennifer M. and {Medrano}, Aubrey and {Morales}, Alexa M. and {Pentericci}, Laura and {Ravindranath}, Swara and {Trump}, Jonathan R. and {Wilkins}, Stephen M. and {Calabr{\`o}}, Antonello and {Cooper}, M.~C. and {Costantin}, Luca and {de la Vega}, Alexander and {Hilbert}, Bryan and {Hutchison}, Taylor A. and {Larson}, Rebecca L. and {Lucas}, Ray A. and {McGrath}, Elizabeth J. and {Ryan}, Russell and {Wang}, Xin and {Wuyts}, Stijn},
        title = "{CEERS Epoch 1 NIRCam Imaging: Reduction Methods and Simulations Enabling Early JWST Science Results}",
      journal = {\apjl},
         year = 2023,
        month = mar,
       volume = {946},
       number = {1},
          eid = {L12},
        pages = {L12},
          doi = {10.3847/2041-8213/acbb08},
archivePrefix = {arXiv},
       eprint = {2211.02495},
 primaryClass = {astro-ph.IM},
       adsurl = {https://ui.adsabs.harvard.edu/abs/2023ApJ...946L..12B}
}

@ARTICLE{Koekemoer2007,
       author = {{Koekemoer}, A.~M. and {Aussel}, H. and {Calzetti}, D. and {Capak}, P. and {Giavalisco}, M. and {Kneib}, J. -P. and {Leauthaud}, A. and {Le F{\`e}vre}, O. and {McCracken}, H.~J. and {Massey}, R. and {Mobasher}, B. and {Rhodes}, J. and {Scoville}, N. and {Shopbell}, P.~L.},
        title = "{The COSMOS Survey: Hubble Space Telescope Advanced Camera for Surveys Observations and Data Processing}",
      journal = {\apjs},
         year = 2007,
        month = sep,
       volume = {172},
       number = {1},
        pages = {196-202},
          doi = {10.1086/520086},
archivePrefix = {arXiv},
       eprint = {astro-ph/0703095},
 primaryClass = {astro-ph},
       adsurl = {https://ui.adsabs.harvard.edu/abs/2007ApJS..172..196K}
}

@ARTICLE{gaia21,
       author = {{Gaia Collaboration} and {Brown}, A.~G.~A. and {Vallenari}, A. and {Prusti}, T. and {de Bruijne}, J.~H.~J. and {Babusiaux}, C. and {Biermann}, M. and {Creevey}, O.~L. and {Evans}, D.~W. and {Eyer}, L. and {Hutton}, A. and {Jansen}, F. and {Jordi}, C. and {Klioner}, S.~A. and {Lammers}, U. and {Lindegren}, L. and {Luri}, X. and {Mignard}, F. and {Panem}, C. and {Pourbaix}, D. and {Randich}, S. and {Sartoretti}, P. and {Soubiran}, C. and {Walton}, N.~A. and {Arenou}, F. and {Bailer-Jones}, C.~A.~L. and {Bastian}, U. and {Cropper}, M. and {Drimmel}, R. and {Katz}, D. and {Lattanzi}, M.~G. and {van Leeuwen}, F. and {Bakker}, J. and {Cacciari}, C. and {Casta{\~n}eda}, J. and {De Angeli}, F. and {Ducourant}, C. and {Fabricius}, C. and {Fouesneau}, M. and {Fr{\'e}mat}, Y. and {Guerra}, R. and {Guerrier}, A. and {Guiraud}, J. and {Jean-Antoine Piccolo}, A. and {Masana}, E. and {Messineo}, R. and {Mowlavi}, N. and {Nicolas}, C. and {Nienartowicz}, K. and {Pailler}, F. and {Panuzzo}, P. and {Riclet}, F. and {Roux}, W. and {Seabroke}, G.~M. and {Sordo}, R. and {Tanga}, P. and {Th{\'e}venin}, F. and {Gracia-Abril}, G. and {Portell}, J. and {Teyssier}, D. and {Altmann}, M. and {Andrae}, R. and {Bellas-Velidis}, I. and {Benson}, K. and {Berthier}, J. and {Blomme}, R. and {Brugaletta}, E. and {Burgess}, P.~W. and {Busso}, G. and {Carry}, B. and {Cellino}, A. and {Cheek}, N. and {Clementini}, G. and {Damerdji}, Y. and {Davidson}, M. and {Delchambre}, L. and {Dell'Oro}, A. and {Fern{\'a}ndez-Hern{\'a}ndez}, J. and {Galluccio}, L. and {Garc{\'\i}a-Lario}, P. and {Garcia-Reinaldos}, M. and {Gonz{\'a}lez-N{\'u}{\~n}ez}, J. and {Gosset}, E. and {Haigron}, R. and {Halbwachs}, J. -L. and {Hambly}, N.~C. and {Harrison}, D.~L. and {Hatzidimitriou}, D. and {Heiter}, U. and {Hern{\'a}ndez}, J. and {Hestroffer}, D. and {Hodgkin}, S.~T. and {Holl}, B. and {Jan{\ss}en}, K. and {Jevardat de Fombelle}, G. and {Jordan}, S. and {Krone-Martins}, A. and {Lanzafame}, A.~C. and {L{\"o}ffler}, W. and {Lorca}, A. and {Manteiga}, M. and {Marchal}, O. and {Marrese}, P.~M. and {Moitinho}, A. and {Mora}, A. and {Muinonen}, K. and {Osborne}, P. and {Pancino}, E. and {Pauwels}, T. and {Petit}, J. -M. and {Recio-Blanco}, A. and {Richards}, P.~J. and {Riello}, M. and {Rimoldini}, L. and {Robin}, A.~C. and {Roegiers}, T. and {Rybizki}, J. and {Sarro}, L.~M. and {Siopis}, C. and {Smith}, M. and {Sozzetti}, A. and {Ulla}, A. and {Utrilla}, E. and {van Leeuwen}, M. and {van Reeven}, W. and {Abbas}, U. and {Abreu Aramburu}, A. and {Accart}, S. and {Aerts}, C. and {Aguado}, J.~J. and {Ajaj}, M. and {Altavilla}, G. and {{\'A}lvarez}, M.~A. and {{\'A}lvarez Cid-Fuentes}, J. and {Alves}, J. and {Anderson}, R.~I. and {Anglada Varela}, E. and {Antoja}, T. and {Audard}, M. and {Baines}, D. and {Baker}, S.~G. and {Balaguer-N{\'u}{\~n}ez}, L. and {Balbinot}, E. and {Balog}, Z. and {Barache}, C. and {Barbato}, D. and {Barros}, M. and {Barstow}, M.~A. and {Bartolom{\'e}}, S. and {Bassilana}, J. -L. and {Bauchet}, N. and {Baudesson-Stella}, A. and {Becciani}, U. and {Bellazzini}, M. and {Bernet}, M. and {Bertone}, S. and {Bianchi}, L. and {Blanco-Cuaresma}, S. and {Boch}, T. and {Bombrun}, A. and {Bossini}, D. and {Bouquillon}, S. and {Bragaglia}, A. and {Bramante}, L. and {Breedt}, E. and {Bressan}, A. and {Brouillet}, N. and {Bucciarelli}, B. and {Burlacu}, A. and {Busonero}, D. and {Butkevich}, A.~G. and {Buzzi}, R. and {Caffau}, E. and {Cancelliere}, R. and {C{\'a}novas}, H. and {Cantat-Gaudin}, T. and {Carballo}, R. and {Carlucci}, T. and {Carnerero}, M.~I. and {Carrasco}, J.~M. and {Casamiquela}, L. and {Castellani}, M. and {Castro-Ginard}, A. and {Castro Sampol}, P. and {Chaoul}, L. and {Charlot}, P. and {Chemin}, L. and {Chiavassa}, A. and {Cioni}, M. -R.~L. and {Comoretto}, G. and {Cooper}, W.~J. and {Cornez}, T. and {Cowell}, S. and {Crifo}, F. and {Crosta}, M. and {Crowley}, C. and {Dafonte}, C. and {Dapergolas}, A. and {David}, M. and {David}, P. and {de Laverny}, P. and {De Luise}, F. and {De March}, R. and {De Ridder}, J. and {de Souza}, R. and {de Teodoro}, P. and {de Torres}, A. and {del Peloso}, E.~F. and {del Pozo}, E. and {Delbo}, M. and {Delgado}, A. and {Delgado}, H.~E. and {Delisle}, J. -B. and {Di Matteo}, P. and {Diakite}, S. and {Diener}, C. and {Distefano}, E. and {Dolding}, C. and {Eappachen}, D. and {Edvardsson}, B. and {Enke}, H. and {Esquej}, P. and {Fabre}, C. and {Fabrizio}, M. and {Faigler}, S. and {Fedorets}, G. and {Fernique}, P. and {Fienga}, A. and {Figueras}, F. and {Fouron}, C. and {Fragkoudi}, F. and {Fraile}, E. and {Franke}, F. and {Gai}, M. and {Garabato}, D. and {Garcia-Gutierrez}, A. and {Garc{\'\i}a-Torres}, M. and {Garofalo}, A. and {Gavras}, P. and {Gerlach}, E. and {Geyer}, R. and {Giacobbe}, P. and {Gilmore}, G. and {Girona}, S. and {Giuffrida}, G. and {Gomel}, R. and {Gomez}, A. and {Gonzalez-Santamaria}, I. and {Gonz{\'a}lez-Vidal}, J.~J. and {Granvik}, M. and {Guti{\'e}rrez-S{\'a}nchez}, R. and {Guy}, L.~P. and {Hauser}, M. and {Haywood}, M. and {Helmi}, A. and {Hidalgo}, S.~L. and {Hilger}, T. and {H{\l}adczuk}, N. and {Hobbs}, D. and {Holland}, G. and {Huckle}, H.~E. and {Jasniewicz}, G. and {Jonker}, P.~G. and {Juaristi Campillo}, J. and {Julbe}, F. and {Karbevska}, L. and {Kervella}, P. and {Khanna}, S. and {Kochoska}, A. and {Kontizas}, M. and {Kordopatis}, G. and {Korn}, A.~J. and {Kostrzewa-Rutkowska}, Z. and {Kruszy{\'n}ska}, K. and {Lambert}, S. and {Lanza}, A.~F. and {Lasne}, Y. and {Le Campion}, J. -F. and {Le Fustec}, Y. and {Lebreton}, Y. and {Lebzelter}, T. and {Leccia}, S. and {Leclerc}, N. and {Lecoeur-Taibi}, I. and {Liao}, S. and {Licata}, E. and {Lindstr{\o}m}, E.~P. and {Lister}, T.~A. and {Livanou}, E. and {Lobel}, A. and {Madrero Pardo}, P. and {Managau}, S. and {Mann}, R.~G. and {Marchant}, J.~M. and {Marconi}, M. and {Marcos Santos}, M.~M.~S. and {Marinoni}, S. and {Marocco}, F. and {Marshall}, D.~J. and {Martin Polo}, L. and {Mart{\'\i}n-Fleitas}, J.~M. and {Masip}, A. and {Massari}, D. and {Mastrobuono-Battisti}, A. and {Mazeh}, T. and {McMillan}, P.~J. and {Messina}, S. and {Michalik}, D. and {Millar}, N.~R. and {Mints}, A. and {Molina}, D. and {Molinaro}, R. and {Moln{\'a}r}, L. and {Montegriffo}, P. and {Mor}, R. and {Morbidelli}, R. and {Morel}, T. and {Morris}, D. and {Mulone}, A.~F. and {Munoz}, D. and {Muraveva}, T. and {Murphy}, C.~P. and {Musella}, I. and {Noval}, L. and {Ord{\'e}novic}, C. and {Orr{\`u}}, G. and {Osinde}, J. and {Pagani}, C. and {Pagano}, I. and {Palaversa}, L. and {Palicio}, P.~A. and {Panahi}, A. and {Pawlak}, M. and {Pe{\~n}alosa Esteller}, X. and {Penttil{\"a}}, A. and {Piersimoni}, A.~M. and {Pineau}, F. -X. and {Plachy}, E. and {Plum}, G. and {Poggio}, E. and {Poretti}, E. and {Poujoulet}, E. and {Pr{\v{s}}a}, A. and {Pulone}, L. and {Racero}, E. and {Ragaini}, S. and {Rainer}, M. and {Raiteri}, C.~M. and {Rambaux}, N. and {Ramos}, P. and {Ramos-Lerate}, M. and {Re Fiorentin}, P. and {Regibo}, S. and {Reyl{\'e}}, C. and {Ripepi}, V. and {Riva}, A. and {Rixon}, G. and {Robichon}, N. and {Robin}, C. and {Roelens}, M. and {Rohrbasser}, L. and {Romero-G{\'o}mez}, M. and {Rowell}, N. and {Royer}, F. and {Rybicki}, K.~A. and {Sadowski}, G. and {Sagrist{\`a} Sell{\'e}s}, A. and {Sahlmann}, J. and {Salgado}, J. and {Salguero}, E. and {Samaras}, N. and {Sanchez Gimenez}, V. and {Sanna}, N. and {Santove{\~n}a}, R. and {Sarasso}, M. and {Schultheis}, M. and {Sciacca}, E. and {Segol}, M. and {Segovia}, J.~C. and {S{\'e}gransan}, D. and {Semeux}, D. and {Shahaf}, S. and {Siddiqui}, H.~I. and {Siebert}, A. and {Siltala}, L. and {Slezak}, E. and {Smart}, R.~L. and {Solano}, E. and {Solitro}, F. and {Souami}, D. and {Souchay}, J. and {Spagna}, A. and {Spoto}, F. and {Steele}, I.~A. and {Steidelm{\"u}ller}, H. and {Stephenson}, C.~A. and {S{\"u}veges}, M. and {Szabados}, L. and {Szegedi-Elek}, E. and {Taris}, F. and {Tauran}, G. and {Taylor}, M.~B. and {Teixeira}, R. and {Thuillot}, W. and {Tonello}, N. and {Torra}, F. and {Torra}, J. and {Turon}, C. and {Unger}, N. and {Vaillant}, M. and {van Dillen}, E. and {Vanel}, O. and {Vecchiato}, A. and {Viala}, Y. and {Vicente}, D. and {Voutsinas}, S. and {Weiler}, M. and {Wevers}, T. and {Wyrzykowski}, {\L}. and {Yoldas}, A. and {Yvard}, P. and {Zhao}, H. and {Zorec}, J. and {Zucker}, S. and {Zurbach}, C. and {Zwitter}, T.},
        title = "{Gaia Early Data Release 3. Summary of the contents and survey properties}",
      journal = {\aap},
         year = 2021,
        month = may,
       volume = {649},
          eid = {A1},
        pages = {A1},
          doi = {10.1051/0004-6361/202039657},
archivePrefix = {arXiv},
       eprint = {2012.01533},
 primaryClass = {astro-ph.GA},
       adsurl = {https://ui.adsabs.harvard.edu/abs/2021A&A...649A...1G}
}

@ARTICLE{Angulo25,
       author = {{Angulo}, Rodrigo and {Rest}, Armin and {Blair}, William P. and {Jencson}, Jacob and {Coulter}, David A. and {Wang}, Qinan and {Foley}, Ryan J. and {Kilpatrick}, Charles D. and {Li}, Xiaolong and {Piro}, Anthony L. and {Rojas-Bravo}, C{\'e}sar},
        title = "{Optimizing Convolution Direction and Template Selection for Difference Image Analysis}",
      journal = {\apjs},
         year = 2025,
        month = sep,
       volume = {280},
       number = {1},
          eid = {29},
        pages = {29},
          doi = {10.3847/1538-4365/adf05b},
archivePrefix = {arXiv},
       eprint = {2508.10155},
 primaryClass = {astro-ph.IM},
       adsurl = {https://ui.adsabs.harvard.edu/abs/2025ApJS..280...29A}
}

@software{Bradley2024Astropy/photutils:1.12.0,
       author = {{Bradley}, Larry and {Sip{\H{o}}cz}, Brigitta and {Robitaille}, Thomas and {Tollerud}, Erik and {Vin{\'\i}cius}, Z{\'e} and {Deil}, Christoph and {Barbary}, Kyle and {Wilson}, Tom J and {Busko}, Ivo and {Donath}, Axel and {G{\"u}nther}, Hans Moritz and {Cara}, Mihai and {Lim}, P.~L. and {Me{\ss}linger}, Sebastian and {Burnett}, Zach and {Conseil}, Simon and {Droettboom}, Michael and {Bostroem}, Azalee and {Bray}, E.~M. and {Andersen Bratholm}, Lars and {Jamieson}, William and {Ginsburg}, Adam and {Barentsen}, Geert and {Craig}, Matt and {Pascual}, Sergio and {Rathi}, Shivangee and {Perrin}, Marshall and {Morris}, Brett M. and {Perren}, Gabriel},
        title = "{astropy/photutils: 1.12.0}",
         year = 2024,
        month = apr,
          eid = {10.5281/zenodo.596036},
          NOOPdoi = {10.5281/zenodo.596036},
      version = {1.12.0},
    publisher = {Zenodo},
       NOOPadsurl = {https://ui.adsabs.harvard.edu/abs/2022zndo....596036B}
}

@ARTICLE{Stetson1987DAOPHOT:Photometry,
       author = {{Stetson}, Peter B.},
        title = "{DAOPHOT: A Computer Program for Crowded-Field Stellar Photometry}",
      journal = {\pasp},
         year = 1987,
        month = mar,
       volume = {99},
        pages = {191},
          NOOPdoi = {10.1086/131977},
       NOOPadsurl = {https://ui.adsabs.harvard.edu/abs/1987PASP...99..191S}
}

@ARTICLE{Franco25,
       author = {{Franco}, Maximilien and {Casey}, Caitlin M. and {Koekemoer}, Anton M. and {Liu}, Daizhong and {Bagley}, Micaela B. and {McCracken}, Henry Joy and {Kartaltepe}, Jeyhan S. and {Akins}, Hollis B. and {Ilbert}, Olivier and {Shuntov}, Marko and {Harish}, Santosh and {Robertson}, Brant E. and {Arango-Toro}, Rafael C. and {Battisti}, Andrew J. and {Chartab}, Nima and {Drakos}, Nicole E. and {Faisst}, Andreas L. and {Flayhart}, Carter and {Gozaliasl}, Ghassem and {Hirschmann}, Michaela and {Massey}, Richard and {Rhodes}, Jason and {Sattari}, Zahra and {Scognamiglio}, Diana and {Weaver}, John R. and {Yang}, Lilan and {Zavala}, Jorge A. and {Berman}, Edward M. and {Gentile}, Fabrizio and {Gillman}, Steven and {Long}, Arianna S. and {Magdis}, Georgios and {McCleary}, Jacqueline E. and {McKinney}, Jed and {Mobasher}, Bahram and {Paquereau}, Louise and {Rest}, Armin and {Sanders}, David B. and {Toft}, Sune and {Yu}, Si-Yue},
        title = "{COSMOS-Web: Comprehensive Data Reduction for Wide-Area JWST NIRCam Imaging}",
      journal = {arXiv e-prints},
         year = 2025,
        month = jun,
          eid = {arXiv:2506.03256},
        pages = {arXiv:2506.03256},
          doi = {10.48550/arXiv.2506.03256},
archivePrefix = {arXiv},
       eprint = {2506.03256},
 primaryClass = {astro-ph.IM},
       adsurl = {https://ui.adsabs.harvard.edu/abs/2025arXiv250603256F}
}

@inproceedings{Morrell2012CarnegieSupernovae,
    title = {{Carnegie supernova project: Spectroscopic observations of core collapse supernovae}},
    year = {2012},
    booktitle = {Death of massive stars: Supernovae and gamma-ray bursts},
    author = {Morrell, Nidia I},
    editor = {Roming, P and Kawai, N and Pian, E},
    month = {9},
    pages = {361--362},
    series = {IAU symposium},
    volume = {279},
    doi = {10.1017/S174392131201335X}
}

@article{Pierel2018ExtendingObservations,
    title = {{Extending Supernova Spectral Templates for Next-generation Space Telescope Observations}},
    year = {2018},
    journal = {Publications of the Astronomical Society of the Pacific},
    author = {Pierel, J D R and Rodney, S and Avelino, A and Bianco, F and Filippenko, A V and Foley, R J and Friedman, A and Hicken, M and Hounsell, R and Jha, S W and Kessler, R and Kirshner, R P and Mandel, K and Narayan, G and Scolnic, D and Strolger, L},
    number = {993},
    month = {11},
    pages = {114504},
    volume = {130},
    url = {http://stacks.iop.org/1538-3873/130/i=993/a=114504?key=crossref.44ee315a73acf8a52666778cf6cc1881},
    doi = {10.1088/1538-3873/aadb7a},
    issn = {0004-6280, 1538-3873},
    language = {en}
}
\bibliographystyle{aasjournalv7}

\end{document}